\documentclass[twocolumn]{aastex62}
\usepackage{natbib}
\graphicspath{{./}{figures/}}
\usepackage{color,subfigure,lineno,booktabs} 

\newcommand{\OIIIab}{[O{\sc iii}]\,$\lambda\lambda$4959,5007}

\newcommand{\hbeta}{H{$\beta$}}
\newcommand{\halpha}{H{$\alpha$}}

\def\CIV{C\,{\sc iv}}
\def\CIII{C\,{\sc iii]}}
\def\HeII{He\,{\sc ii}}
\def\CIV{C\,{\sc iv}}
\def\MgII{Mg\,{\sc ii}}
\def\SiIV{Si\,{\sc iv}}
\def\FeII{Fe\,{\sc ii}}
\def\OIII{[O\,{\sc iii}]\,5007}

\def\kms{{\rm km\,s^{-1}}}
\def\pyroa{{\tt PyROA}\ }
\def\javelin{{\tt Javelin}\ }

\shorttitle{SDSS-RM: Key Results} 
\shortauthors{Shen et~al.}

\begin{document}

\title{The Sloan Digital Sky Survey Reverberation Mapping Project: Key Results}


\author[0000-0003-1659-7035]{Yue Shen}
\affiliation{Department of Astronomy, University of Illinois at Urbana-Champaign, Urbana, IL 61801, USA}
\affiliation{National Center for Supercomputing Applications, University of Illinois at Urbana-Champaign, Urbana, IL 61801, USA}

\author[0000-0001-9920-6057]{Catherine J. Grier}
\affiliation{Department of Astronomy, University of Wisconsin-Madison, 475 North Charter Street, Madison, WI 53703, USA}

\author[0000-0003-1728-0304]{Keith Horne}
\affiliation{SUPA Physics and Astronomy, University of St Andrews, Fife, KY16 9SS, UK}

\author[0000-0002-8501-3518]{Zachary Stone}
\affiliation{Department of Astronomy, University of Illinois at Urbana-Champaign, Urbana, IL 61801, USA}
\affiliation{Center for AstroPhysical Surveys, National Center for Supercomputing Applications, University of Illinois at Urbana-Champaign, Urbana, IL 61801, USA}

\author[0000-0002-0311-2812]{Jennifer~I.~Li}
\affiliation{Michigan Institute for Data Science, University of Michigan, Ann Arbor, MI, 48109, USA}
\affiliation{Department of Astronomy, University of Michigan, Ann Arbor, MI, 48109, USA}

\author[0000-0002-6893-3742]{Qian Yang}
\affiliation{Center for Astrophysics $\vert$ Harvard \& Smithsonian, 60 Garden Street, Cambridge, MA 02138, USA}
\affiliation{Department of Astronomy, University of Illinois at Urbana-Champaign, Urbana, IL 61801, USA}

\author[0000-0002-0957-7151]{Yasaman Homayouni}
\affiliation{Space Telescope Science Institute, 3700 San Martin Drive, Baltimore, MD 21218, USA}
\affiliation{Department of Astronomy and Astrophysics, 525 Davey Lab, The Pennsylvania State University, University Park, PA 16802, USA}

\author[0000-0002-1410-0470]{Jonathan R. Trump}
\affil{Department of Physics, 196A Auditorium Road, Unit 3046, University of Connecticut, Storrs, CT 06269, USA}

\author[0000-0002-6404-9562]{Scott F. Anderson}
\affiliation{Astronomy Department, University of Washington, Box
351580, Seattle, WA 98195, USA}

\author[0000-0002-0167-2453]{W.~N. Brandt}
\affiliation{Department of Astronomy and Astrophysics, 525 Davey Lab, The Pennsylvania State University, University Park, PA 16802, USA}
\affiliation{Institute for Gravitation and the Cosmos, The Pennsylvania State University, University Park, PA 16802, USA}
\affiliation{Department of Physics, 104 Davey Laboratory, The Pennsylvania State University, University Park, PA 16802, USA}

\author[0000-0002-1763-5825]{Patrick B. Hall}
\affiliation{Department of Physics \& Astronomy, York University, 4700 Keele St., Toronto, ON M3J 1P3, Canada}

\author[0000-0001-6947-5846]{Luis C. Ho}
\affiliation{Kavli Institute for Astronomy and Astrophysics, Peking University, Beijing 100871, China}
\affiliation{Department of Astronomy, School of Physics, Peking University, Beijing 100871, China}

\author[0000-0003-4176-6486]{Linhua Jiang}
\affiliation{Kavli Institute for Astronomy and Astrophysics, Peking University, Beijing 100871, China}
\affiliation{Department of Astronomy, School of Physics, Peking University, Beijing 100871, China}

\author{Patrick Petitjean}
\affiliation{Institut d'Astrophysique de Paris \& Sorbonne Universit\'e, 98bis Boulevard Arago 75014, Paris, France}

\author[0000-0001-7240-7449]{Donald P. Schneider}
\affiliation{Department of Astronomy and Astrophysics, Penn State, University Park, PA, 16802}
\affiliation{Institute for Gravitation and the Cosmos, Penn State, University Park, PA, 16802}

\author[0000-0001-7961-8177]{Charling Tao}
\affiliation{Center for Astrophysics, Department of Astronomy, Tsinghua University, Beijing, China}
\affiliation{Centre de Physique des Particules de Marseille, IN2P3, CNRS, France}

\author{Fergus. R. Donnan}
\affiliation{Department of Physics, University of Oxford, Keble Road, Oxford, OX1 3RH, UK}

\author[0009-0008-9216-7516]{Yusra AlSayyad}
\affiliation{Department of Astrophysical Sciences, Princeton University, Princeton, NJ, 08544, USA}

\author[0000-0002-3131-4374]{Matthew A. Bershady}
\affiliation{Department of Astronomy, University of Wisconsin-Madison, 475 North Charter Street, Madison, WI 53703, USA}
\affiliation{South African Astronomical Observatory, P.O. Box 9, Observatory 7935, Cape Town, South Africa}
\affiliation{Department of Astronomy, University of Cape Town, Private Bag X3, Rondebosch 7701, South Africa}

\author[0000-0003-1641-6222]{Michael R. Blanton}
\affiliation{Center for Cosmology and Particle Physics, Department of Physics, 726 Broadway, Room 1005, New York University, New York, NY 10003, USA}

\author{Dmitry Bizyaev}
\affiliation{Apache Point Observatory and New Mexico State University, P.O. Box 59, Sunspot, NM, 88349-0059, USA}
\affiliation{Sternberg Astronomical Institute, Moscow State University, Moscow}

\author[0000-0001-9742-3138]{Kevin Bundy}
\affiliation{UCO/Lick Observatory, University of California, Santa Cruz, 1156 High Street, Santa Cruz, CA 95064, USA}

\author[0000-0003-4520-5395]{Yuguang Chen}
\affiliation{Department of Physics \& Astronomy, University of California, Davis, CA 95616, USA}

\author[0000-0001-9776-9227]{Megan C. Davis}
\affiliation{Department of Physics, 196A Auditorium Road, Unit 3046, University of Connecticut, Storrs, CT 06269, USA}

\author[0000-0002-0553-3805]{Kyle Dawson}
\affiliation{Department of Physics and Astronomy, University of Utah, 115 South 1400 East, Salt Lake City, UT 84112, USA}

\author[0000-0003-3310-0131]{Xiaohui Fan}
\affil{Steward Observatory, University of Arizona, Tucson, AZ 85721, USA}

\author[0000-0002-5612-3427]{Jenny E. Greene}
\affiliation{Department of Astrophysical Sciences, Princeton University, Princeton, NJ 08544, USA}

\author[0000-0001-6327-2757]{Hannes Gr\"oller}
\affiliation{Lunar and Planetary Laboratory, University of Arizona, Tucson, AZ 85721, USA}

\author[0000-0002-6137-0422]{Yucheng Guo}
\affiliation{Department of Astronomy, School of Physics, Peking University, Beijing 100871, China}
\affiliation{Univ Lyon, Univ Lyon1, Ens de Lyon, CNRS, Centre de Recherche Astrophysique de Lyon UMR5574, F-69230, Saint-Genis-Laval, France}

\author[0000-0002-9790-6313]{H\'ector Ibarra-Medel}
\affiliation{Escuela Superior de F\'{\i}sica y Matem\'aticas, Instituto Polit\'ecnico Nacional, U.P. Adolfo L\'opez Mateos, C.P. 07738, Ciudad de M\'exico, M\'exico}
\affiliation{Universidad Nacional Aut\'onoma de M\'exico, Instituto de Astronom\'ia, AP 70-264, CDMX 04510, Mexico}

\author{Yuanzhe Jiang}
\affiliation{Department of Astronomy and Astrophysics, 525 Davey Lab, The Pennsylvania State University, University Park, PA 16802, USA}

\author[0000-0003-1859-9640]{Ryan P. Keenan}
\affiliation{Steward Observatory, University of Arizona, Tucson, AZ 85721, USA}

\author{Juna A. Kollmeier}
\affiliation{The Observatories of the Carnegie Institution for Science, 813 Santa Barbara Street, Pasadena, CA 91101, USA}
\affiliation{Canadian Institute for Theoretical Astrophysics, University of Toronto, Toronto, ON M5S-98H, Canada}

\author[0000-0003-0165-7701]{Cassandra Lejoly}
\affiliation{Lunar and Planetary Laboratory, University of Arizona, Tucson, AZ 85721, USA}

\author[0000-0001-7373-3115]{Zefeng Li}
\affiliation{Research School of Astronomy \& Astrophysics, Australian National University, Weston Creek, ACT 2611, Australia}

\author{Axel de la Macorra}
\affiliation{Instituto de F\'{i}sica, Universidad Nacional Autónoma de M\'{e}xico, Apdo. Postal 20-364, M\'{e}xico}

\author{Maxwell Moe} 
\affiliation{Steward Observatory, University of Arizona, Tucson, AZ 85721, USA}
\affiliation{Department of Physics \& Astronomy, University of Wyoming, Laramie, WY 82071, USA}

\author{Jundan Nie}
\affiliation{Key Laboratory of Optical Astronomy, National Astronomical Observatories, Chinese Academy of Sciences, Beijing100012, People's Republic of China} 

\author{Graziano Rossi}
\affiliation{Department of Astronomy and Space Science, Sejong University, 209, Neungdong-ro, Gwangjin-gu, Seoul, South Korea}

\author[0000-0002-5083-3663]{Paul S. Smith}
\affiliation{Steward Observatory, University of Arizona, Tucson, AZ 85721, USA}

\author[0000-0003-0747-1780]{Wei Leong Tee}
\affiliation{Steward Observatory, University of Arizona, Tucson, AZ 85721, USA}

\author[0000-0002-5908-6852]{Anne-Marie Weijmans}
\affiliation{SUPA Physics and Astronomy, University of St Andrews, Fife, KY16 9SS, UK}

\author[0000-0003-0871-8941]{Jiachuan Xu}
\affiliation{Steward Observatory, University of Arizona, Tucson, AZ 85721, USA}

\author{Minghao Yue} 
\affiliation{Steward Observatory, University of Arizona, Tucson, AZ 85721, USA}

\author{Xu Zhou}
\affiliation{Key Laboratory of Optical Astronomy, National Astronomical Observatories, Chinese Academy of Sciences, Beijing100012, People's Republic of China} 

\author{Zhimin Zhou}
\affiliation{Key Laboratory of Optical Astronomy, National Astronomical Observatories, Chinese Academy of Sciences, Beijing100012, People's Republic of China}

\author{Hu Zou}
\affiliation{Key Laboratory of Optical Astronomy, National Astronomical Observatories, Chinese Academy of Sciences, Beijing100012, People's Republic of China} 

\begin{abstract}
We present the final data from the Sloan Digital Sky Survey Reverberation Mapping (SDSS-RM) project, a precursor to the SDSS-V Black Hole Mapper Reverberation Mapping program. This data set includes 11-year photometric and 7-year spectroscopic light curves for 849 broad-line quasars over a redshift range of $0.1<z< 4.5$ and a luminosity range of $L_{\rm bol}=10^{44-47.5}\,{\rm erg\,s^{-1}}$, along with spectral and variability measurements. We report 23, 81, 125, and 110 reverberation mapping lags (relative to optical continuum variability) for broad \halpha, \hbeta, \MgII\ and \CIV\ using the SDSS-RM sample, spanning much of the luminosity and redshift ranges of the sample. Using 30 low-redshift RM AGNs with dynamical-modeling black hole masses, we derive a new estimate of the average virial factor of $\left<\log f\right>=0.62\pm0.07$ for the line dispersion measured from the RMS spectrum. The intrinsic scatter of individual virial factors is $0.31\pm0.07$\,dex, indicating a factor of two systematic uncertainty in RM black hole masses. Our lag measurements reveal significant $R-L$ relations for \hbeta\ and \MgII\ at high redshift, consistent with the latest measurements based on heterogeneous samples. While we are unable to robustly constrain the slope of the $R-L$ relation for \CIV\ given the limited dynamic range in luminosity, we found substantially larger scatter in \CIV\ lags at fixed $L_{1350}$. Using the SDSS-RM lag sample, we derive improved single-epoch (SE) mass recipes for \hbeta, \MgII\ and \CIV, which are consistent with their respective RM masses as well as between the SE recipes from two different lines, over the luminosity range probed by our sample. The new \hbeta\ and \MgII\ recipes are approximately unbiased estimators at given RM masses, but there are systematic biases in the \CIV\ recipe. The intrinsic scatter of SE masses around RM masses is $\sim 0.45$~dex for \hbeta\ and \MgII, increasing to $\sim 0.58$~dex for \CIV. 
\end{abstract}
\keywords{black hole physics --- galaxies: active --- quasars: general --- surveys}

\section{Introduction}\label{sec:introduction}

Accurate measurements of supermassive black hole (hereafter SMBH or BH for short) masses are cornerstones for understanding the cosmic assembly of SMBHs, their potential co-evolution with host galaxies, and fundamental accretion physics. Far beyond the nearby universe, it is difficult to use spatially resolved gas or stellar kinematics to measure a dynamical mass of the accreting SMBH. Currently, the primary method to measure black hole masses in distant AGNs or quasars is reverberation mapping \citep[e.g.,][]{Blandford_McKee_1982,Peterson_1993,Cackett_etal_2021}. RM measures a characteristic size of the broad-line region (BLR) from the time lag between continuum variability and the response in the broad-line flux. By combining the size of the BLR (assumed to be viralized) with the broad-line width (as the surrogate for the virial velocity), one can derive a virial black hole mass estimate for the AGN. RM results for local AGN provide the foundation of the so-called ``single-epoch'' method \citep[e.g.,][]{Greene_Ho_2005,Vestergaard_Peterson_2006,Shen_2013} that allows efficient estimation of quasar black-hole masses using single-epoch spectroscopy \citep[e.g.,][]{Shen_etal_2008,Shen_etal_2011,Wu_Shen_2022}. This single-epoch technique is built on the observed tight relation between the \hbeta\ BLR size and the optical luminosity of the AGN for the local RM sample \citep[e.g.,][]{Kaspi_etal_2000,Bentz_etal_2013}, although some recent RM studies targeting a broad range of AGN properties have shown deviations and increased scatter in the $R-L$ relation \citep[e.g.,][]{Du_etal_2016,Fonseca_Alvarez_etal_2020}. The systematic uncertainties of extrapolating the local RM results to distant, luminous quasars, however, are not yet well quantified \citep[e.g., see the review in][]{Shen_2013}.

In the past few years, there have been two major advancements in the RM field. On the one hand, high-quality monitoring data have become available for bright, low-redshift (typically $z\lesssim 0.3$) AGNs to measure broad-line RM lags with unprecedented precision, and to cover a broader range of accretion parameters \citep[e.g.,][]{Barth_etal_2013, Fausnaugh_etal_2016, Du_etal_2016,Du_etal_2018,Brotherton_etal_2020,Kara_etal_2021}. These high-quality RM data further enabled more powerful constraints on the dynamics of the BLR with velocity-resolved RM \citep[e.g.,][]{Grier_etal_2013,DeRosa_etal_2018}, or dynamical modeling of the BLR response to continuum variations \citep[e.g.,][]{Pancoast_etal_2014,LiY_etal_2018,Williams_etal_2020}. These new measurements of high-quality RM data are offering new insights on the structure and kinematics of the BLR gas, as well as high-fidelity BH masses in AGNs. In particular, the SEAMBH project \citep{Du_etal_2016,Du_etal_2018} targeting high-accretion-rate low-redshift AGNs is extending RM studies to systems with extreme accretion parameters. However, most of these high-quality RM measurements are for \hbeta, with only two cases for \CIV\ using intensive UV spectroscopy from HST \citep[][]{DeRosa_etal_2015, Kara_etal_2021}.  

On the other hand, there have been considerable efforts to push RM to the high-redshift regime, and for additional broad lines in AGN spectra. For example, there have been monitoring programs to measure RM lags in high-redshift and high-luminosity quasars \citep[e.g.,][]{Kaspi_etal_2007,Czerny_etal_2019,Lira_etal_2018,Kaspi_etal_2021}, with decade-long baselines to capture the anticipated multi-year lags. These most luminous quasars are rare and sparsely distributed on the sky, thus requiring individual monitoring campaigns. In most of these cases, the data quality is only sufficient to derive a mean lag. One notable exception is the gravitationally lensed quasar SDSS J2222+2745 at $z=2.8$, for which Gemini optical monitoring spectroscopy combined with ground-based photometric light curves enabled velocity-resolved RM measurements \citep{Williams_etal_2021}. The success of RM measurements in SDSS J2222+2745 rests on anticipated variability features from prior photometric monitoring for this lensed quasar. For other monitoring programs of high-redshift quasars, however, the variability patterns are unknown ahead of time, resulting in a reduced success rate of lag measurements. Therefore individual object monitoring becomes expensive and inefficient in building up the statistics of high-redshift RM measurements. One solution to circumvent these observational difficulties is to perform multi-object spectroscopic monitoring that piggybacks on a survey program to greatly improve the efficiency of RM monitoring. 

The Sloan Digital Sky Survey Reverberation Mapping (SDSS-RM) project \citep{Shen_etal_2015a} and the OzDES-RM project \citep{King_etal_2015} recently emerged as the first two multi-object spectroscopic RM programs (MOS-RM) to perform RM measurements for large statistical samples of quasars far beyond the low-redshift universe. Both MOS-RM programs target all quasars within small areas of the sky that can be simultaneously monitored with wide-field optical MOS facilities. Because the target quasars are typically fainter than $m\sim 18$ in the optical, the spectral S/N is on average much lower than those achieved for low-redshift RM campaigns, a situation to be improved with future larger-aperture MOS facilities for similar target brightness \citep[e.g.,][]{MSE,Swann_etal_2019}. Nevertheless, these MOS-RM programs are starting to produce large numbers of lag measurements at $z>0.3$ that cover major rest-frame UV to optical broad emission lines \citep[e.g.,][]{Shen_etal_2016a,Grier_etal_2017,Grier_etal_2019,Shen_etal_2019c,Homayouni_etal_2020,Hoormann_etal_2019,Yu_etal_2023,Malik_etal_2023}.

In this work we present the final data set from the SDSS-RM project, which includes 11 years of photometric monitoring (2010--2020) and 7 years of spectroscopic monitoring (2014--2020). Using this data set, we measure RM lags for four major broad lines in quasar spectra: \halpha, \hbeta, \MgII, and \CIV. We describe the data in \S\ref{sec:data}, and the preparation of light curves in \S\ref{sec:lc}. We detail our methodology of lag measurements in \S\ref{sec:lag_mea}, and present our RM results along with discussions in \S\ref{sec:disc}. We conclude in \S\ref{sec:con} with future prospects. In the appendices, we provide additional information about data access and lag detection efficiencies. Throughout this paper we adopt a flat $\Lambda$CDM cosmology with $\Omega_{\Lambda}=0.7$ and $H_{0}=70\,{\rm km\,s^{-1}Mpc^{-1}}$. By default time lags refer to those in the observed frame unless otherwise specified. We adopt the convention that a positive lag means line variability lags behind continuum variability. 

\section{Data}\label{sec:data}

\subsection{The SDSS-RM sample}

The SDSS-RM sample contains 849 broad-line quasars at $0.1<z<4.5$ spectroscopically confirmed in a single 7 ${\rm deg}^2$ field (R.A. J2000$=213.704$\,deg, decl. J2000$=+53.083$\,deg). These quasars were discovered in previous SDSS surveys and were selected by different algorithms \citep[see details in][]{Shen_etal_2019b}. The sample is flux limited to $i_{\rm psf}=21.7$, although the completeness near this flux limit is not well quantified and is particularly low at $z<0.5$ \citep[see fig.~1 in][]{Shen_etal_2019b}. No additional selection cuts, such as variability or emission-line strength, were imposed on the SDSS-RM sample. The SDSS-RM sample covers a broad and contiguous range in redshift-luminosity space, and is representative of luminous quasars with $L_{\rm bol}\gtrsim 10^{45}\,{\rm erg\,s^{-1}}$. The detailed sample characterization is provided in \citet{Shen_etal_2019b}, where spectral variability properties are based on the 2014 SDSS-RM spectra alone.  

\subsection{Spectroscopy}

Optical spectroscopic monitoring was obtained with the BOSS multi-object spectrographs \citep{Smee_etal_2013} on the SDSS telescope \citep{Gunn_etal_2006} during 2014--2020, as an ancillary program within SDSS-III \citep{Eisenstein_etal_2011} and SDSS-IV \citep{Blanton_etal_2017}. The spectroscopy had an average cadence of $\sim 4$\,days in 2014, with a total of 32 epochs at a nominal exposure time of 2\,hr each. In subsequent years, SDSS-RM obtained $\sim 12$ epochs per year (2 per month) with a nominal exposure time of 1 hr each during 2015--2017 and $\sim 6$ epochs per year (monthly cadence) during 2018--2020. The final spectroscopic baseline covers 7 years, with a total of 90 spectroscopic epochs. The wavelength coverage of BOSS spectroscopy is $\sim 3650-10400$\,\AA, with a spectral resolution of $R\sim 2000$. The typical signal-to-noise ratio (S/N) per 69\,$\kms$ pixel averaged over the $g$ band in a 2 hr exposure is $\sim 4.5$ at $g_{\rm psf}=21.2$, but could be lower for epochs observed with poor observing conditions. The nominal broad-band spectrophotometry accuracy is $\sim 5\%$ in $gri$ bands, but can be worse than $\sim 10\%$ below 4000\,\AA\ \citep{Shen_etal_2015a}.  

\subsection{Photometry}

Optical monitoring photometry provides continuum light curves to facilitate RM lag measurements. SDSS-RM obtained photometry in the $g$ and $i$ bands with the Steward Observatory Bok 2.3~m telescope on Kitt Peak and the 3.6~m Canada-France-Hawaii Telescope (CFHT) on Maunakea. The Bok and CFHT imaging roughly covers the same monitoring period as the SDSS-RM spectroscopy, with the highest cadence in 2014 and reduced cadences in subsequent years. The details of these photometric observations and data processing are described in \citet{Kinemuchi_etal_2020}. In addition to dedicated Bok and CFHT photometry, the SDSS-RM field coincides with the MD07 Medium-Deep Field in the PanSTARRS-1 survey \citep{Kaiser_etal_2010}, with few-night-cadence, multi-band ($grizY$) photometric light curves during 2010--2013. These PS1 data were published in \citet{Shen_etal_2019b}, and provide early continuum light curves to extend the effective baseline for our RM measurements. Finally, we compile available photometric light curves for the SDSS-RM sample from the Zwicky Transient Facility public survey \citep{Bellm_etal_2019} during 2018--2020. The combined photometric light curves span a baseline of 11 years (2010--2020), and were merged to provide a uniform continuum light curve for each quasar (\S\ref{sec:merge_lc}). 

\section{Light curve compilation}\label{sec:lc}

\subsection{PrepSpec}

PrepSpec is a spectral refining procedure performed on the flux-calibrated multi-epoch SDSS spectra in order to reduce further the scatter in the flux calibration \citep{Shen_etal_2016a}. PrepSpec rescales the flux levels of each individual epoch by optimizing model fits (in parameterized functional forms) to describe the continuum and broad-line variability patterns as functions of time and wavelength, using the fluxes of the narrow emission lines (in particular \OIIIab) as an internal calibrator \citep[e.g.,][]{VW1992}, which are assumed to remain constant over the relatively short monitoring period compared with the light-crossing time of the narrow-line region. This procedure improves the calibration of the relative spectrophotometry to $\sim 2\%$ for low-redshift quasars with strong narrow emission lines. This refining procedure is particularly useful in calibrating a small number of spectroscopic fiber+epochs that suffered significant flux losses in the fiber due to unknown systematics \citep[][]{Shen_etal_2015a}.

PrepSpec also generates model light curves (and uncertainties) for a given set of broad emission lines and continuum fluxes, as well as model mean and RMS profiles of the broad lines. We use these broad-line light curves in our lag measurements, and the line widths in the calculation of RM BH masses. The full technical details of PrepSpec are described in \citet{Shen_etal_2016a}. All PrepSpec outputs are released as part of the final data products of SDSS-RM. 

Compared with earlier PrepSpec results used for intermediate SDSS-RM lag measurements \citep[][]{Shen_etal_2016a,Shen_etal_2019b,Grier_etal_2017,Grier_etal_2019,Homayouni_etal_2020}, there are several improvements. First of all, we have run PrepSpec on the full 7-year spectroscopic data, which provide better constraints on variability models than earlier shorter light curves. Second, PrepSpec measured more reliable broad-line widths in the mean spectrum by subtracting the emission of the \FeII\ complex. Third, PrepSpec adopted more accurate systemic redshifts provided in \citet{Shen_etal_2019b} and refined line windows for better line width and flux measurements. 

The broad-line RMS spectrum constructed by PrepSpec is the true RMS profile for the variable line component. This is also the standard approach used by most of the recent RM campaigns on low-$z$ AGNs \citep[e.g., the LAMP project,][]{Barth_etal_2015}, where the individual-epoch spectra are decomposed into separate emission-line and continuum components. Some earlier RM studies \citep[e.g.,][]{Peterson_etal_2004,Grier_etal_2012} construct the RMS spectrum from the total (emission line+continuum) epoch spectra, and subtract a local continuum to derive a line-only RMS spectrum. There are systematic biases in the measured RMS line widths from the latter approach when the monitoring period is not much longer than the lag \citep[e.g.,][]{Barth_etal_2015,Wang_etal_2019}. We will revisit this issue in \S\ref{sec:lag_mass}. 

\begin{figure*}
  \centering
    \includegraphics[width=0.98\textwidth]{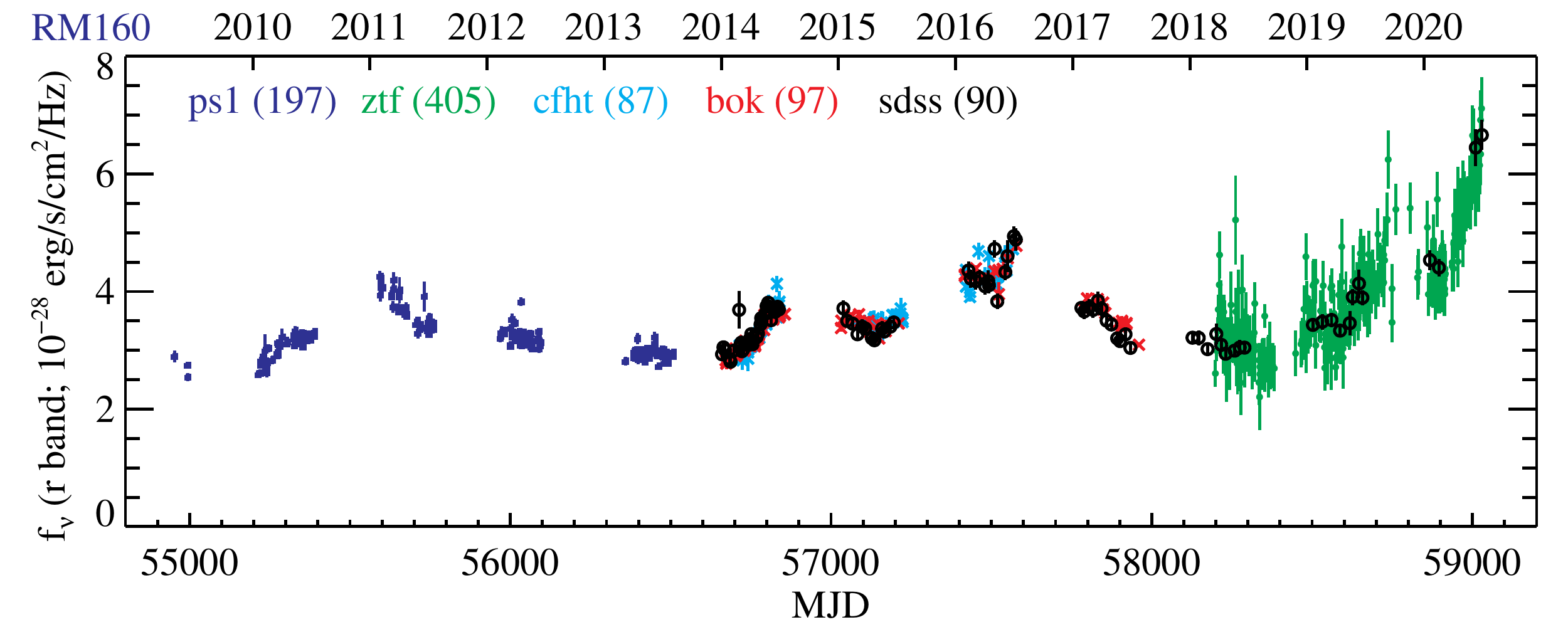}
    \caption{An example of merged 11-yr photometric light curves for the SDSS-RM sample (see \S\ref{sec:merge_lc} for details). Light curves from different surveys are calibrated to the SDSS $r$ band. For each SDSS-RM quasar, there are typically hundreds of nightly-averaged photometric epochs, to go with 90 spectroscopic epochs (black open circles) from SDSS. }
    \label{fig:lc_examp}
\end{figure*}

\subsection{Merged continuum light curves}\label{sec:merge_lc}

We combine the photometric light curves from different facilities/surveys to form a merged continuum light curve for lag measurements, as commonly done for RM studies. For the PS1 data, we use observations in $gri$ filters to improve the sampling. The $zY$ data are not used since the continuum variability there is small and more contaminated by host-galaxy emission. We use the PS1 data published in \citet{Shen_etal_2019b}, and coadd the nightly observations weighted by the inverse variance of the individual flux measurements. 

For the purpose of measuring broad-line lags, we ignore the inter-band continuum lags. Such continuum lags trace the structure of AGN accretion disks, which are typically much more compact than the BLR, hence these continuum lags are on much shorter timescales \citep[e.g., less than a few rest-frame days, but can be longer for more luminous quasar accretion disks;][]{Fausnaugh_etal_2016}. A complication is that a fraction (e.g., $\sim 10-20\%$) of the optical continuum emission may come from the diffuse BLR emission \citep[e.g.,][]{Guo_etal_2022b}, and cause inter-band continuum lags that are non-negligible compared with broad-line lags. This is still an ongoing investigation, and without a robust method to correct for inter-band continuum lags for SDSS-RM quasars, we simply follow previous RM work and ignore this detail. We scale the merged light curves (in $f_\nu$) to physical flux scales in $r$ band. This band was chosen because it offers the best spectrophotometry in SDSS spectra \citep[][]{Shen_etal_2015a}. We use the synthetic broad-band flux computed from the SDSS spectroscopy as the baseline to calibrate and merge other photometric light curve sets. 

The continuum light curve merging is performed with the public code {\tt PyCali} \citep{pycali}. {\tt PyCali} is a Bayesian method to calibrate the light curves obtained with different facilities at different times to a reference set of light curves, during which the flux uncertainties of individual light curve sets can also be adjusted to account for overestimated or underestimated flux uncertainties in the original data set. Before running {\tt PyCali} we coadd (weighted by inverse variance) intra-night photometry points from the same facility and same band. However, for CFHT and Bok observations, some objects are covered on multiple detector chips, and we only coadd the nightly photometry from the same chip, using the light curves compiled in \citet{Kinemuchi_etal_2020}.

We show an example of the merged 11-year continuum light curves in Fig.~\ref{fig:lc_examp}. The full set of {\tt PyCali}-merged continuum light curves is released as part of the data products from this work.

\subsection{Light curve properties}\label{sec:lc_var}

We quantify the variability characteristics of the light curves and compile the results in the summary Table \ref{tab:sum}. We use a maximum-likelihood estimator \citep{Shen_etal_2019b} to measure the intrinsic variability (rms) of the light curve, corrected for flux uncertainties. These rms variability measurements are performed for both the 11-yr photometric light curves and the 7-yr emission-line light curves from spectroscopy+PrepSpec analysis. In addition to the rms variability metric, we use another empirical metric to assess evidence for intrinsic variability of the light curve \citep{Shen_etal_2019b}, defined as\footnote{In rare cases within the SDSS-RM sample, if the light curve is consistent with noise or the flux uncertainties are overestimated, $\chi^2-{\rm dof}$ may become negative. These cases will not yield a meaningful lag detection. In such cases PrepSpec calculates ${\rm SNR2}=-\sqrt{|\chi^2-{\rm dof}|}$, which is negative.} ${\rm SNR2}=\sqrt{\chi^2 - {\rm dof}}$. Here $\chi^2\equiv \Sigma (y_i-\left<y_i\right>)^2/\sigma_i^2$ is relative to the optimal average of the light curve $\left<y_i\right>$, using flux measurement errors $\sigma_i$ only; dof is the number of degrees of freedom $(N-1)$, where $N$ is the number of light curve points. Larger intrinsic variability (with respect to flux uncertainties) tends to have a larger SNR2.  

\begin{figure}
  \centering
    \includegraphics[width=0.48\textwidth]{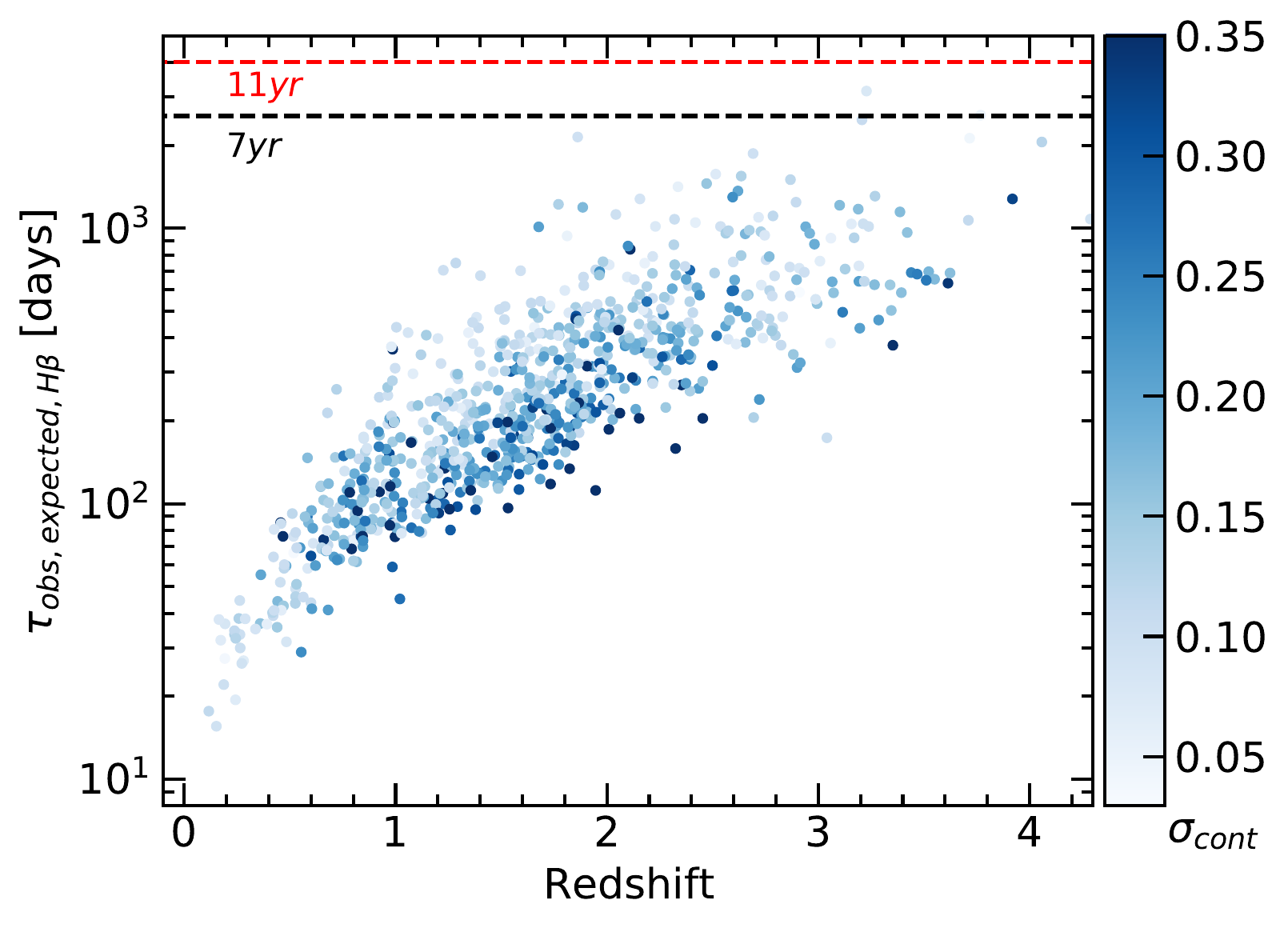}
    \caption{Expected \hbeta\ lag in the observed-frame for the SDSS-RM sample, estimated using the mean \hbeta\ $R-L$ relation from \citet{Bentz_etal_2013}. The points are color-coded by the fractional RMS variability $\sigma_{\rm cont}$ from the 11-yr photometric light curve, corrected for flux measurement uncertainties. The black and red dashed lines are the 7-yr spectroscopic baseline and the 11-yr photometric baseline, respectively. There is a general tendency of increasing variability toward lower luminosities (shorter lags) at fixed redshift. }
    \label{fig:z_lag}
\end{figure}

\begin{figure}
  \centering
    \includegraphics[width=0.48\textwidth]{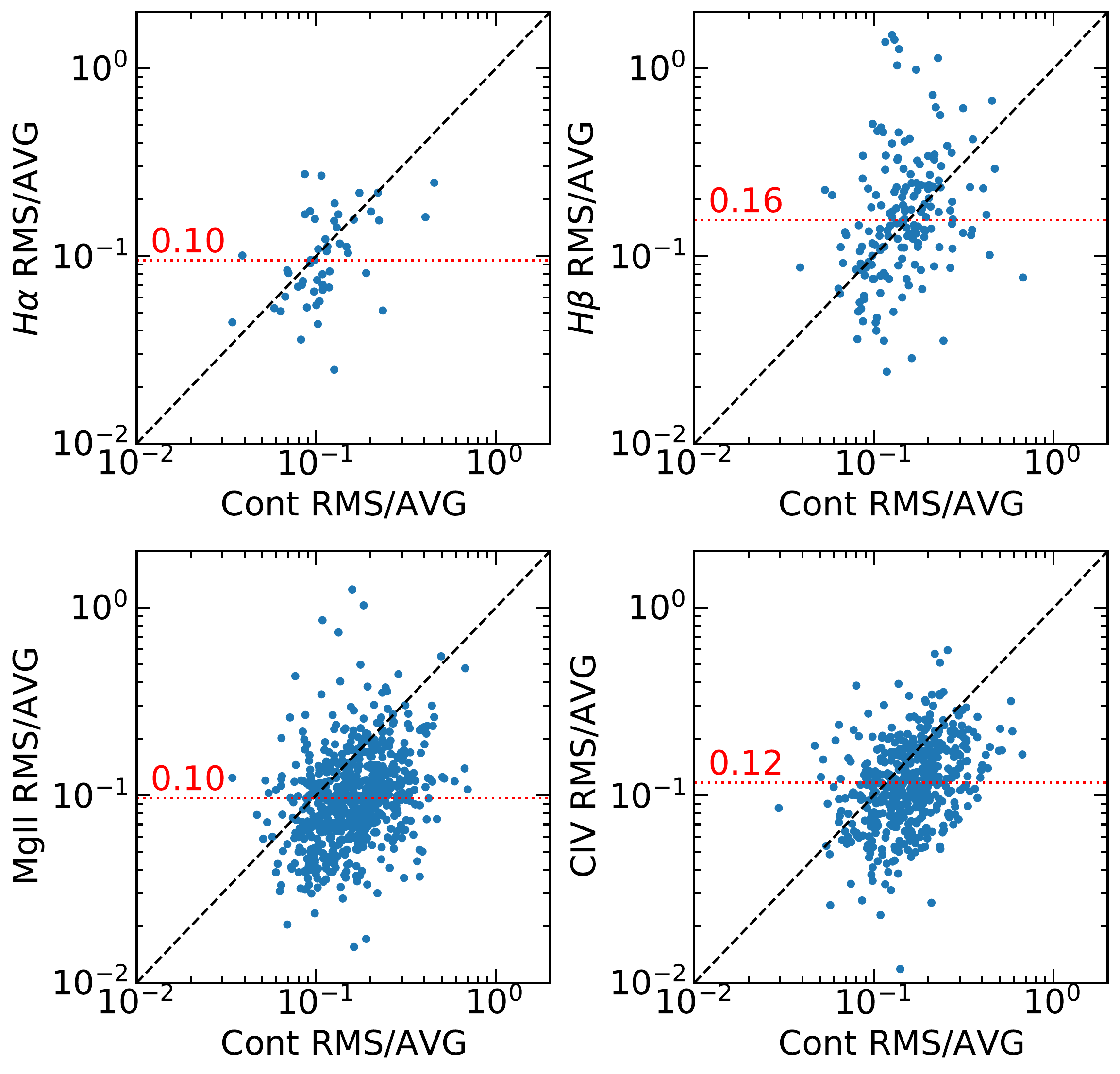}
    \caption{Fractional (intrinsic) RMS variability for the four broad lines considered in this work versus that of the continuum light curve. The dashed diagonal lines show the unity relation, and the horizontal red lines (with numbers) indicate the median fractional line variability. There are reasonably good correlations between the line and continuum RMS variability, as expected from reverberation. }
    \label{fig:rms_var}
\end{figure}

Fig.~\ref{fig:z_lag} displays the distribution of the SDSS-RM sample in the redshift versus expected (\hbeta) lag space, color-coded by the intrinsic fractional rms variability of the continuum. The continuum rms variability has a sample median of $\sim 15\%$, and broadly increases with decreasing luminosity at a given redshift \citep[e.g.,][]{VandenBerk_etal_2006,MacLeod_etal_2012}. Fig.~\ref{fig:rms_var} displays the comparisons between the continuum and broad-line (fractional) rms variabilities. Given these typical fractional variability amplitudes in the continuum and in the responding lines, measuring a lag requires good flux calibration and spectral S/N. Many targets in the SDSS-RM sample do not have the adequate S/N in the emission-line light curves to allow a robust lag detection. In addition, to measure a lag, the light-curve variability must contain distinctive inflection ``features'', as a monotonically changing light curve will produce broad correlations over a range of lags. Nevertheless, given the large sample size, there are many SDSS-RM quasars with well-detected variability patterns to measure the lags. A handful of the best-quality light curves are even good enough to measure velocity-resolved lags, which will be presented elsewhere \citep[e.g.,][]{Fries_etal_2023}. 

Compared with the optical continuum variability, different broad lines exhibit different levels of fractional variability. Fig.~\ref{fig:rms_var} shows that while there are broad correlations between the continuum and line fractional RMS variability, \MgII\ and \halpha\ have slightly lower average variability compared with \CIV\ and \hbeta. The lower variability for \MgII, as seen in earlier studies \citep[e.g.,][]{Sun_etal_2015,Yang_etal_2020a}, can be understood if the \MgII-emitting gas is further out than the BLRs of other species, and/or the response of \MgII\ is in general weaker than \hbeta, as shown in photoionization calculations \citep[e.g.,][]{Goad_etal_2012, Guo_etal_2020}. The high-ionization \CIV\ line usually shows larger RMS variability than \hbeta\ \citep{Peterson_etal_2004} in local low-luminosity AGNs. However, for high-luminosity quasars, it is argued that a portion of the \CIV\ line does not reverberate to continuum variability, thus diluting the line fractional RMS variability \citep[e.g.,][]{Denney_2012, Wang_etal_2020}. The situation may be further complicated by the fact that the ionizing continuum (powering the lines) and the optical continuum have different RMS variability, as the variable accretion flow onto the SMBH results in varying shapes of the spectral energy distribution (SED). 

\begin{figure}
  \centering
    \includegraphics[width=0.48\textwidth]{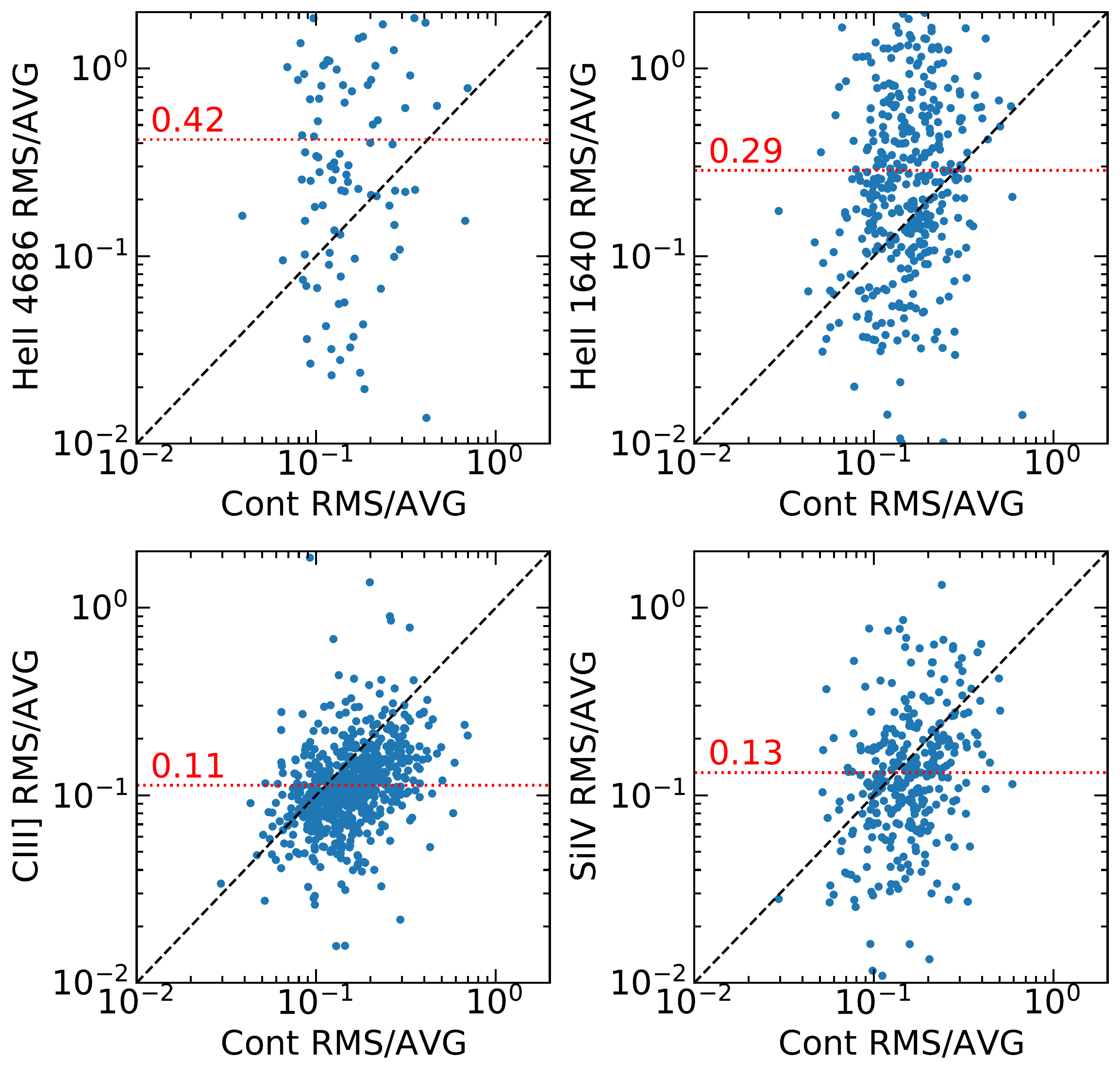}
    \caption{Fractional (intrinsic) RMS variability for additional broad lines versus that of the continuum light curve. We do not report lag measurements for these lines, but their variability properties are compiled in Table~\ref{tab:sum}. }
    \label{fig:rms_var_morelines}
\end{figure}

Although not covered in our lag measurements, we show the fractional RMS variability for additional broad lines in Fig.~\ref{fig:rms_var_morelines}. The two strong UV broad lines, \SiIV\ and \CIII, show similar RMS variability as \CIV, while the two high-ionization lines \HeII\,1640 and \HeII\,4686 show notably stronger RMS variability than other lines. The behaviors of the \HeII\ broad lines are consistent with local RM observations \citep[e.g.,][]{Peterson_etal_2004}.

Finally, we show the comparisons between the FWHM measured from the mean spectrum and the line dispersion $\sigma_{\rm line,rms}$ measured from the RMS spectrum in Fig.~\ref{fig:line_width_corr}, using PrepSpec outputs. A strong correlation between these two line widths for the same line is a necessary (but not sufficient) condition for the validity of single-epoch virial mass estimators. The average ratios of ${\rm FWHM/\sigma_{line}}$ for all four broad lines are smaller than 2.35, implying the line profile is more Lorentzian \citep[e.g.,][]{Collin_etal_2006,Villafana_etal_2023}. 

\section{Lag Measurements}\label{sec:lag_mea}

\subsection{Methodology}\label{sec:lag_methods}

Most RM studies measure a reverberation-weighted ``average'' lag across the BLR between the continuum and broad-line light curves. This is mainly limited by the quality of the data (i.e., cadence, baseline and S/N), relative to the intrinsic variability of the quasar. For high-quality RM light curves, it is possible to measure the 1D transfer function $\Psi(\tau)$ between the continuum and integrated line fluxes \citep[e.g.,][]{Horne_etal_2004}. If the spectroscopic data are of sufficient quality to reveal the velocity-resolved BLR responses, it becomes feasible to measure the 2D transfer function ($\Psi(\tau, v)$; or velocity-delay map) and constrain the dynamical structure of the BLR. The ultimate goal of RM is to measure high-fidelity velocity-delay maps and to constrain the geometry and kinematics of the BLR, along with a dynamical measurement of the black hole mass. 

\begin{figure}
  \centering
    \includegraphics[width=0.48\textwidth]{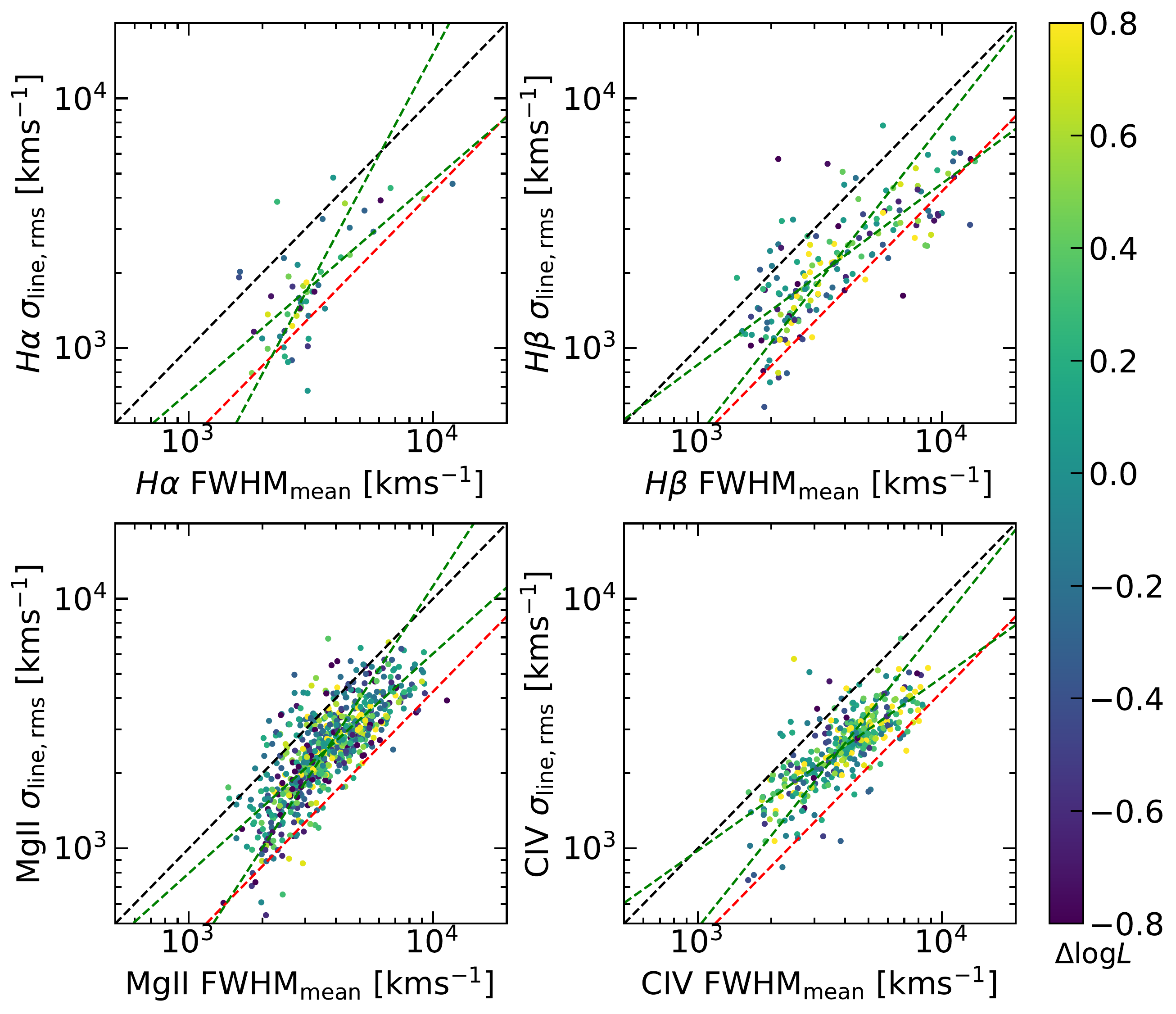}
    \caption{Comparisons between the FWHM from the mean spectrum (for SE estimators) and the $\sigma_{\rm line,rms}$ from the rms spectrum (for RM masses) for \halpha, \hbeta, \MgII\ and \CIV, color-coded by the deviation from the sample median luminosity for the corresponding line (e.g., $L5100$, $L3000$, etc). The \MgII\ $\sigma_{\rm line,rms}$ here {has been} corrected for the velocity split of the doublet, but not for the SDSS spectral resolution (negligible). In each panel, the black dashed line is the unity relation, and the red dashed line represents the FWHM-$\sigma_{\rm line}$ relation for a Gaussian profile. The best-fit linear regressions of $(Y|X)$ and $(X|Y)$ are also shown. There is no clear luminosity trend (measured with the deviations from the sample median luminosity, $\Delta\log L$) in the line width correlation for all lines.  }
    \label{fig:line_width_corr}
\end{figure}

SDSS-RM is a pioneering program to measure BLR lags in a large sample of distant and luminous quasars. As such, the quality of the monitoring data is generally insufficient to robustly measure the 1D or 2D transfer functions in individual objects, except for a small number of objects with very large variability amplitudes. Rather, the main science goal of SDSS-RM is to measure an average lag for quasars in a luminosity-redshift regime uncharted by earlier RM programs. While an average lag does not inform us about the detailed geometry and kinematics of the BLR, it allows us to measure an approximate mass of the black hole, assuming the BLR is virialized. The uncertainty of the inferred RM mass in individual objects is dominated by the assumed geometric factor (the virial factor, see \S\ref{sec:f_factor}) and the uncertainties in the measured average lag and line width. 

\begin{figure*}
  \centering
    \includegraphics[width=0.98\textwidth]{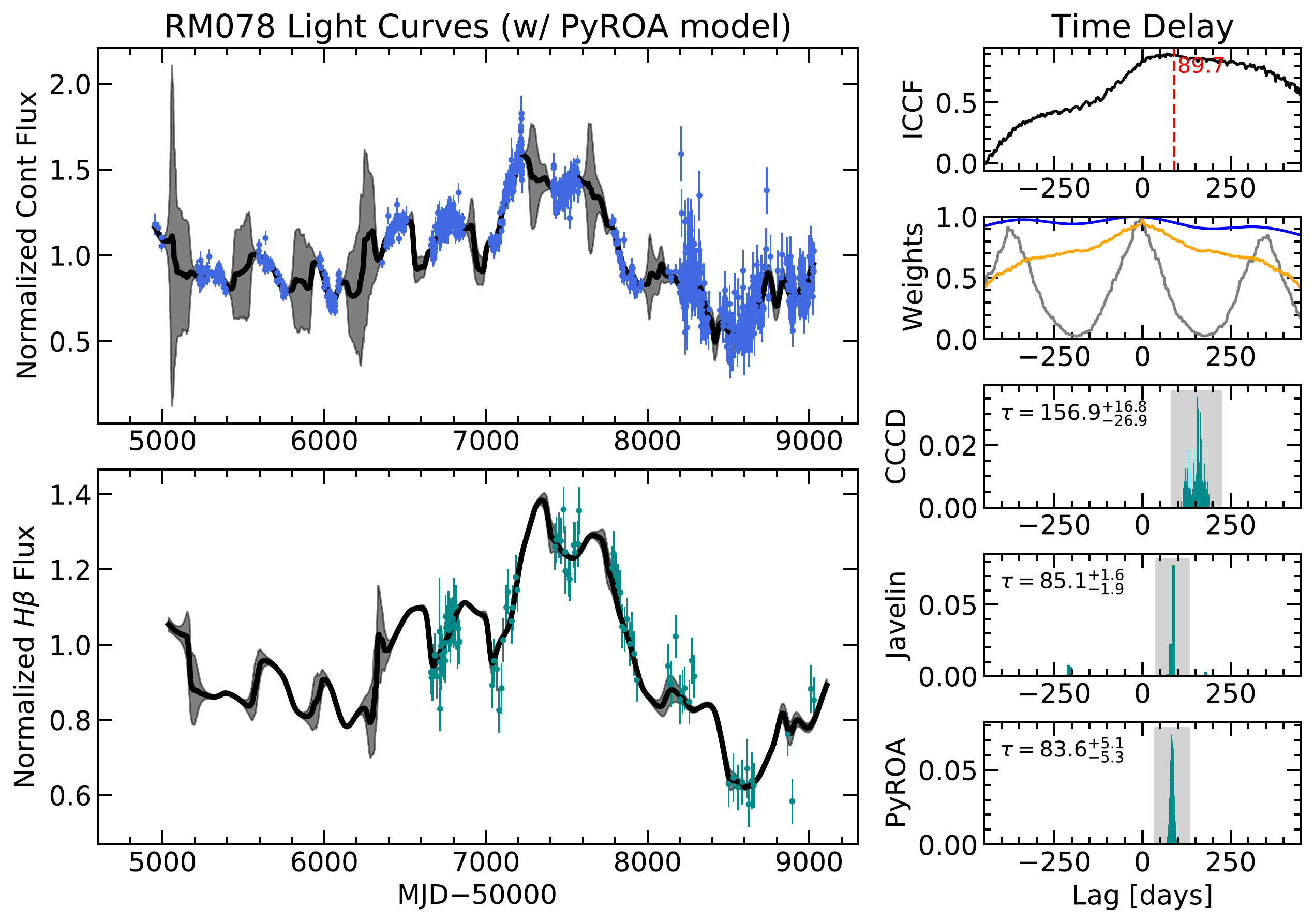}
    \caption{An illustration of lag measurements using ICCF, {\tt Javelin}, and {\tt PyROA}, as well as our alias mitigation scheme using lag posterior weights. The left two panels show the light curve data in points, and the \pyroa model light curves in solid lines and shaded area (best model and $1\sigma$ uncertainty). The right panels show the lag measurements. The ``ICCF'' panel shows the original ICCF, with the red vertical line (and number) indicating the expected (observed-frame) lag for \hbeta. The ``Weights'' panel shows the ACF (orange), periodic overlapping light curve data points $(N(\tau)/N_{\rm max})^2$ (gray), and the final weights (blue). The next three panels show the lag results for ICCF, \javelin and {\tt PyROA}, with the lag posterior displayed in the normalized histogram. The gray shaded band shows the boundaries of the identified primary peak, from which we measure the lag (median) and 1$\sigma$ uncertainties (16th and 84th percentiles). The final lag and uncertainties are reported in each panel. For this particular object, the ICCF is relatively broad, resulting in a different lag using the cross-correlation centroid distribution (CCCD). However, the lag inferred from the cross-correlation peak distribution (CCPD; also see the top-right panel of this figure) is in excellent agreement with the \javelin and \pyroa lags. }
    \label{fig:lag_examp}
\end{figure*}

To measure this average BLR lag, RM studies traditionally deploy non-parametric (or with minimal parameterization) cross-correlation analysis methods, such as the interpolated cross-correlation function \citep[ICCF,][]{Gaskell_Peterson_1987}, discrete correlation function \citep{Edelson_Krolik_1988}, or the z-transform discrete correlation function \citep[ZDCF,][]{Alexander_2013}. Other non-parametric methods are also available to measure the average time delay between two light curves \citep[see, e.g.,][for some examples]{Czerny_etal_2019}. Alternatively, recent approaches such as {\tt Javelin} \citep{Zu_etal_2011} and {\tt CREAM} \citep{Starkey_etal_2016} use a stochastic variability model to interpolate the light curves and measure the lag using MCMC to sample the posterior lag distribution. {\tt PyROA} \citep{pyroa} takes a slightly different approach: it performs a Bayesian analysis on the light curves with a running optimal average, from which the mean lag and the shape and width of the transfer function can be simultaneously constrained. The later lag measuring techniques are more reliable in recovering the lag and its uncertainties when the light-curve quality is insufficient for the other non-parametric methods to measure a significant lag \citep[e.g.,][]{Li_etal_2019a, Yu_etal_2020b}. Nevertheless, earlier correlation analysis methods are less model-dependent, and provide important confirmation of the mean lag measurements. In particular, we found that the empirical ICCF correlation coefficient provides an efficient and intuitive metric on the successful detection of a lag. For this reason, information in the original ICCF will be used in determining the success of a lag measurement, even if the best lag is taken from more sophisticated approaches such as {\tt Javelin} or {\tt PyROA}. 

Each pair of continuum and broad-line light curves is passed through a lag detection pipeline (wrapper) that collects publicly available packages \citep[][]{pyccf,Zu_etal_2011,pyroa} to perform ICCF, {\tt Javelin (v0.33)} and {\tt PyROA (v3.1.0)} analyses. The pipeline first flags and removes outliers $5\sigma$ away from a Damped Random Walk (DRW) model fit to the entire continuum light curve, using the package {\it celerite} \citep{celerite}. This outlier rejection is only performed for the continuum, given the numerous epochs combined from different surveys. The best-fit (taken as the median of the posterior distribution) DRW parameters are adopted (and fixed) in the subsequent {\tt Javelin} analysis for self-consistency. We report these best-fit continuum DRW parameters in Table~\ref{tab:sum}. The best-fit damping timescales $\tau_{\rm DRW}$ for the SDSS-RM sample have a median value of $\sim 270$\,days in the quasar rest frame, which is consistent with the findings from past studies using $\sim 10$~yr long light curves \citep[e.g.,][]{MacLeod_etal_2012,Suberlak_etal_2021,Stone_etal_2022}. However, there is evidence that the damping timescale may be even longer if measured from light curves with much longer baselines \citep[e.g.,][]{Stone_etal_2022}.

For the {\tt Javelin} run we use MCMC parameters $n_{\rm chain}=n_{\rm walkers}=n_{\rm burn}=200$ to provide enough sampling of the posterior over the lag search range. For {\tt PyROA} we use a similar amount of MCMC samples, and turn on the extra variance and delay distribution (adopting a Gaussian kernel) options in {\tt PyROA}. The pipeline computes the relevant lag measurement properties in each method and determines the best-fit lag and its uncertainty following the rules specified in \S\ref{sec:alias}. The pipeline (wrapper) is publicly available at https://pypetal.readthedocs.io/en/latest/ 

Our long-term baseline allows the exploration of lags up to as much as $\sim 11$\,years in the observed frame (but the success rate will drop dramatically when approaching the maximum baseline), given the leading photometric light curves from PS1. However, quasars in the SDSS-RM sample have a wide range of observed-frame lags, from a few days to a few years. Searching the entire possible lag range enabled by the baseline is expensive (given the required MCMC samples) and unnecessary for objects with much shorter lags, and greatly increases the chance of aliases that become difficult to mitigate with our methodology described in \S\ref{sec:alias}. We therefore limit the lag search range on an object-by-object basis. 

We first estimate the expected observed-frame lag, $\tau_{\rm obs,exp}$, based on the median continuum luminosity over the spectroscopic baseline, using the \hbeta\ $R-L$ relation from \citet{Bentz_etal_2013}. The \CIV\ lag is typically shorter than the \hbeta\ lag, and the \halpha\ and \MgII\ lags are slightly longer than that of \hbeta\ based on past RM results \citep[e.g.,][]{Peterson_etal_2004, Homayouni_etal_2020}. We then limit the lag search range within $\pm 5\times \tau_{\rm obs,exp}$, where negative lags refer to the line variability leading the continuum variability. However, we also require a minimum search range of $\pm 250\,$days and a maximum search range of $\pm 2500\,$days for each object. The latter maximum lag search range is defined to be roughly the same as the spectroscopic baseline. Importantly, as we are imposing a symmetric lag search range, we do not expect a systematic bias toward aliased (positive) lag values. In a small number of cases, the lag approaches $90\%$ of the search range, and we rerun the lag detection with expanded search ranges to avoid introducing an artificial boundary in the measured lags. 

There is a small fraction ($\sim 4\%$) of objects for which the expected $\tau_{\rm obs,exp}$ exceeds 1000 days (Fig.~\ref{fig:z_lag}). For these objects, we relax the symmetry requirement on the lag search range, and utilize the full range of detectable lags given the leading years of photometry. The lag search range for this subset of quasars is $[-2500, 4000]\,$days in the observed frame. 

\begin{figure}
  \centering
    \includegraphics[width=0.48\textwidth]{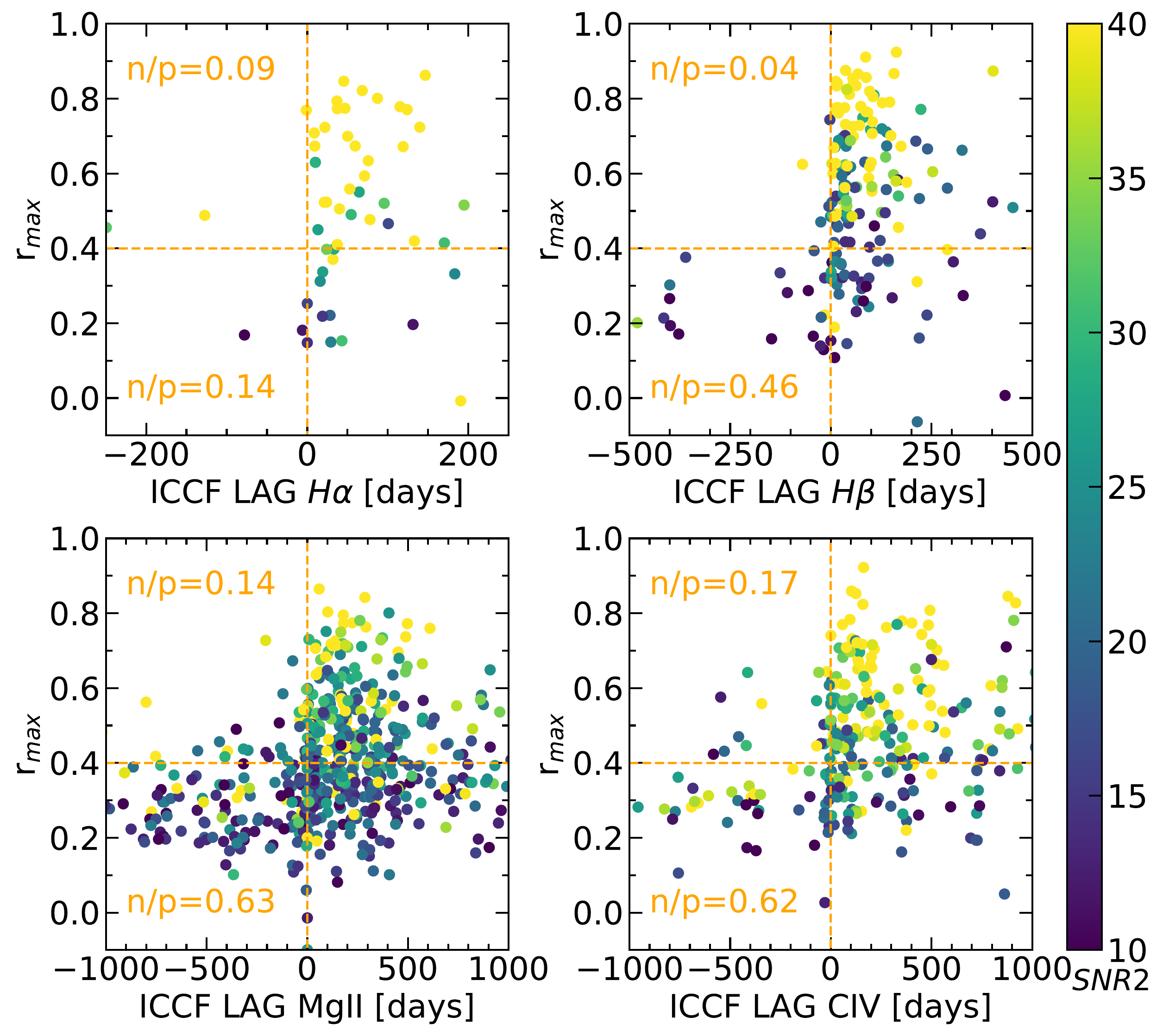}
    \caption{Statistical detection of lags for the SDSS-RM sample using ICCF. As the $r_{\rm max}$ value decreases, there are increased incidents of negative lags. While overall the asymmetry toward positive lags is clear, below $r_{\rm max}=0.4$, the negative to positive lag ratio is substantially higher than that at $r_{\rm max}>0.4$, indicating a high false-positive rate for the positive lags measured with low $r_{\rm max}$ values. The points are color-coded by the line variability SNR2 (see \S\ref{sec:lc_var}). Objects with higher SNR2 values tend to cluster in the upper-right quadrant, where a positive lag detection is more likely to happen, along with a high $r_{\rm max}$ value. }
    \label{fig:stat_lag_det}
\end{figure}

\subsection{Alias mitigation and detection of lags}\label{sec:alias}

The standard ICCF, \javelin and \pyroa analyses do not take into account the number of LC pairs contributing to the correlation calculation at each lag. For typical SDSS-RM quality light curves, there are often a number of aliasing peaks in the lag posterior distribution that impact the measurement of the true lag. We mitigate the impact from aliasing peaks in the lag posterior following the scheme introduced in earlier work \citep[][]{Grier_etal_2017,Grier_etal_2019}. In short, we calculate a weight $P(\tau)=[N(\tau)/N(0)]^2$, where $N(\tau)$ is the number of shifted emission-line data points that overlap in the date ranges with the continuum and $N(0)$ is the number of overlapping points at $\tau=0$. Shifted emission-line data points falling into seasonal gaps are not counted. Since our continuum light curve encompasses the emission-line light curve by design, $N(0)$ is the maximum number of spectroscopic epochs. We then derive a final weight $P_{f}(\tau)=P(\tau)\ast {\rm ACF(\tau)}$, where ${\rm ACF}(\tau)$ is the auto correlation function of the continuum LC and ``$\ast$'' is the convolution operation. The weight $P(\tau)$ accounts for the reduction in the statistical constraint on the lag due to the fewer data points. While the exponent of 2 in the $P(\tau)$ weight is somewhat arbitrary, we find it results in optimal rejection of aliasing peaks \citep[e.g.,][]{Grier_etal_2019}. The convolution with the continuum ACF accounts for the effect of seasonal gaps on the detection of certain lags. If the ACF declines rapidly, the annual seasonal gaps will have a significant effect on our detection because we are less likely to account correctly for the light-curve behavior during the gaps. On the other hand, if the ACF declines slowly away from zero lag, it implies it is straightforward to interpolate across the seasonal gaps as the variability is slow; in this case the seasonal gaps are less likely to have an effect on our lag measurements. In the latter case, it is possible to measure a lag even if most of the shifted emission-line data points fall into seasonal gaps in the continuum light curves \citep[see examples in, e.g.,][]{Shen_etal_2019c}.

Once we have the posterior lag distribution (for \javelin or {\tt PyROA}, this is the posterior lag distribution; for ICCF, this is the cross-correlation centroid distribution; CCCD), we weight the distribution with $P_f(\tau)$. The weighted lag distribution is then smoothed by a Gaussian kernel with a width of 15 days, and we identify the tallest peak within this smoothed distribution as the ``primary'' peak. We identify local minima in the distribution to either side of the peak and adopt these minima as the minimum and maximum lags to be included in our final lag calculation. We then return to the original {\em unweighted} posterior, reject all lag samples that lie outside the determined range, and use the remaining samples to calculate the final lag (median of the truncated distribution) and its $1\sigma$ uncertainties (16th and 84th percentiles of the distribution). Fig.~\ref{fig:lag_examp} demonstrates this procedure with one example of our targets (RM078). We did not detect an \hbeta\ lag for this object using the 2014 data \citep{Grier_etal_2017}, but the final SDSS-RM data, with multi-year coverage, are able to firmly detect a lag. {In Appendix~\ref{sec:lag_examples}, we show several additional examples that sample different lag detection qualities, and all lag diagnostic figures are available from our public server (see Appendix~\ref{sec:data_format}).}

\begin{table}
\caption{Lag Detection Summary}\label{tab:lag_sum}
\centering
\begin{tabular}{lcccc}
\hline\hline
\# of Lags & \halpha & \hbeta & \MgII & \CIV  \\
\hline
All attempted & 53 & 186 & 714 & 494 \\
\pyroa ($>2\sigma$ det) & 23 & 81 & 125 & 110 \\
\pyroa ($>1\sigma$ det) & 31 & 108 & 189 & 163 \\
\hline
\end{tabular}
\tablecomments{We report all lag measurements in Table~\ref{tab:sum}. But we recommend to only use lags with $>2\sigma$ detection in statistical analyses. }
\end{table}

\begin{figure*}
  \centering
    \includegraphics[width=0.48\textwidth]{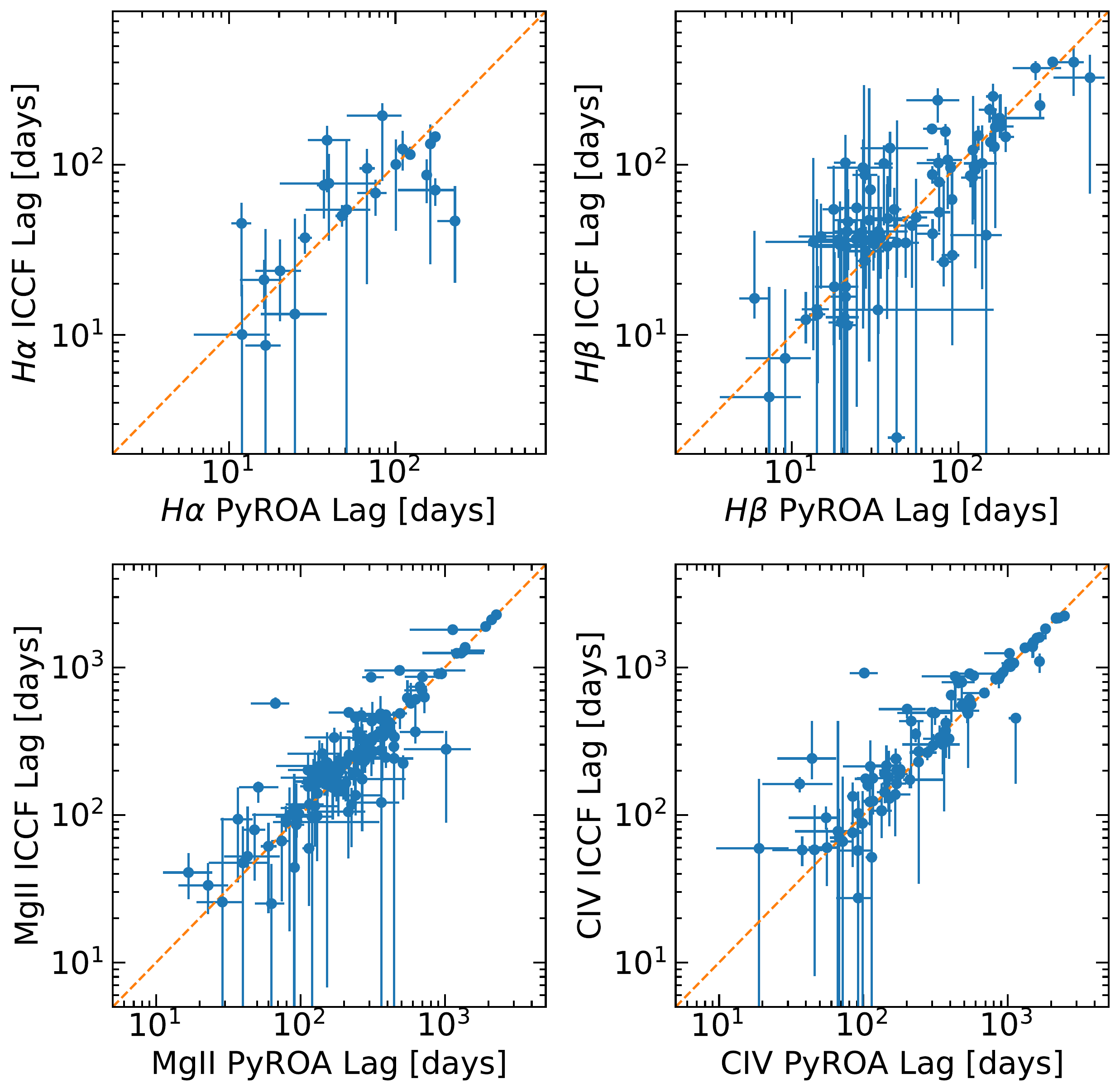}\quad
    \includegraphics[width=0.48\textwidth]{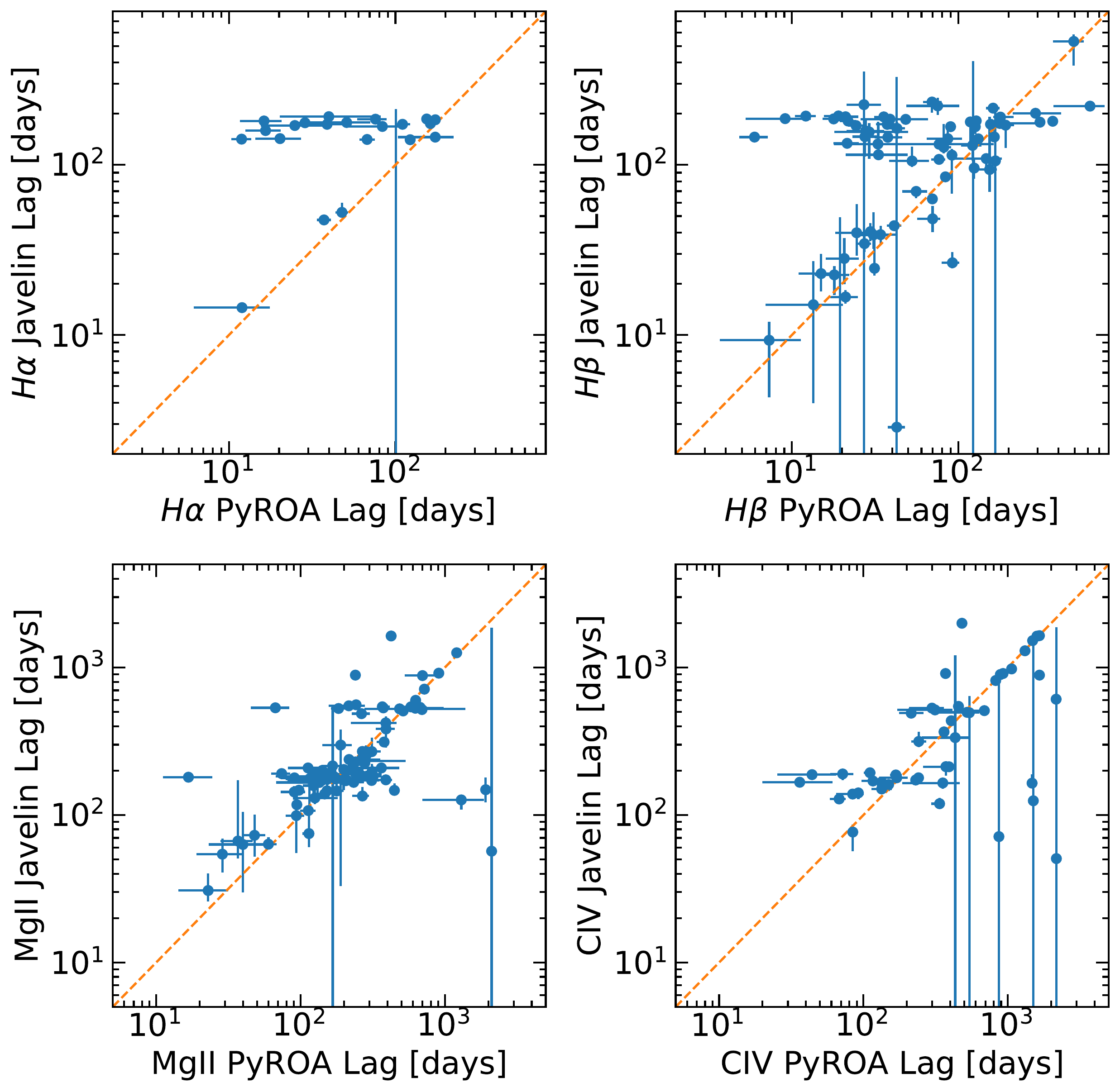}
    \caption{Comparisons of the detected lags between two different methods (ICCF vs \pyroa at left and \javelin vs \pyroa at right). In most cases, our fiducial \pyroa lags are consistent with the ICCF lags. However, for our multi-year light curve data, the \javelin method often traps the lag in the seasonal gaps, leading to artificial clustering of lags there. }
    \label{fig:lag_det_comp}
\end{figure*}

Fig.~\ref{fig:stat_lag_det} shows the distribution of ICCF lag measurements in the lag versus $r_{\rm max}$ plane \citep[e.g.,][]{Shen_etal_2016a,Grier_etal_2017,Grier_etal_2019,Homayouni_etal_2020}, where $r_{\rm max}$ is the maximum Pearson correlation coefficient within $\pm 1\sigma$ from the reported ICCF lag, which can be positive or negative. For all four lines, there is a clear asymmetry toward positive lags (i.e., line lagging continuum), especially when $r_{\rm max}$ is high. This figure demonstrates the statistical detection of RM lags for all four lines, given the symmetrical lag search ranges. When the light-curve quality is poor (or variability insignificant), the significance of correlation $r_{\rm max}$ is low, and there are more random incidents of positive and negative lags. At $r_{\rm max}>0.4$, the overall negative-to-positive lag ratio within the plotted lag bounds is $\sim 0.04-0.17$, indicating a false positive rate of $\sim 4-17\%$ in the reported (positive) lags. At $r_{\rm max}<0.4$, although there are still more positive lags than negative ones, the false positive rate is substantially higher, and the measured positive lags there are of lower significance. Similar lag asymmetries are observed when we replace $r_{\rm max}$ with the line variability SNR2, which is expected as successful lag recoveries rely on well-detected line variability. However, there is no rigid boundary on SNR2 that would guarantee the detection or non-detection of a lag. 

Fig.~\ref{fig:stat_lag_det} also dictates that using a positive lag search range in ICCF will not introduce a large confirmation bias if $r_{\rm max}>0.4$. This may justify the usage of a positive lag search range for MOS-RM samples to reduce computational time. However, it is important to use the $r_{\rm max}$ value derived from the original ICCF to pre-filter low-significance lag detections. 

During our lag measurements with ICCF, \javelin and {\tt PyROA}, we discovered that many {\tt Javelin} fits have lag posteriors that peak in the seasonal gaps. Because the ACF for SDSS-RM quasars is generally broad enough to allow the DRW model to interpolate into seasonal gaps, our weighting scheme is often unable to suppress these gap peaks. The situation is different from our previous analyses using shorter SDSS-RM light curves \citep[e.g.,][]{Grier_etal_2017,Grier_etal_2019,Homayouni_etal_2020}. Now with our full multi-year baseline, it becomes more common for the {\tt Javelin} posterior to be trapped in these seasonal gaps, resulting in spurious lag measurements. 

In these failed {\tt Javelin} fits, the favored model is where the line light curve is shifted into an annual gap, with minimal overlap with the continuum light curve in order to produce a good match between the two light curves. One particular reason for this gap alignment may be underestimated systematic uncertainties in the flux measurements that cause tension in matching up the continuum and line light curves. To remedy the situation, we deploy {\tt PyROA}, which includes an excess variance term to account for unknown systematic uncertainties in flux measurements. Like ICCF, {\tt PyROA} is an empirical method that does not rely on a physical DRW model to interpolate the light curves. However, {\tt PyROA} implements modules to account for the smearing of the line light curve due to the transfer function and for excess variance in flux measurements, with MCMC sampling to produce the lag posterior and rigorous lag uncertainties. Therefore, {\tt PyROA} preserves the advantages of being model-independent, while being less prone to underestimated flux uncertainties than {\tt Javelin}.

We visually inspect the ICCF, \javelin and \pyroa results for each object, and conclude that indeed \pyroa returns the most reasonable lag posterior and model light curves among all three methods. We therefore adopt the \pyroa lag results as our fiducial results in the following analyses. 

We now quantify the significance of individual lag measurements, and define a set of criteria to declare successful detection of a positive lag. These cases have \$line\$\_LAG\_DET$>0$ in Table~\ref{tab:sum}. To do so, we rely on a hybrid set of parameters from the ICCF and {\tt PyROA} results: 
\begin{enumerate}

\item[$\bullet$] The \pyroa lag must be positive at $\ge 2\sigma$ significance. Although lower-significance detection might still be valid, the large uncertainties will render their measurements less useful, and could complicate our comparative analyses. 

\item[$\bullet$] The original ICCF must reach Pearson correlation coefficient $r_{\rm max}>0.4$ within $\pm 1\sigma$ of the reported \pyroa lag.

\item[$\bullet$] Less than half of the lag posterior samples can be removed by our alias-removal procedure. If this procedure eliminates more than half of the samples, it indicates we lack a solid measurement of the lag as most of the samples are rejected. For \pyroa lag posterior, this criterion is usually always satisfied. 

\end{enumerate}

As a final check, we visually inspect all detected lags (see \S\ref{sec:inspection}) and confirm there are no peculiarities in the light curves (e.g., predominantly corrupted data) or in the measured lag posterior. The choice of a threshold of $r_{\rm max}>0.4$ is somewhat arbitrary, but provides an efficient means to remove a large number of false positives. If we lower this threshold, we would recover more lags, but the contamination rate will also rise quickly. On the other hand, we caution that imposing more stringent selection criteria may introduce additional selection biases in the resulting lag sample, and artificially reduce the observed (intrinsic) scatter or bias the slope in the $R-L$ relation. 

 {In Appendix~\ref{sec:quality_cuts}, we show the results with more stringent quality cuts on detected lags, which are generally consistent with our fiducial results using the full sample. However, there are noticeable differences when we limit to the manually inspected subset with the highest grades. These highest-grade lags, while being the most reliable, are limited to well-detected lags much shorter than the baseline. Thus using this subset of lags imposes severe selection biases in the $R-L$ plane. }

In Fig.~\ref{fig:lag_det_comp}, we compare the detected lags for the four lines using the three methods. As expected, overall the \pyroa lags agree with ICCF lags, with the latter producing larger lag uncertainties. On the other hand, \javelin produces too many artificial lags within the seasonal gaps, although there is agreement between \javelin and \pyroa for many objects. The underperformance of \javelin on large MOS-RM samples with multi-year light curves is unexpected.\footnote{We tested limited cases where we inflate the flux uncertainties in the \javelin run, and find that the \javelin lags are still incorrectly placed in seasonal gaps. We plan to further investigate this issue in future work. } Considering the flexibility of adding extra variance in the light curves with {\tt PyROA}, and the excellent agreement between \pyroa results and the empirical, model-independent ICCF results, we recommend \pyroa as an efficient method to measure lags for MOS-RM data. 

\subsection{Manual inspection of lags}\label{sec:inspection}

{For the manual inspection procedure, we follow the convention of our earlier work \citep{Grier_etal_2017,Grier_etal_2019,Homayouni_etal_2020} to assign an integer grade ($1-5$; higher grades correspond to better quality) to each detected lag (see below). This manual inspection process is of course subjective, but nevertheless provides a useful means to select the most reliable lags. Subsequently we consider the subset with lag grade$=4$ or 5 the ``Gold'' sample. These lag grades are included in Table~\ref{tab:sum}.}

{For detailed manual inspection, the grading criteria are as follows:}

\begin{enumerate}
    \item Spectral Variability: We use the RMS spectrum to ensure that the emission line of interest is variable over 7 years of observation. Therefore, we inspect the 7-year RMS spectrum to ensure that the line variability signal is present, and is not overlapping with the edges of the spectrum.
    \item Light Curve Variability: The continuum and line light curves demonstrate sufficient variability, and features that deviate from a smooth trend, suggesting underlying dynamic processes.
    \item Lag Posterior Peak: As an additional check on the quality of the measured lags, we require the PyROA lag pdf has a well-defined primary peak away from the lag search range boundaries.
    \item Lag Validation: We inspect the superposition of the shifted line light curve onto the continuum light curve to validate the measured lag against the continuum and line light curves. This is particularly important due to the seasonal gaps in SDSS-RM observations. This could coincide with lags of approximately 180, 540, etc. days, which could introduce false positive lag detections in our 7-year measurements. This validation process consists of two steps:
    \begin{itemize}
        \item Good alignment between the continuum and line light curves suggest that the measured lag represents a reliable estimate of the underlying light travel time delay, characteristic of physical RM lag.
        \item If the measured lag coincides with the seasonal gap, indicating fewer overlapping data points between the continuum and the shifted line light curve, then we examine to see if the overall trends between the two light curves and correlated variations match a reverberation picture. 
    \end{itemize}
    In either case, if the superimposed light curves show excessive overlap due to high data variance, we consider the light curves are an inadequate match for reliable lag determination.
\end{enumerate}

{The visual inspection procedure introduces complex selection functions on the resulting best-quality lags. As a consequence, we caution on the danger of removing certain lags using visual inspection from the sample in measuring the $R-L$ relation. We do not recommend on using the ``Gold'' lag sample alone or in combination with other heterogeneous lag samples to measure the $R-L$ relation. On the other hand, studies that are not affected by lag selection effects, such as the investigation of host galaxy properties, can utilize this ``Gold'' sample for best-measured RM masses. }

\subsection{False positive rate for individual lags}\label{sec:fap}

Using \javelin or \pyroa to estimate the false positive rate (FPR) for individual lags would be prohibitive, given the large number of lag detections in the SDSS-RM sample and the computational demands of running \javelin or \pyroa for the full light curve sets. However, Fig.~\ref{fig:stat_lag_det} demonstrates that using the much faster ICCF method to estimate the FPR is a reasonable approach. Here we pair the observed line light curve with randomly generated continuum light curves for mock lag measurements. To capture the variability characteristics of individual objects, we use the best-fit DRW parameters for each object to generate random continuum light curves.

For each detected positive lag, we generate 100 mock continuum light curves, and perform the ICCF analysis on each realization. If a mock lag detection from the CCCD has $r_{\rm max}>0.4$ (within $1\sigma$ range of the CCCD median) and recovers a positive lag at $\ge 2\sigma$, we count it as a false positive. The fraction of such incidents among the 100 mock realizations is recorded as the FPR for the observed lag. 

For the detected lags (i.e., \$line\$\_LAG\_DET$>0$ in Table~\ref{tab:sum}), the median FPR for each line ranges from 6\% to 14\%, which is in line with the expected rates using our $r_{\rm max}$ cut (\S\ref{sec:alias}). There is a small fraction of detections (0\%, 5\%, 5\% and 15\% for \halpha, \hbeta, \MgII\ and \CIV) for which the estimated ${\rm FPR}>0.3$. However, visually inspecting these cases, the light curves are often significantly variable and the lags robustly detected. Some examples are RM078 (\hbeta\ lag), RM101 (\halpha\ and \hbeta\ lags), RM578 (\MgII), RM256 (\CIV), etc. We argue that the FPR has been significantly overestimated in these cases for the following reasons: 

\begin{enumerate}
\item[{\it (i).}] The dominant variability feature in the light curve is a broad bump or dip. Given the typical DRW damping timescale, the mock continuum light curve often contains multiple peaks/dips over the 11-year baseline, leading to aliasing lags in the cross-correlation, some of which will be identified as positive detections. In that sense, even for some of the best measured lags in the local RM AGN sample, this FPR estimation approach would produce a non-negligible fraction of false positives, if we use simulated continuum light curves that are much longer than the measured lag.  

\item[{\it (ii).}] Our lag search range is generous (e.g., $\pm 5$ times the expected lag). When combined with the long light curve baseline, we will often recover an alias lag with the mock continuum light curve, boosting the FPR. In other words, the prior information on the anticipated lag is not used in this FPR estimation. 

\item[{\it (iii).}] Using ICCF in the FPR estimation is efficient, but not as robust as using \pyroa in measuring the lag, which may also tend to overestimate the FPR. 

\end{enumerate}

Given these reasons, and our visual inspection of detected lags with high FPR estimates, we conclude that the individual FPR estimates should not be used as the sole metric to assess the robustness of individual lag measurements. Some high-fidelity lag detections (e.g., Fig.~\ref{fig:lag_examp}) may have a high FPR. We do not recommend to use the reported FPR to select a refined lag sample, unless the lag measurement quality (e.g., $r_{\rm max}$ and lag errors) and diagnostic plots are checked simultaneously. 

Nevertheless, the most important utility of our FPR test is to ensure consistency. The statistical behaviors of our FPR estimates confirm that our reported lags have a low overall false positive rate of $\sim 10\%$, in concordance with our estimates in \S\ref{sec:alias} based on the positive/negative lag asymmetry (e.g., Fig.~\ref{fig:stat_lag_det}). 

\subsection{The virial coefficient $f$}\label{sec:f_factor}

To scale the measured virial product, ${\rm VP}\equiv c\tau_{\rm rest} \Delta V^2/G$, to the RM BH mass we need to multiply by the virial coefficient $f$. The exact value of $f$ depends on the choice of velocity indicator $\Delta V$, which can be the FWHM or line dispersion ($\sigma_{\rm line}$) measured either from the mean spectrum or from the rms spectrum \citep[e.g.,][]{Collin_etal_2006,Wang_etal_2020}. Throughout this paper we focus on the line dispersion measured from the rms spectrum, $\sigma_{\rm line, rms}$, as commonly adopted in RM studies. Given a specified line width, this scale factor depends on the geometry and kinematics of the BLR gas, and can differ significantly from object to object. For example, if the BLR gas is primarily distributed in a flattened structure as suggested in earlier observations of broad Balmer lines \citep[e.g.,][]{Wills_Browne_1986,Shen_Ho_2014,Mejia-Restrepo_etal_2018a}, then more face-on BLRs require larger $f$-factors than more edge-on BLRs. The situations may be more complicated for realistic kinematics of the BLR gas that could involve non-virial motions, e.g., for the high-ionization \CIV\ line \citep[e.g.,][]{Proga_etal_2000,Waters_etal_2016}. 

For traditional RM work that utilizes the average lag, the virial coefficient $f$ is usually determined by comparing the virial products to the expected BH masses from the local $M_{\rm BH}-\sigma_*$ relation \citep[e.g.,][]{Onken_etal_2004, Graham_etal_2011,Park_etal_2012,Woo_etal_2013}. These calibrations can only provide an average viral coefficient $\langle f\rangle$ for the sample (because the $M_{\rm BH}-\sigma_*$ relation itself has significant intrinsic scatter), varying from $\langle f\rangle=2.8$ \citep{Graham_etal_2011} to $\langle f\rangle=5.9$ \citep{Woo_etal_2013} if using $\sigma_{\rm line,rms}$ as the velocity indicator. Because the best-fit local $M_{\rm BH}-\sigma_*$ relation has been updated over the years \citep[see discussions in, e.g..][]{Kormendy_Ho_2013}, the calibration of the virial coefficient has also evolved. Moreover, because the $M_{\rm BH}-\sigma_*$ relation depends on galaxy bulge properties, in principle one should use different virial coefficients based on the bulge classification of the host galaxy \citep[e.g.,][]{Ho_Kim_2014} -- such bulge classifications, however, are usually difficult for broad-line AGNs and their accuracy largely depends on the expertise of the classifier.

Recent progress on dynamical modeling of the BLR with RM data has made it possible to derive the individual virial coefficient for AGNs \citep[e.g.,][]{Pancoast_etal_2014,Grier_etal_2017b,Williams_etal_2018,LiY_etal_2018,Williams_etal_2020,Bentz_etal_2021,LiY_etal_2022,Villafana_etal_2022,Bentz_etal_2022}. By dynamically modeling the BLR and constraining the model parameters with high-quality RM data, this approach can derive the underlying BH mass along with the geometrical and dynamical properties of the BLR. Comparing the dynamical BH mass with the average-lag-based virial product, one can derive the required virial coefficient for individual systems, rather than an average virial coefficient from the $M_{\rm BH}-\sigma_*$ relation. This approach, while still dependent on the details of the dynamical modeling, eliminates the systematics and limitations in the estimated average $f$-factor with BH-host relations. 

Based on $16$ local RM AGNs with available dynamical-modeling BH masses, \citet{Williams_etal_2018} found an average $\langle \log f\rangle=0.57$ dex with a dispersion of 0.19 dex, which is consistent (within 1$\sigma$) with the average $f$-factor derived from the $M_{\rm BH}-\sigma_*$ relation. Several studies \citep[e.g.,][]{Pancoast_etal_2014,Williams_etal_2018} also studied the correlations between the individual $f$-factor and AGN properties and found no significant trend \citep[but see][for the latest results on potential correlations]{Linzer_etal_2022,Villafana_etal_2023}. However, the individual $f$-factor strongly depends on the inclination of the BLR derived from dynamical modeling, where the BLR gas is often distributed in a thick-disk geometry. 

\begin{table*}
\caption{Format of the Summary Table}\label{tab:sum}
\centering
\begin{tabular}{lccc}
\hline\hline
Name & Format & Units & Description  \\
(1) & (2)  & (3) & (4)  \\
\hline
RMID & LONG & - & RMID of the target \\
ZSYS & DOUBLE & - & Systemic redshift from \citet{Shen_etal_2019b} \\
RA & DOUBLE & deg & J2000 right ascension \\
DEC & DOUBLE & deg & J2000 declination \\
DL\_MPC & DOUBLE & Mpc & Luminosity distance \\
LOGL5100\_YR1 & DOUBLE & [erg/s] & Monochromic luminosity at rest-frame 5100\,\AA\ from 2014 \\
LOGL3000\_YR1 & DOUBLE & [erg/s] & Monochromic luminosity at rest-frame 3000\,\AA\ from 2014 \\
LOGL1350\_YR1 & DOUBLE & [erg/s] & Monochromic luminosity at rest-frame 5100\,\AA\ from 2014 \\
F\_H\_5100\_YR1 & DOUBLE & & Host-to-total fraction (within SDSS fiber) from 2014\\
L5100\_MEAN & DOUBLE & erg/s & Mean rest-frame 5100\,\AA\ luminosity during 2014-2020\\
L5100\_MEDIAN & DOUBLE & erg/s & Median rest-frame 5100\,\AA\ luminosity during 2014-2020\\
L3000\_MEAN & DOUBLE & erg/s & Mean rest-frame 3000\,\AA\ luminosity during 2014-2020\\
L3000\_MEDIAN & DOUBLE & erg/s & Median rest-frame 3000\,\AA\ luminosity during 2014-2020\\
L2500\_MEAN & DOUBLE & erg/s & Mean rest-frame 2500\,\AA\ luminosity during 2014-2020\\
L2500\_MEDIAN & DOUBLE & erg/s & Median rest-frame 2500\,\AA\ luminosity during 2014-2020\\
L1350\_MEAN & DOUBLE & erg/s & Mean rest-frame 1350\,\AA\ luminosity during 2014-2020\\
L1350\_MEDIAN & DOUBLE & erg/s & Median rest-frame 1350\,\AA\ luminosity during 2014-2020\\
HB\_LOGMSE & DOUBLE & [$M_\odot$] & Single-epoch BH mass based on \hbeta\ \citep{Vestergaard_Peterson_2006} \\
MG2\_LOGMSE & DOUBLE & [$M_\odot$] & Single-epoch BH mass based on \MgII\ \citep{Shen_etal_2011} \\
C4\_LOGMSE & DOUBLE & [$M_\odot$] & Single-epoch BH mass based on \CIV\ \citep{Vestergaard_Peterson_2006} \\
LAG\_OBS\_EXP & DOUBLE & days & Expected observed-frame lag for \hbeta \\
\hline 
\$line\$\_W\_MEAN & DOUBLE[2] & km/s & [FWHM, line dispersion] from the mean spectrum \\ 
\$line\$\_W\_RMS & DOUBLE[2] & km/s & [FWHM, line dispersion] from the RMS spectrum \\ 
\$line\$\_W\_MEAN\_ERR & DOUBLE[2] & km/s & Uncertainty in \$line\$\_W\_MEAN \\ 
\$line\$\_W\_RMS\_ERR & DOUBLE[2] & km/s & Uncertainty in \$line\$\_W\_RMS \\
\$line\$\_STAT\_NAME & STRING[24] &   & Names of line light curve statistics \\ 
\$line\$\_STAT\_VALUE & DOUBLE[24] &   & Values of line light curve statistics \\ 
\hline
DRW\_SIGMA & DOUBLE & & Amplitude from continuum DRW fit \\
DRW\_SIGMA\_ERR & DOUBLE[2] & & Uncertainty in DRW\_SIGMA \\
DRW\_TAU & DOUBLE & days & Observed-frame damping timescale from continuum DRW fit \\
DRW\_TAU\_ERR & DOUBLE[2] & days & Uncertainty in DRW\_TAU \\
CONT\_FRAC\_RMS & DOUBLE & & Intrinsic fractional RMS variability from 11-yr continuum light curve\\
CONT\_FRAC\_RMS\_ERR & DOUBLE & & Uncertainty in CONT\_FRAC\_RMS \\
\hline
LAG\_SEARCH & DOUBLE[2] & days & Lag search range (in observed frame)\\
\$line\$\_LAG\_DONE & INT & & $=1$ if a lag measurement is attempted\\
\$line\$\_LAG\_DET & DOUBLE &  days & Fiducial observed-frame lag measurement \\
\$line\$\_LAG\_DET\_ERR & DOUBLE[2] & days & Uncertainty in \$line\$\_LAG\_DET $[-,+]$ \\
\$line\$\_LAG\_FPR & DOUBLE &   & False positive rate of the lag; $-1$ if the lag is not detected \\
\$line\$\_MRM & DOUBLE & $M_{\odot}$ & RM BH mass based on the fiducial lag \\
\$line\$\_MRM\_ERR & DOUBLE[2] & $M_{\odot}$ & Uncertainty in \$line\$\_MRM $[-,+]$\\
\$line\$\_ICCF\_* & DOUBLE & & ICCF results \\ 
\$line\$\_JAVELIN\_* & DOUBLE & & \javelin results \\ 
\$line\$\_PYROA\_* & DOUBLE & & PyROA results \\ 
\$line\$\_LAG\_GRADE & LONG & & Lag quality grade by visual inspection (1: poorest; 5: best) \\ 
\hline
\hline
\end{tabular}
\tablecomments{For each of the lag measuring methods (ICCF, \javelin and \pyroa), we provide the best lag (*\_LAG), lag uncertainties (*\_LAG\_ERR), bounds of the identified primary peak in the lag posterior (*\_PEAK\_BOUNDS), the fraction of rejected samples (*\_FRAC\_REJ), and the maximum Pearson correlation coefficient within $\pm 1\sigma$ of the best lag (*\_LAG\_RMAX). In a handful of cases, the reported \pyroa lag may lie outside the LAG\_SEARCH range. These are cases where the original \pyroa lag measurement reached 90\% of the search boundary, and we have expanded the search range (for that specific line) to the maximum allowed range ([$-2500$, 4000]\,days). Finally, we caution that the false positive rate estimate for individual lag detections (\$line\$\_LAG\_FPR) is often an overestimation, and should be combined with the lag measurement diagnostic plot to assess the fidelity of the reported lag. }
\end{table*}

\begin{figure}
  \centering
    \includegraphics[width=0.48\textwidth]{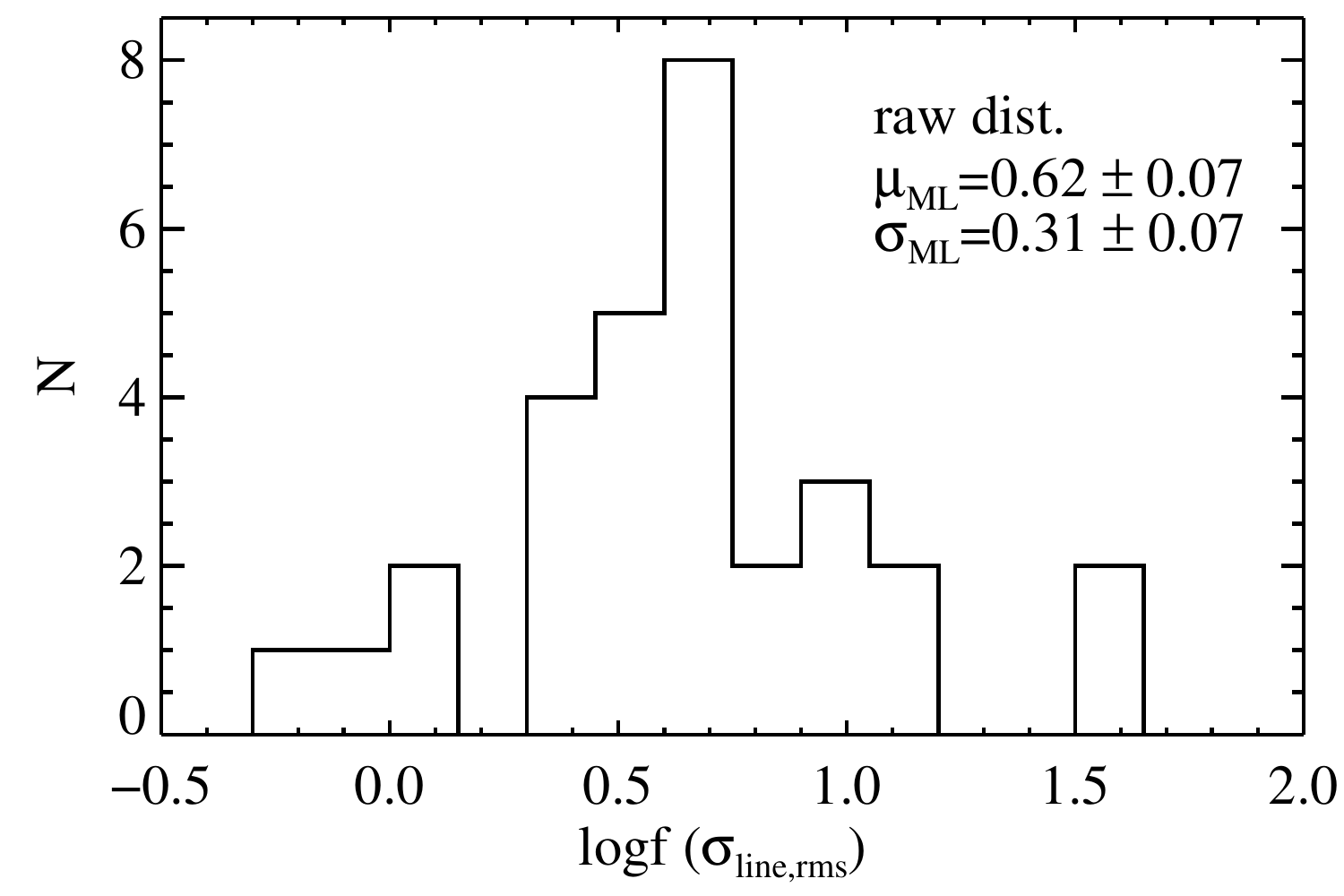}
    \includegraphics[width=0.48\textwidth]{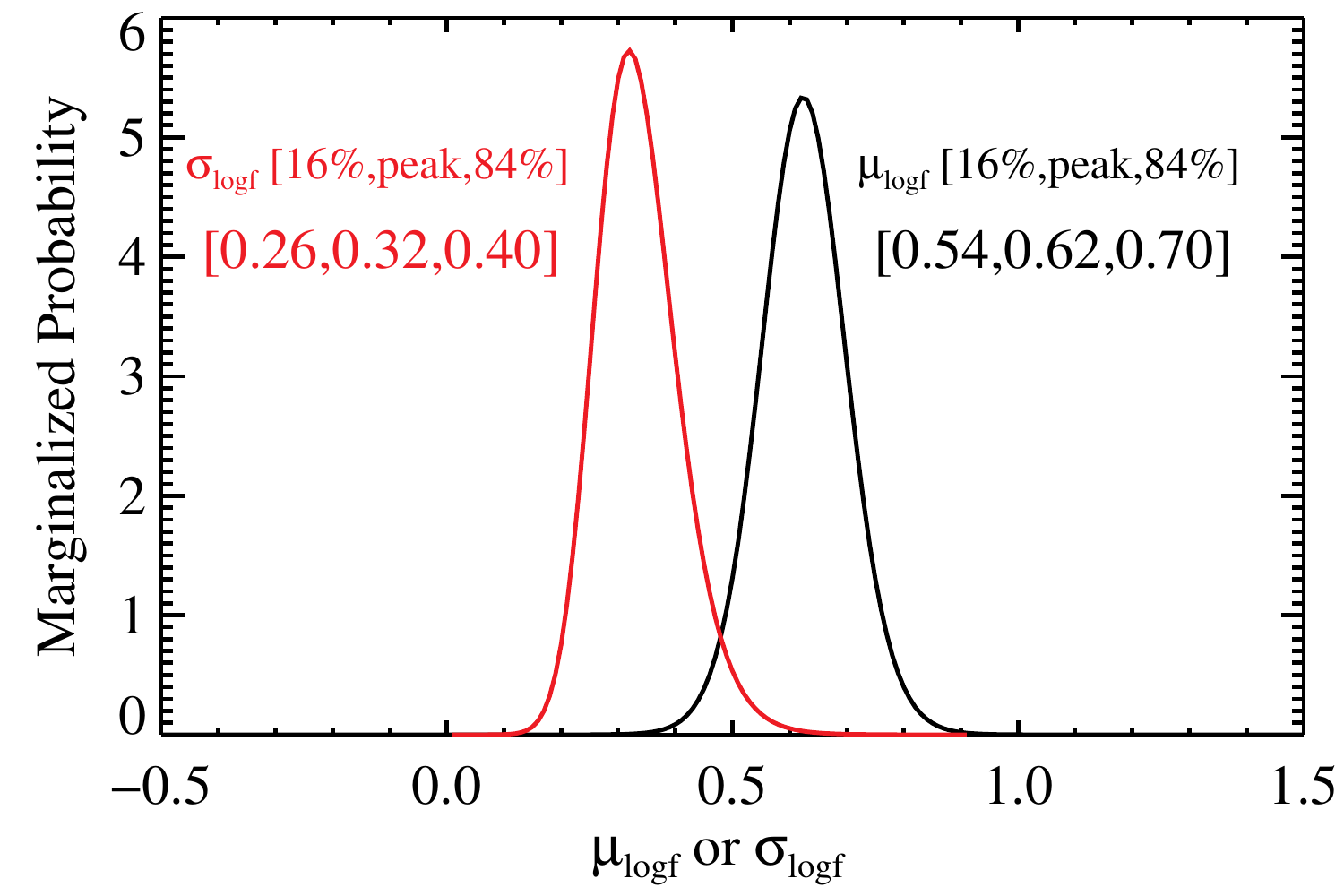}
    \caption{Constraints on the mean and dispersion of the distribution of individual $f$ factors based on the line dispersion measured from the RMS spectrum, using a sample of 30 objects with dynamical-modeling and RM BH masses as compiled in Table~\ref{tab:f_sample}. {\em Top:} raw distribution of individual $f$ factors. The ML estimates of the sample mean and intrinsic 1$\sigma$ dispersion are marked. {\em Bottom:} Marginalized probability distributions of the mean and intrinsic 1$\sigma$ dispersion of $f$ factors. The 16\%, peak and 84\% locations of the distributions are marked. }
    \label{fig:f_factor}
\end{figure}

Here we compile the latest sample of RM AGNs with dynamical modeling BH masses from the literature. This sample contains 30 objects (NGC5548 has two independent measurements) and the details are summarized in Table \ref{tab:f_sample}. We only compile the virial product computed using the line dispersion measured from the RMS spectrum, since this is the most commonly adopted line width in RM work. Fig.~\ref{fig:f_factor} (top) shows the distribution of the individual $f$-factors ($\log f (\sigma_{\rm line,rms})$) in the sample, which is peaked around $\log f\approx 0.6$. 

To quantify the mean and intrinsic dispersion in $\log f (\sigma_{\rm line,rms})$, we follow the approach in, e.g., \citet{Pancoast_etal_2014}. We assume the intrinsic $\log f$ distribution is a Gaussian distribution with mean value $\mu_{\rm logf}$ and dispersion $\sigma_{\rm logf}$. The posterior probability distributions of parameters $\mu_{\rm logf}$ and $\sigma_{\rm logf}$, given the sample of measured $\log f$, are calculated following appendix B in \citet{Pancoast_etal_2014} that takes into account the posterior probability distribution function (PDF) of individual $\log f$ measurements. Because these dynamical modeling papers typically do not publish the full PDF of $\log f$, we simply adopt symmetric Gaussian errors for these $\log f$ measurements. The constrained marginalized PDFs for $\mu_{\rm logf}$ and $\sigma_{\rm logf}$ are shown as black and red lines in Fig.~\ref{fig:f_factor} (bottom), where the peak and 16th/84th percentile of the distribution are marked. The constrained mean $\log f$ is $\mu_{\rm logf}=0.62\pm 0.08$, and the constrained intrinsic dispersion $\sigma_{\rm logf}=0.32^{+0.08}_{-0.06}$, where uncertainties are $1\sigma$. 

Alternatively, given the set of $\log f$ measurements and their uncertainties, we can calculate the mean and intrinsic dispersion in $\log f$ using a maximum-likelihood (ML) estimator described in \citet{Shen_etal_2019b}. This ML approach also assumes that the intrinsic $\log f$ values are Gaussian distributed around mean value $\mu_{\rm ML}$ with a dispersion of $\sigma_{\rm ML}$. The results are $\mu_{\rm ML}=0.62\pm 0.07$ and $\sigma_{\rm ML}=0.31\pm 0.07$, which are nearly identical to the results using the PDF approach above. 

Thus we conclude, based on a sample of 30 independent measurements of $\log f(\sigma_{\rm line,rms})$ with both VPs and dynamical-modelling BH masses, that the mean logarithmic virial factor is $\left<\log f\right>=0.62\pm 0.07$. This value is consistent with most of the recent determinations of the virial factors either using BH-host scaling relations or using dynamical masses from RM \citep{Onken_etal_2004, Graham_etal_2011,Park_etal_2012,Woo_etal_2013,Ho_Kim_2014,Pancoast_etal_2014,Grier_etal_2017b,Williams_etal_2018}. Importantly, this literature sample compiled in Table \ref{tab:f_sample} allows us to robustly constrain the intrinsic dispersion of the individual virial factor to be $\sim 0.3$~dex (or a factor of two in $f$). This dispersion in individual $f$ factors sets a typical systematic uncertainty of RM-based BH masses derived using a single, average $f$ factor. This intrinsic scatter in $\log f$ is notably larger than (but consistent within $\sim 2\sigma$) the value of $\sigma_{\rm logf}=0.14\pm0.10$ reported in \citet{Williams_etal_2018} based on 16 objects with RM data and dynamical masses. The additional objects in our sample require a larger $\sigma_{\rm log f}$ to account for their dispersion. 

In our analysis we have only focused on the $f$-factor based on $\sigma_{\rm line,rms}$, and ignored potential dependences of the $f$-factor on AGN properties. \citet{Villafana_etal_2023} used a similar AGN sample with dynamical BH masses to investigate correlations between the virial factors based on different line width definitions and AGN properties, for \hbeta\ only. Such correlations, if confirmed with future larger samples with dynamical BH masses, will further shed light on the structure of AGN BLRs and improve BH mass estimation using average RM lags. 

In the following discussion, we adopt the same average $f$-factor for all four broad lines. As shown in \S\ref{sec:disc_rm}, the overall agreement between the RM BH masses from two different lines indicates adopting the same $f$-factor for all lines is a reasonable approach. However, we acknowledge that the SDSS-RM sample is different from the local RM AGN sample in terms of luminosity and BH mass. Whether or not the average $f$-factors differ among different samples is a topic for future studies.  

\subsection{Lags and RM masses}\label{sec:lag_mass}

Our fiducial RM masses are computed from the measured broad-line lags combined with the line dispersion measured from the RMS spectrum of the monitoring data. In mathematical form:
\begin{equation}
M_{\rm RM} \equiv \frac{\left<f\right>c\tau_{\rm rest}\sigma_{\rm line,rms}^2}{G}\ .
\end{equation}
where $\left<f\right>$ is the geometric mean $f$ factor when using $\sigma_{\rm line, rms}$ for the line width. 

The PrepSpec $\sigma_{\rm line,rms}$ measurements are based on the continuum-subtracted RMS spectrum, which corresponds to the true line-only variability. Some earlier approaches \citep[e.g.,][]{Peterson_etal_2004} generate the RMS spectrum from the (total) emission line+continuum spectra and measure the line dispersion by removing a local continuum from the RMS spectrum. The latter approach can produce line dispersions that are on average biased low by $\sim 20-30\%$ if the monitoring period is not much longer than the lag \citep[e.g., by a factor of $\lesssim 3$, ][]{Barth_etal_2015,Wang_etal_2019}. Fortunately, this monitoring duration criterion is satisfied in most of the recent low-$z$ RM campaigns listed in Table~\ref{tab:f_sample}, and many of these campaigns (such as the LAMP project) measured $\sigma_{\rm rms}$ from the properly constructed line-only RMS spectrum. Therefore we can safely apply the mean $f$ factor derived in \S\ref{sec:f_factor} to the SDSS-RM sample. 

\begin{figure}
  \centering
    \includegraphics[width=0.48\textwidth]{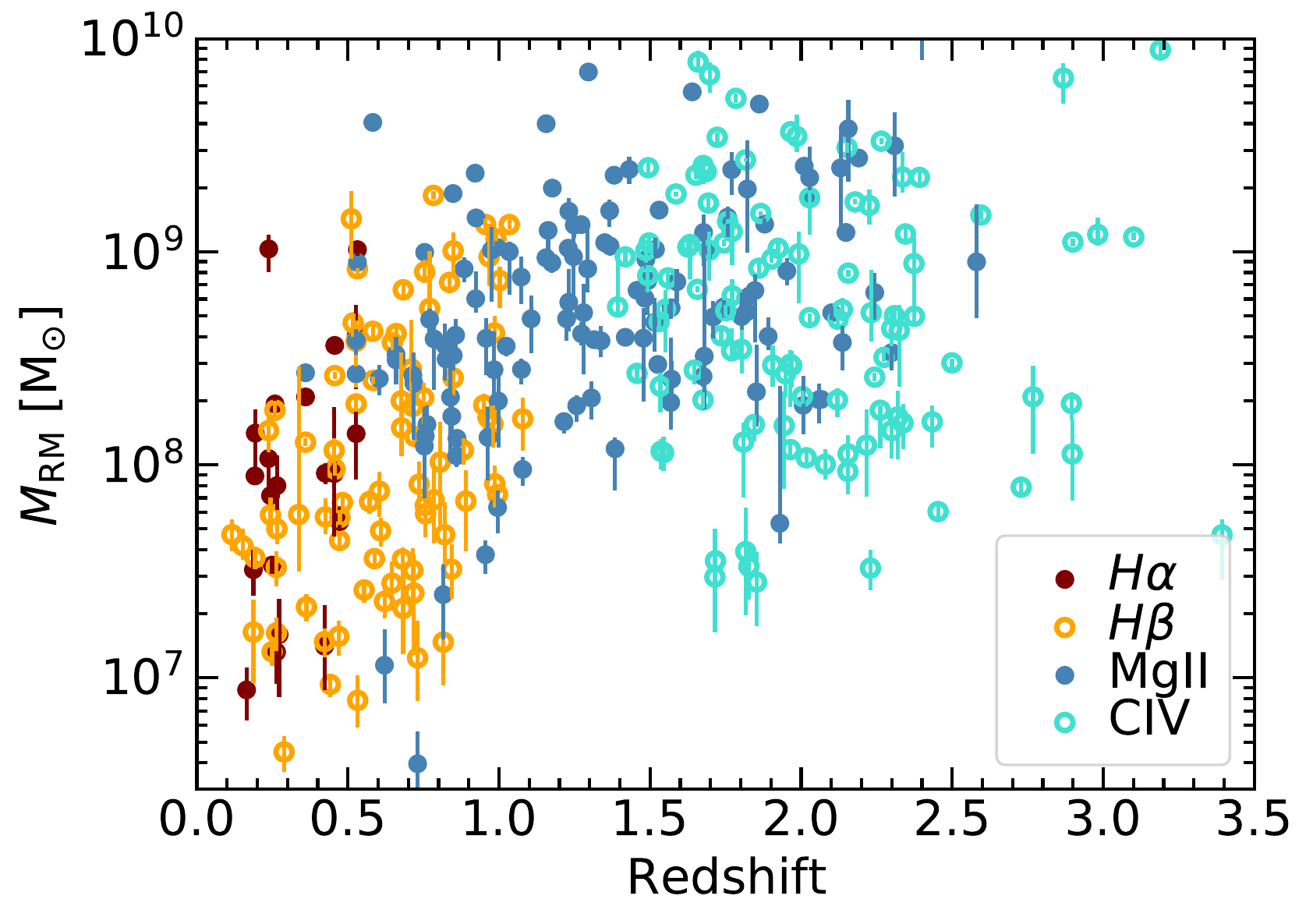}
    \caption{Redshift distribution of RM masses derived from \halpha, \hbeta, \MgII\ and \CIV\ lags for the SDSS-RM sample. Error bars are measurement uncertainties only and do not include the $\sim 0.3$~dex systematic uncertainty in $M_{\rm RM}$. }
    \label{fig:z_Mrm}
\end{figure}

We compile our lags and RM masses in Table~\ref{tab:sum}. In Fig.~\ref{fig:z_Mrm} we show the distribution of RM masses and redshift for the SDSS-RM sample. With $\sim 300$ unique objects, this sample represents the largest sample of quasars with direct RM-based BH masses up to $z\sim 3.5$.

\begin{table*}
\caption{Compiled VPs and dynamical-modeling BH masses from previous RM studies. For simplicity, we have symmetrized all uncertainties in order to use these uncertainties in our PDF calculation (\S\ref{sec:f_factor}). Reference keys are: Barth11b \citep{Barth_etal_2011b}, Barth13 \citep{Barth_etal_2013}, Barth15 \citep{Barth_etal_2015}, Bentz09b \citep{Bentz_etal_2009b}, Bentz21 \citep{Bentz_etal_2021}, Bentz22 \citep{Bentz_etal_2022}, Busch14 \citep{Busch_etal_2014}, DeRosa18 \citep{DeRosa_etal_2018}, Grier12 \citep{Grier_etal_2012}, Grier17 \citep{Grier_etal_2017b}, Li18 \citep{LiY_etal_2018}, Li22 \citep{LiY_etal_2022}, Pancoast14 \citep{Pancoast_etal_2014}, Pei17 \citep{Pei_etal_2017}, Peterson04 \citep{Peterson_etal_2004}, U22 \citep{U_etal_2022}, Villafana22 \citep{Villafana_etal_2022}, Williams18 \citep{Williams_etal_2018}, Williams20 \citep{Williams_etal_2020}, Zhang19 \citep{Zhang_etal_2019}.  }\label{tab:f_sample}
\centering
\begin{tabular}{lcccccccc}
\hline\hline
Name & logVP & $\sigma_{\log {\rm VP}}$ & VP ref & $\log M_{\rm dyn}$ &  $\sigma_{\log M_{\rm dyn}}$ & $M_{\rm dyn}$ ref & $\log f$ & $\sigma_{\log f}$ \\
(1) & (2)  & (3) & (4)  & (5) & (6) & (7) & (8) & (9) \\
\hline
NGC4151 & 6.70 & 0.04 & DeRosa18 & 7.22 & 0.11 & Bentz22 & 0.52 & 0.11 \\
PG2209+184 & 6.81 & 0.09 & U22 & 7.53 & 0.20 & Villafana22 & 0.72 & 0.21 \\
RBS1917 & 6.50 & 0.17 & U22 & 7.04 & 0.29 & Villafana22 & 0.54 & 0.34 \\
MCG+04-22-042 & 6.53 & 0.07 & U22 & 7.59 & 0.35 & Villafana22 & 1.06 & 0.36 \\
NPM1G+27.0587 & 6.71 & 0.27 & U22 & 7.64 & 0.38 & Villafana22 & 0.93 & 0.47 \\
Mrk1392 & 7.15 & 0.06 & U22 & 8.16 & 0.12 & Villafana22 & 1.01 & 0.13 \\
RBS1303 & 6.75 & 0.09 & U22 & 6.79 & 0.15 & Villafana22 & 0.04 & 0.17 \\
Mrk1048 & 6.69 & 0.55 & U22 & 7.79 & 0.46 & Villafana22 & 1.10 & 0.72 \\
RXJ2044.0+2833 & 6.43 & 0.07 & U22 & 7.09 & 0.17 & Villafana22 & 0.66 & 0.18 \\
Mrk841 & 7.02 & 0.19 & U22 & 7.62 & 0.40 & Villafana22 & 0.60 & 0.44 \\
Mrk335 & 6.66 & 0.05 & Grier12 & 7.25 & 0.10 & Grier17 & 0.59 & 0.11 \\
Mrk1501 & 7.52 & 0.06 & Grier12 & 7.86 & 0.18 & Grier17 & 0.34 & 0.20 \\
3C120 & 7.09 & 0.04 & Grier12 & 7.84 & 0.17 & Grier17 & 0.75 & 0.17 \\
PG2130+099 & 6.92 & 0.04 & Grier12 & 6.92 & 0.23 & Grier17 & 0.00 & 0.24 \\
Mrk50 & 6.78 & 0.07 & Barth11 & 7.50 & 0.21 & Williams18 & 0.72 & 0.23 \\
Mrk141 & 6.76 & 0.38 & Williams18;Barth15 & 7.46 & 0.18 & Williams18 & 0.70 & 0.42 \\
Mrk279 & 6.80 & 0.12 & Peterson04 & 7.58 & 0.08 & Williams18 & 0.78 & 0.14 \\
Mrk1511 & 6.48 & 0.08 & Barth13;Barth15 & 7.11 & 0.18 & Williams18 & 0.63 & 0.20 \\
NGC4593/Mrk1330 & 6.24 & 0.14 & Barth13;Barth15 & 6.65 & 0.21 & Williams18 & 0.41 & 0.25 \\
Zw229--015 & 6.28 & 0.09 & Barth11b & 6.94 & 0.14 & Williams18 & 0.66 & 0.17 \\
PG1310--108$^b$ & 6.74 & 0.30 & Busch14 & 6.48 & 0.20 & Williams18 & $-0.26$ & 0.36 \\
Mrk142 & 5.49 & 0.17 & Bentz09b & 6.23 & 0.35 & Li18 & 0.74 & 0.39 \\
NGC5548 & 7.17 & 0.14 & Pei17 & 7.54 & 0.29 & Williams20 & 0.37 & 0.32 \\
Arp151/Mrk40$^a$ & - & - & Pancoast14 & 6.62 & 0.12 & Pancoast14 & 0.51 & 0.12 \\
Mrk1310$^a$ & - & - & Pancoast14 & 7.42 & 0.27 & Pancoast14 & 1.63 & 0.27 \\
NGC5548$^a$ & - & - & Pancoast14 & 7.51 & 0.18 & Pancoast14 & 0.42 & 0.18 \\
NGC6814$^a$ & - & - & Pancoast14 & 6.42 & 0.21 & Pancoast14 & $-0.14$ & 0.21 \\
SBS1116+583A$^a$ & - & - & Pancoast14 & 6.99 & 0.28 & Pancoast14 & 0.96 & 0.28 \\
NGC3783 & 6.69 & 0.08 & Bentz21 & 7.51 & 0.20 & Bentz21 & 0.82 & 0.21 \\
3C273 & 7.54 & 0.04 & Zhang19 & 9.06 & 0.24 & Li22 & 1.52 & 0.24 \\\hline
\hline
\end{tabular}
\tablecomments{ $^a$$\log f$ values directly taken from Pancoast14, which did not include errors in $\log {\rm VP}$. $^b$VP in Busch14 was computed using $\sigma_{\rm line}$ from the mean spectrum and an estimated lag from the mean $R-L$ relation \citep{Bentz_etal_2013}. Note that NGC5548 has two independent $\log f$ measurements, both of which are included in our analysis. }
\end{table*}

\section{Discussion}\label{sec:disc}

\begin{figure*}[!h]
  \centering
    \includegraphics[width=0.98\textwidth]{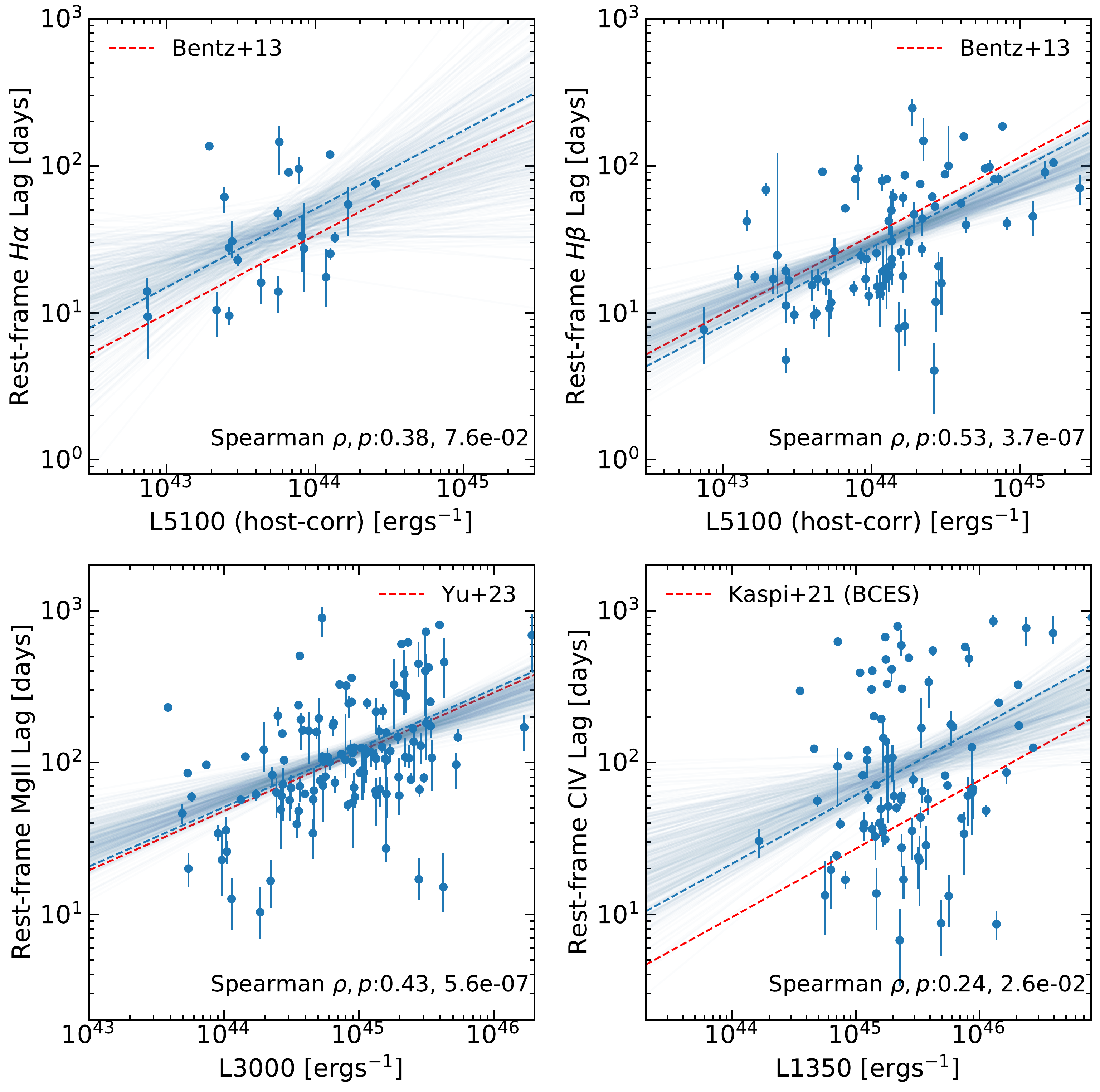}
    \caption{$R-L$ relations for the four broad lines in this work. In each panel, the red dashed line shows the most recent literature relation \citep{Bentz_etal_2013,Yu_etal_2023,Kaspi_etal_2021}. The light blue lines are random draws from the linear regression of predicting $\log\tau$ with $\log L$ using the Bayesian approach in \citet{Kelly_2007}. The blue dashed line is a forced regression fit with slope fixed to that measured in earlier work. Overall, measured SDSS-RM lags follow these earlier relations, but there are notable deviations (in particular for \CIV). Our lag measurements suggest that the \hbeta\ and \MgII\ $R-L$ relations are reasonably tight (with an intrinsic scatter of $\sim 0.3$~dex in lag), while the \CIV\ $R-L$ relation has substantially larger scatter ($\sim 0.5$~dex) than the other two lines. The Spearman rank-order test results on the SDSS-RM sample (correlation coefficient $\rho$ and $p$-value $p$) are marked in each panel. These results are discussed in detail in \S\ref{sec:disc_rm} and Table~\ref{tab:regression}. }
    \label{fig:rl_relation}
\end{figure*}

\begin{figure*}
  \centering
    \includegraphics[width=0.98\textwidth]{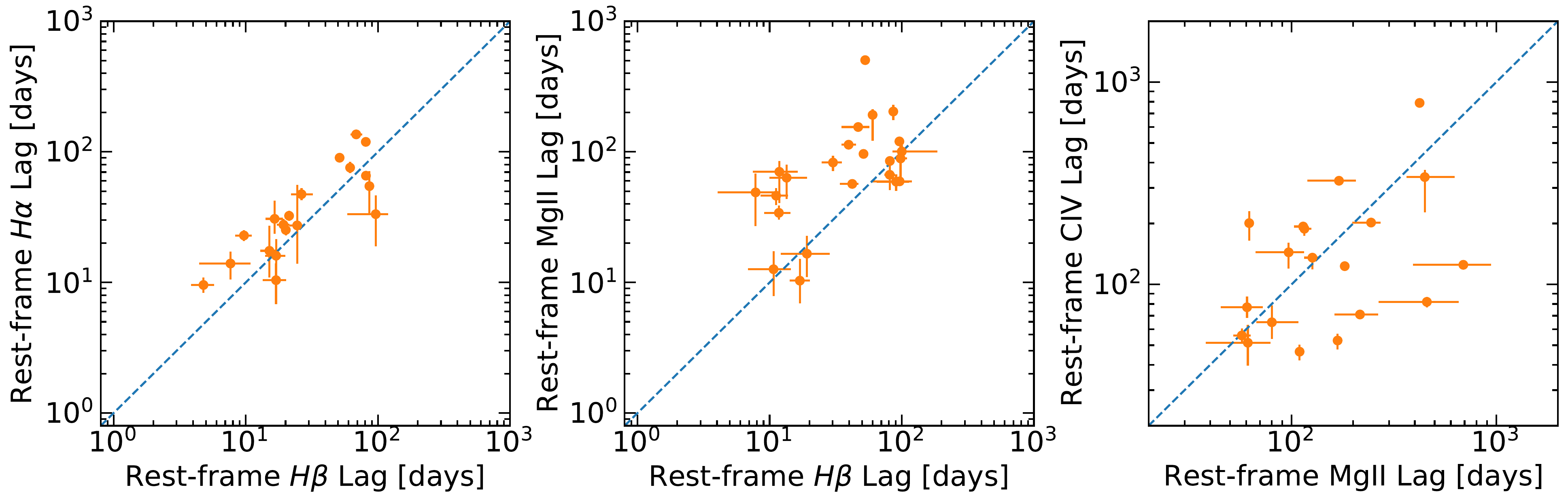}
    \caption{Comparisons between lags from two different lines for the same object. \halpha\ and \MgII\ lags are typically longer than \hbeta\ lags, while \CIV\ lags are somewhat shorter (but the sample statistics are limited and the scatter is large) than \MgII\ lags. }
    \label{fig:lag_comp}
\end{figure*}

\begin{figure*}
  \centering
   \includegraphics[width=0.98\textwidth]{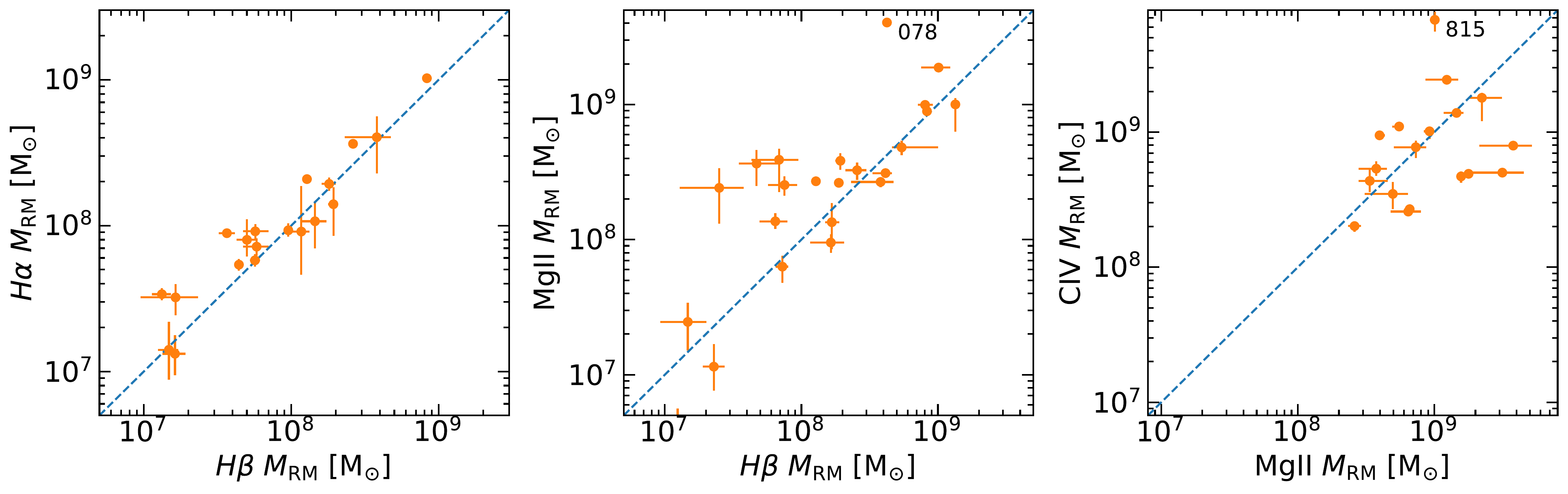}
    \caption{Comparisons between $M_{\rm RM}$ from two lines for the same object. For \MgII, we have corrected the rms line dispersion for the velocity split $V\approx 750\,{\rm km\,s^{-1}}$ of the doublet ($\sigma_{\rm true}=\sqrt{\sigma_{\rm measure}^2 - (V/2)^2}$). There is general agreement between the RM masses measured from two different lines, validating the RM technique of measuring BH masses. The two most significant outliers in the line comparisons are marked with their RMIDs. }
    \label{fig:mrm_comp}
\end{figure*}

\begin{figure*}
  \centering
   \includegraphics[width=0.98\textwidth]{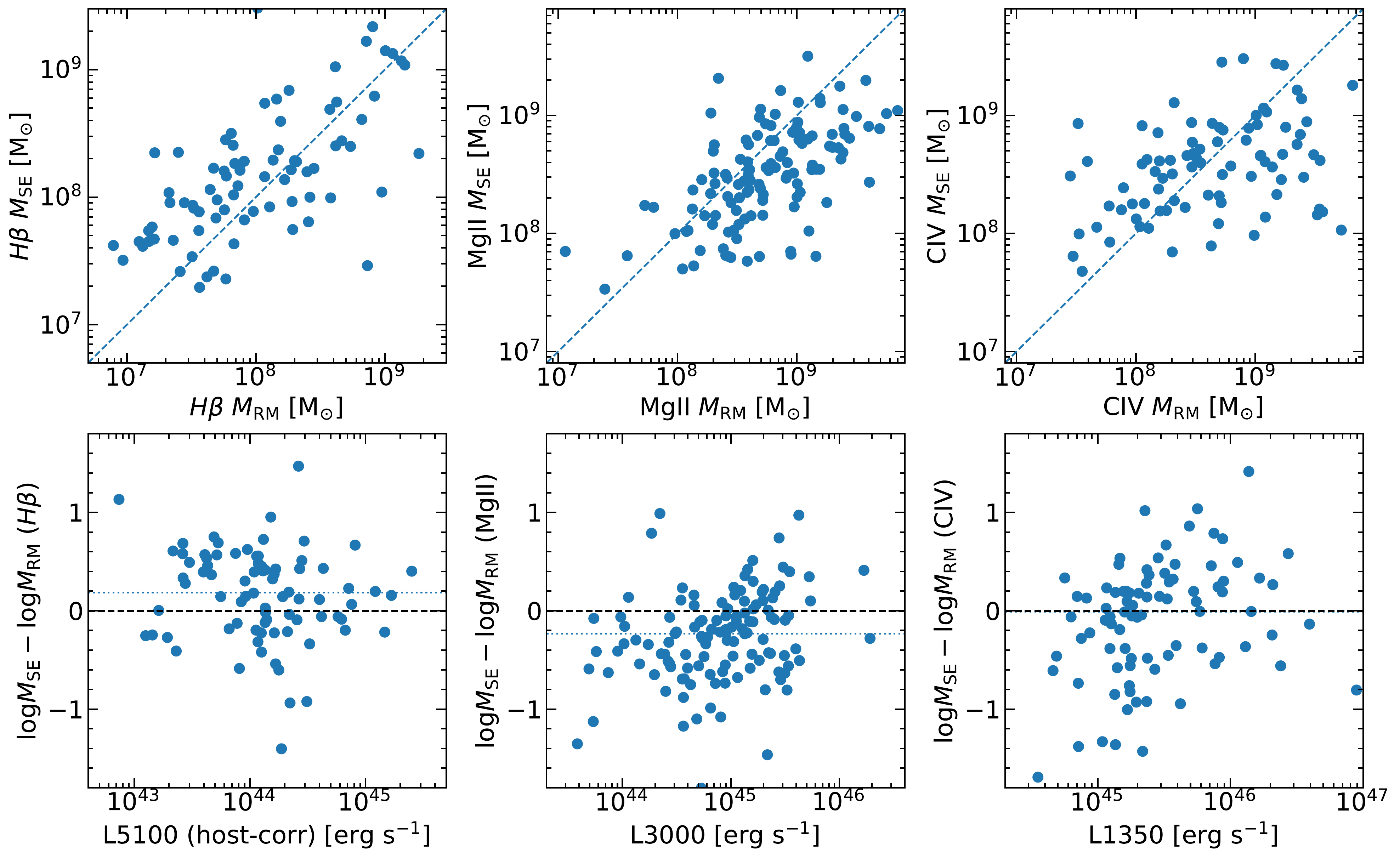}
    \caption{Comparisons between $M_{\rm RM}$ and $M_{\rm SE}$ (using earlier SE recipes) for the same line. \hbeta\ has the best correlation while \CIV\ has the worst correlation. However, for all three lines, there are good correlations between $\sigma_{\rm line, rms}$ and ${\rm FWHM_{mean}}$, so the poor mass correlation for \CIV\ is mainly driven by the poor $R-L$ relation. Because the single-epoch mass is solely determined by continuum luminosity and line width, its dynamic range can be artificially suppressed compared with RM-based masses \citep{Shen_2013}. In the bottom panels, the blue dotted line is a simple median of the residuals, rather than the average offset weighted by measurement uncertainties. }
    \label{fig:mrm_mse_comp}
\end{figure*}

\subsection{RM results}\label{sec:disc_rm}

For a MOS-RM program like ours that targets a flux-limited quasar sample across broad redshift and luminosity ranges, the common baseline of the program may impose significant selection biases in the measured lags, missing long lags that cannot be measured, given the length of the light curves. Fortunately for the SDSS-RM program, the final light curves span an effective baseline of 11 years, sufficient to recover the vast majority of expected lags for the SDSS-RM sample (see Fig.~\ref{fig:z_lag}). 

Fig.~\ref{fig:rl_relation} shows our measured lags for the four lines in their respective $R-L$ planes, where luminosity is taken as the usual local continuum luminosity corresponding to each line and is the median luminosity over the spectroscopic baseline. For $L5100$, we have corrected for host contamination, using the estimated host fraction (within the SDSS fiber) from the 2014 spectroscopy \citep{Shen_etal_2015b}. When fitting an $R-L$ relation to these data, we only use the SDSS-RM sample, rather than combining the SDSS-RM sample with other literature samples. This is because by number the SDSS-RM sample over-represents the population over the probed luminosity range, and would likely dominate the global fit. A joint fit with other lag samples probing different luminosity ranges will be the objective of future work, after selection effects are properly taken into account.  

For \hbeta\ and \MgII, the luminosity range in the SDSS-RM lag sample spans $\sim 2$~dex and a $R-L$ relation is clearly present. Overall the SDSS-RM lag sample follows the reported global $R-L$ relation in the latest studies that derived such relations using heterogeneous samples \citep{Bentz_etal_2013,Yu_etal_2023}. However, the \hbeta\ lags for SDSS-RM quasars on average fall slightly below the local relation by $\sim 0.1$~dex, if we fix the slope to that in the local relation in the regression. A similar, albeit slightly larger, average offset in \hbeta\ lags was reported in earlier SDSS-RM results based on shorter baselines \citep{Grier_etal_2017,Fonseca_Alvarez_etal_2020}. Because the baseline is limited in earlier SDSS-RM lag measurements, this average offset from the local $R-L$ relation might have been overestimated in our earlier papers. Note that the host correction in \citet{Shen_etal_2015b} is on average smaller by $\sim 30-40\%$ than that estimated from image decomposition \citep{Li_etal_2021a,Li_etal_2023}. If we further reduce the 5100\,\AA\ luminosities by $\sim 0.1$~dex, the lag offset in the $R-L$ relation would become negligible. The intrinsic scatter in our \hbeta\ lags around the mean $R-L$ relation is $\sim 0.32$~dex. 

The dispersion around the canonical Bentz et~al.\ $R-L$ relation for \hbeta\ have gained substantial interest in the past few years \citep[e.g.,][]{Du_etal_2016, DallaBonta_etal_2020, Du_Wang_2019, Czerny_etal_2019b,Martinez_Aldama_etal_2019}. It is suggested \citep[e.g.,][]{Du_Wang_2019, Czerny_etal_2019b,Fonseca_Alvarez_etal_2020} that this dispersion is due to different spectral energy distributions between the local RM AGN sample in \citet{Bentz_etal_2013} and other samples. In particular, the SEAMBH collaboration has targeted local AGNs with high accretion rates, and found that the lag deviation from the canonical Bentz et~al.\ $R-L$ relation can reach as much as a factor of few \citep[e.g.,][]{Du_etal_2016,Du_Wang_2019}. Different accretion rates of the AGN lead to different ratios between the ionizing continuum (directly responsible for the lag) and the local continuum near the line, which in turn leads to an offset from the mean $R-L$ relation. From a theoretical point of view, it is expected that there is a dispersion in lags at fixed optical luminosity caused by the diversity in BH mass and accretion rate and subsequently the underlying SED (Wu~et~al., in prep).

For \MgII, we found that there is a strong $R-L$ relation over the luminosity range probed by our lag sample (Fig.~\ref{fig:rl_relation}). The slope is somewhat shallower than 0.5. Combining the OzDES-RM results on \MgII\ lags and literature \MgII\ lag measurements, \citet{Yu_etal_2023} measure a global $R-L_{3000}$ relation for \MgII, that is consistent with our measurements. The intrinsic scatter in the \MgII\ $R-L$ relation is $\sim 0.32$~dex.   

The luminosity range for the SDSS-RM \CIV\ lag sample is limited to $\sim 1$~dex. Thus, similar to \citet{Grier_etal_2019}, we cannot reliably constrain a $R-L$ relation using the SDSS-RM sample alone. \citet{Kaspi_etal_2021} recently compiled literature \CIV\ lag results that span $\sim 8$ orders of magnitude in luminosity and derived a global $R-L$ relation for \CIV. The SDSS-RM lags generally fall on this relation, with a slight offset to longer lags. It is possible that we were able to recover these longer lags given the baseline of our light curves. But more importantly, within the luminosity range probed by the SDSS-RM \CIV\ lag sample, the scatter around a mean $R-L$ relation is $\sim 0.5$~dex, substantially larger than those seen for \hbeta\ and \MgII. This large scatter argues against using the single-epoch \CIV\ BH mass estimator, which assumes there is a tight $R-L$ relation for \CIV. Since \CIV\ is a high-ionization line, the diversity in the SED due to accretion parameters could be the main driver for the large scatter of lags around the mean $R-L$ relation, where we are using the local continuum $L_{\rm 1350}$ near \CIV\ as the luminosity indicator.

Given the large sample size of SDSS-RM, there are subsets of quasars for which we successfully measured lags for two lines. Fig.~\ref{fig:lag_comp} compares the lags measured from two different lines in the same quasar. Compared with \hbeta\ lags, \halpha\ and \MgII\ lags are on average longer, while \CIV\ lags are somewhat shorter than \MgII\ lags overall (but the sample size is small). These results are consistent with earlier RM results for local AGNs \citep{Peterson_etal_2004}, where BLR stratification and radiative transfer effects are invoked to explain these lag differences \citep[e.g.,][]{Goad_etal_2012,Guo_etal_2020}.

We also compare the RM masses from two different lines in the same quasar in Fig.~\ref{fig:mrm_comp}. The RM masses between two lines are generally consistent with each other, suggesting that the line dispersion measurements are adequate. In addition, this consistency implies using the same virial factor (derived from \hbeta) for all lines does not introduce significant mass offsets in the average sense. There are, however, individual objects for which the RM masses from two lines differ significantly. We examined the lag and line-width measurements for these objects, and found overall these measurements are reasonable.\footnote{We found two exceptions where the lag measurements may be problematic. For RM078 (the most significant outlier in the \hbeta-\MgII\ comparison), the \MgII\ lag is much longer than \hbeta, which is potentially due to different BLR structures for the two lines. For RM815 (the most significant outlier in the \MgII-\CIV\ comparison), the variability in \MgII\ and \CIV\ is weak and both lag measurements may be spurious. } It is possible that these objects have different BLR kinematics for different lines and their respective virial factors differ \citep[e.g.,][]{Runnoe_etal_2013a, Fries_etal_2023}. Either way, our results highlight the importance of deriving velocity-resolved RM results for \MgII\ and \CIV\ to fully understand the BLR structure and kinematics for these lines. 

\subsection{Comparison with earlier SDSS-RM results}\label{sec:disc_earlier_result}

Our final lag results are generally consistent with earlier SDSS-RM lag measurements based on shorter baselines \citep[][]{Shen_etal_2016a, Grier_etal_2017,Grier_etal_2019,Homayouni_etal_2020}. However, over the multi-year baseline, some SDSS-RM quasars reported in earlier work have undergone significant luminosity changes, resulting in changes in lags as well. One notable improvement of our final light curves is that we are able to recover many additional lags that were missed in earlier studies (e.g., Fig.~\ref{fig:lag_examp}). Those lags require either longer baselines or the capture of significant variability features for their successful measurements. In terms of RM masses, some quasars have significantly revised values compared with their earlier SDSS-RM results, due to changes in the RMS line width measurements (i.e., PrepSpec has been run on different light-curve baselines), as well as changes in the lag. While there is no overall concern on these final RM masses upon our random inspections, we will perform a more thorough investigation on individual cases and update their RM masses if necessary in the near future. 

On the other hand, the much longer final baselines also led to more aliases in the lag measurements for some objects. One such example is the \CIV\ lag in RM231. In previous work with 4-year SDSS-RM data, \citet{Grier_etal_2019} was able to measure an observed-frame \CIV\ lag around 200~days for this object. Examining the ICCF for the \CIV\ lag measurement in RM231 with the final SDSS-RM data, the peak around 200~days is still there and strong. But there are additional ICCF peaks, and the highest one happened to be in the negative lag regime, leading to a non-detection for this object. Nevertheless, the number of such missed lags is negligible compared with the additional lags recovered with the final SDSS-RM light curves.

\begin{table*} 
\caption{Regression Results }\label{tab:regression}
\centering
$Y=a + bX + {\rm scatter}$
\begin{tabular}{lcccccc}
\hline\hline
Y$|$X & X & X range & $a$ & $b$ & $\sigma_{\rm int}$ & $N_{\rm fit}$ \\
\hline
$\log\tau_{\rm rest,H\beta}$ & $\log L_{\rm 5100,host-corr}$ & [42.87, 45.40]  & $1.458_{-0.038}^{+0.038}$ & $0.41_{-0.07}^{+0.07}$  & $0.32_{-0.03}^{+0.03}$  & 80 \\
$\log\tau_{\rm rest,H\beta}$ & $\log L_{\rm 5100,host-corr}$ & [42.87, 45.40]   & $1.445_{-0.038}^{+0.037}$& $\mathbf{0.533}$  &  $0.32_{-0.03}^{+0.03}$  & 80 \\
$\log\tau_{\rm rest,MgII}$ & $\log L_{\rm 3000}$ & [43.58, 46.28]  & $2.086_{-0.031}^{+0.030}$ & $0.31_{-0.06}^{+0.06}$  & $0.32_{-0.02}^{+0.02}$ & 124 \\
$\log\tau_{\rm rest,MgII}$ & $\log L_{\rm 3000}$ & [43.58, 46.28]  & $2.095_{-0.029}^{+0.030}$ & $\mathbf{0.39}$ & $0.32_{-0.02}^{+0.02}$ & 124 \\
$\log\tau_{\rm rest,CIV}$ & $\log L_{\rm 1350}$ & [44.22, 46.95] & $1.840_{-0.073}^{+0.075}$ & $0.32_{-0.11}^{+0.11}$ & $0.51_{-0.04}^{+0.04}$ & 89 \\
$\log\tau_{\rm rest,CIV}$ & $\log L_{\rm 1350}$ & [44.22, 46.95] & $1.783_{-0.054}^{+0.055}$ & $\mathbf{0.45}$ & $0.51_{-0.04}^{+0.04}$ & 89 \\
\hline
$\log\sigma_{\rm line,rms,H\beta}$ & $\log{\rm FWHM_{mean,H\beta}}$ & [3.16, 4.13] & $0.76_{-0.16}^{+0.17}$ & $0.73_{-0.05}^{+0.05}$  & $0.152_{-0.008}^{+0.009}$ & 183 \\
$\log\sigma_{\rm line,rms,MgII}$ & $\log{\rm FWHM_{mean,MgII}}$ & [3.14, 4.06] & $0.26_{-0.10}^{+0.10}$ & $0.88_{-0.03}^{+0.03}$ & $0.119_{-0.003}^{+0.003}$ & 702 \\
$\log\sigma_{\rm line,rms,CIV}$ & $\log{\rm FWHM_{mean,CIV}}$ & [3.14, 4.05] & $0.91_{-0.10}^{+0.10}$ & $0.69_{-0.03}^{+0.03}$ & $0.099_{-0.003}^{+0.003}$ & 487 \\
\hline
$\log L_{\rm 5100,host-corr}$ & $\log\tau_{\rm rest,H\beta}$ &  [0.61, 2.39]  & $-1.06_{-0.22}^{+0.22}$ & $0.78_{-0.14}^{+0.14}$  & $0.43_{-0.03}^{+0.04}$  & 80 \\
$\log L_{\rm 3000}$ & $\log\tau_{\rm rest,MgII}$ &  [1.01, 2.95]  & $-1.50_{-0.26}^{+0.26}$ & $0.68_{-0.13}^{+0.13}$  & $0.47_{-0.03}^{+0.03}$ & 124 \\
$\log L_{\rm 1350}$ & $\log\tau_{\rm rest,CIV}$ &  [0.83, 3.05] & $-0.15_{-0.20}^{+0.22}$ & $0.30_{-0.10}^{+0.10}$ & $0.49_{-0.04}^{+0.04}$ & 89 \\
\hline
$\log{\rm FWHM_{mean,H\beta}}$ & $\log\sigma_{\rm line,rms,H\beta}$ &  [2.76, 3.89] & $0.88_{-0.17}^{+0.17}$ & $0.80_{-0.05}^{+0.05}$ & $0.160_{-0.008}^{+0.009}$ & 183 \\
$\log{\rm FWHM_{mean,MgII}}$ & $\log\sigma_{\rm line,rms,MgII}$ & [2.73, 3.84] & $1.30_{-0.07}^{+0.07}$  & $0.67_{-0.02}^{+0.02}$  &  $0.103_{-0.003}^{+0.003}$ & 702 \\
$\log{\rm FWHM_{mean,CIV}}$ & $\log\sigma_{\rm line,rms,CIV}$ & [2.87, 3.84] & $0.81_{-0.11}^{+0.11}$  & $0.82_{-0.03}^{+0.03}$  & $0.108_{-0.003}^{+0.004}$  & 487 \\
\hline\\
\end{tabular}

$Y=C + \alpha X1 + \beta X2 + {\rm scatter}$\\
\begin{tabular}{lccccccccc}
\hline\hline
Y$|$X1,X2 & X1 & X2 & $C$ & $\alpha$ & $\beta$ & $\sigma_{\rm int}$ & $N_{\rm fit}$ \\
\hline
$\log M_{\rm SE,H\beta}$ & $\log L_{\rm 5100,host-corr}$ & $\log {\rm FWHM_{mean,H\beta}}$ & 0.85 ($0.75_{-0.05}^{+0.05}$) & $\mathbf{0.5}$ & $\mathbf{2.0}$ & $0.45_{-0.04}^{+0.04}$ & 80 \\
$\log M_{\rm SE,MgII}$ & $\log L_{\rm 3000}$ & $\log {\rm FWHM_{mean,MgII}}$ & $-2.05$ ($-1.97_{-0.04}^{+0.04}$) & $\mathbf{0.6}$ & $\mathbf{3.0}$ & $0.47_{-0.03}^{+0.03}$ & 124 \\
$\log M_{\rm SE,CIV}$ & $\log L_{\rm 1350}$ & $\log {\rm FWHM_{mean,CIV}}$ & 1.40 ($1.31_{-0.06}^{+0.06}$) & $\mathbf{0.5}$ & $\mathbf{2.0}$ & $0.58_{-0.04}^{+0.05}$ & 88 \\
\hline
\end{tabular}
\tablecomments{Default units for $L_{\rm 5100,host-corr}$, $L_{\rm 3000}$, $L_{\rm 1350}$, line widths, time delays, and BH masses are $10^{44}\,{\rm erg\,s^{-1}}$, $10^{45}\,{\rm erg\,s^{-1}}$, $10^{45}\,{\rm erg\,s^{-1}}$, ${\rm km\,s^{-1}}$, days, and $M_\odot$. Linear regressions for the $R-L$ relations and width correlations were performed using the Bayesian approach in \citet{Kelly_2007}. Regressions with fixed slopes were performed using a MCMC model with {\tt emcee}. Line dispersion for \MgII\ has been corrected for the velocity split of the doublet. }
\end{table*}

\subsection{Comparisons with single-epoch estimators}\label{sec:disc_SE}

As mentioned earlier, SE mass estimators are extensively used to estimate quasar black-hole masses near and far \citep[see the review in, e.g.,][]{Shen_2013}. There are a large number of SE calibrations for each line from various groups \citep[e.g.,][]{Greene_Ho_2005,Vestergaard_Peterson_2006,Wang_etal_2009b,Shen_Liu_2012,Ho_Kim_2014,Park_etal_2017,Coatman_etal_2017,DallaBonta_etal_2020,Dix_etal_2023}. The differences in the luminosity and line-width slopes, the adopted virial $f$ factor, and the methodologies of measuring line widths, combine to produce systematic differences among these SE mass recipes even for the same data. There have not been detailed comparisons between SE and RM masses for the same line and for a large statistical sample (e.g., $N>10$), except for \hbeta\ using the local RM AGN sample \citep[e.g.,][]{Vestergaard_Peterson_2006}. 

The validity of single-epoch mass estimators relies on two conditions: (1) a tight $R-L$ relation; (2) a tight correlation between line widths measured from the mean (or single-epoch) spectrum and from the rms spectrum. Condition (2) is generally satisfied for the four major broad lines considered here (cf. Fig.~\ref{fig:line_width_corr}), although anomalous line ``breathing'' behaviors would lead to luminosity-dependent biases in the derived SE masses \citep[e.g.,][]{Wang_etal_2020}. 

However, Condition (1) has only been tested for \hbeta, but not for \MgII\ and \CIV, despite recent efforts in deriving a global $R-L$ relation for \MgII\ and \CIV\ using heterogeneous samples, for which selection functions are not well quantified. As discussed in \S\ref{sec:disc_rm}, there is evidence for a global $R-L$ relation for \hbeta\ and possibly for \MgII\ as well, although it seems additional physical parameters are involved to introduce scatter in these two $R-L$ relations. For \CIV, an $R-L$ relation is only apparent when the dynamic range in luminosity is sufficiently large \citep{Kaspi_etal_2021}. Compared with \hbeta\ and \MgII, there is substantially larger scatter in \CIV\ lags around a mean $R-L$ relation, as probed by the SDSS-RM sample (Fig.~\ref{fig:rl_relation}). As a result, we expect the correlation between RM mass and SE mass would be the worst for \CIV. 

Fig.~\ref{fig:mrm_mse_comp} compares RM and SE masses for \hbeta, \MgII\ and \CIV. For this comparison, we use fiducial SE estimators from \citet{Vestergaard_Peterson_2006} (\hbeta\ and \CIV) and from \citet{Shen_etal_2011} (\MgII).\footnote{The reason that we adopt the \MgII\ SE recipe in \citet{Shen_etal_2011} instead of that in, e.g., \citet{Vestergaard_Osmer_2009}, is that the former yields consistent results compared with both \hbeta\ and \CIV\ SE recipes in \citet{Vestergaard_Peterson_2006} for luminous SDSS quasars. This apparent agreement is likely a coincidence from the methodologies of measuring the mean FWHMs from the spectra. There is no consensus on what specific SE mass recipes are the best, since the calibration sample and spectral fitting methodologies vary from study to study.} As for the luminosity and line width in the SE mass calculation, we adopt the median luminosity over the 90 spectroscopic epochs, and the FWHM measured from the mean spectrum generated from the 90 epochs. By doing so we are ignoring additional scatter in SE masses from anomalous line ``breathing'', where for the same quasar, varying luminosity does not lead to anticipated changes in the single-epoch FWHM value \citep{Wang_etal_2020}. 


Fig.~\ref{fig:mrm_mse_comp} demonstrates that these SE mass recipes can produce average values that are within $\sim 0.2$~dex of the RM masses, with typical scatter of $\sim 0.4-0.5$~dex. This systematic offset can be explained by the differences in the adopted $R-L$ relation, the average $f$ factor, or systematic differences in the line-width measurements. As expected, the general correlation between SE and RM masses is the best for \hbeta, and the worst for \CIV. In fact, using \MgII\ and \CIV\ SE masses could artificially narrow the mass dynamic range. Namely, for a flux-limited quasar sample like the SDSS-RM sample, the dynamic range in mean FWHM is smaller than that in $\sigma_{\rm line,rms}$, and the dynamic range in $L_{3000}$ or $L_{1350}$ is smaller than that in BLR lags. Thus the dynamic range in SE masses would be smaller than the dynamic range in RM (true) masses. These caveats of SE masses for flux-limited samples were discussed in detail in \citet{Shen_2013}, and procedures to account for the resulting statistical biases in the BH mass distribution are outlined in e.g., \citet{Shen_Kelly_2012} and \citet{Kelly_Shen_2013}. 

\subsection{Improved SE mass recipes}\label{sec:se_mass}

Next, we consider potential refinements of SE mass recipes using our new measurements of RM masses for \hbeta, \MgII\ and \CIV. The SE estimator is defined as:
\begin{equation}\label{eqn:SE}
\log M_{\rm SE}\equiv \alpha \log L + \beta\log {\rm FWHM_{mean}} + C\ ,
\end{equation}
where $L$ is the corresponding local continuum luminosity for each line. By default, luminosity is in units of $10^{44}\,{\rm erg\,s^{-1}}$ for $L_{5100}$ and  $10^{45}\,{\rm erg\,s^{-1}}$ for $L_{3000}$ and $L_{1350}$, roughly the midpoint of the luminosity distribution for each line sample. Here, both continuum luminosity and FWHM are measured with negligible measurement uncertainties, since they are the average values during the monitoring period. We only consider ${\rm FWHM_{mean}}$ for the SE mass recipes, given its greater reproducibility compared with the more challenging $\sigma_{\rm line, mean}$ measurements \citep[see discussions in, e.g.,][]{Wang_etal_2019}. As shown in Fig.~\ref{fig:line_width_corr}, there are good correlations between ${\rm FWHM_{mean}}$ and $\sigma_{\rm line, rms}$ used for RM mass calculations, justifying the use of FWHM in SE recipes. Nevertheless, there are physical motivations to use $\sigma_{\rm line}$ instead of FWHM in SE recipes, as reasoned in, e.g., \citet[][]{DallaBonta_etal_2020}.

\begin{figure*}
  \centering
   \includegraphics[width=0.98\textwidth]{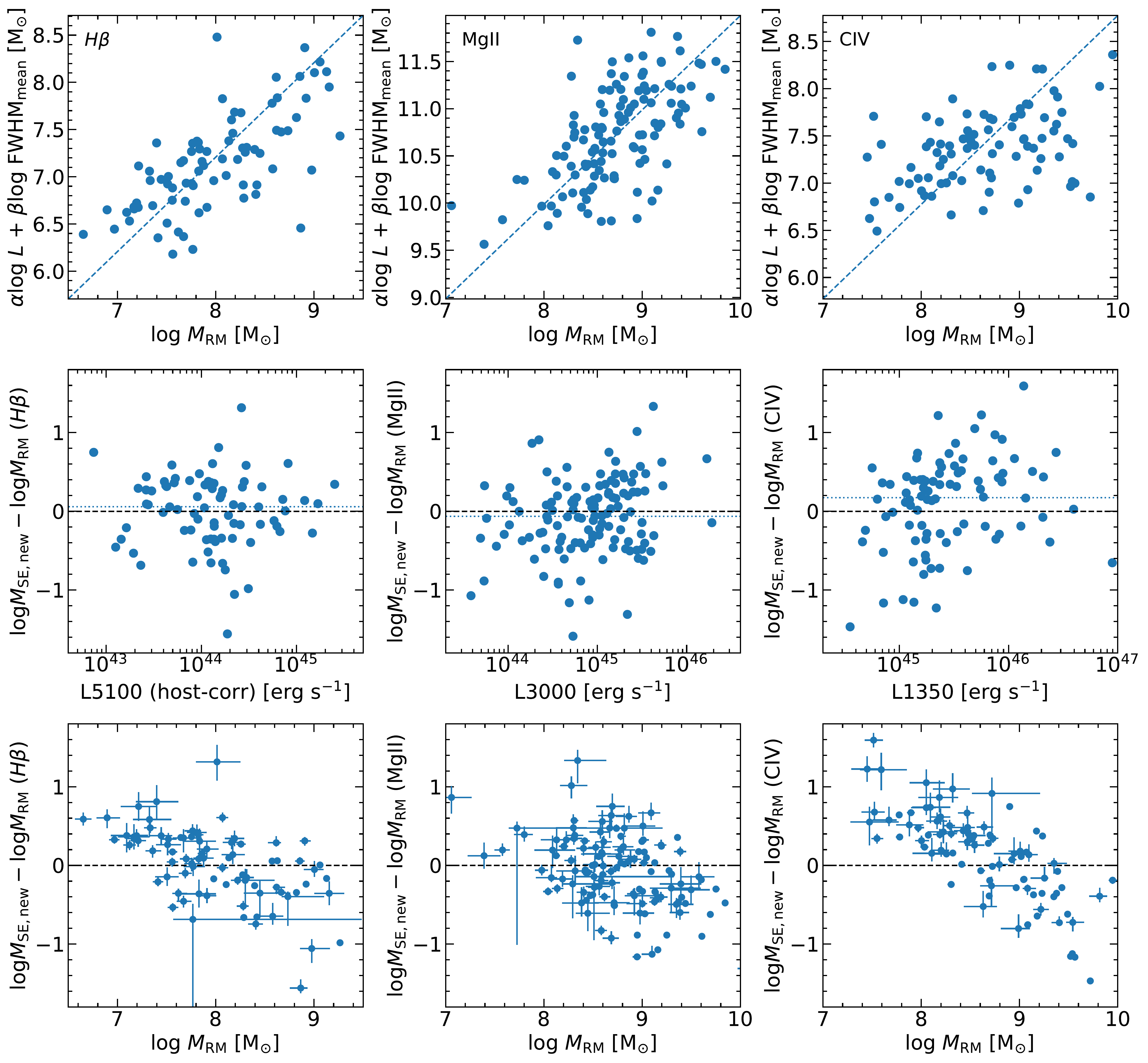}
    \caption{Same as Fig.~\ref{fig:mrm_mse_comp}, but for the comparisons between $M_{\rm RM}$ and updated $M_{\rm SE}$ estimates (Table~\ref{tab:regression}) for the same line. The new \MgII\ SE recipe is now an approximately unbiased estimator (i.e., at a given RM mass, they return an unbiased expectation value), while the \hbeta\ and \CIV\ SE recipes are still biased estimators (as we adopted similar slopes on $L$ and mean FWHM as the earlier recipe in \citet{Vestergaard_Peterson_2006}. However, the normalizations have been recalculated using our RM samples. }
    \label{fig:mrm_mse_comp_new}
\end{figure*}

Because there are correlations between FWHM$_{\rm mean}$ and $\sigma_{\rm line,rms}$, and correlations between continuum luminosity and BLR size for each line, we expect $M_{\rm SE}$ as defined in Eqn.~(\ref{eqn:SE}) will be correlated with $M_{\rm RM}$ in general. With the right choices of slopes $\alpha$ and $\beta$, the calibration of SE mass recipes becomes a 1-parameter fitting problem to constrain the normalization $C$ in Eqn.~(\ref{eqn:SE}) using a sample of quasars with $M_{\rm RM}$ measurements \citep[e.g.,][]{Vestergaard_Peterson_2006}. One option is to adopt slopes that are similar to those in the measured $R-L$ relation and ${\rm FWHM_{\rm mean}}-\sigma_{\rm line,rms}$ correlations. In the work of \citet{Vestergaard_Peterson_2006}, they adopt luminosity slopes of $\alpha\approx 0.5$, and $\beta=2$. The adopted luminosity slopes are consistent with the expected canonical value, and the adopted line-width slopes assume there is a linear relation between ${\rm FWHM_{mean}}$ and $\sigma_{\rm line, rms}$.  

One subtlety in the above calibration procedure is that the regression is performed as matching the observed $\log M_{\rm RM}$ ($Y$) against the model $\log M_{\rm SE}$ ($X$), such that the distribution of $(Y|X)$ is unbiased (i.e., the expectation value $\left<\log M_{\rm RM}\right>=\log M_{\rm SE}$ at fixed $\log M_{\rm SE}$). However, to design an unbiased SE mass estimator, we want $\left<\log M_{\rm SE}\right>=\log M_{\rm RM}$ at fixed true mass $\log M_{\rm RM}$ instead. If the scatter in the $R-L$ relation and in the ${\rm FWHM_{\rm mean}}-\sigma_{\rm line, rms}$ relation is small, it would be adequate to directly use the slope in the $R-L$ relation (predicting $R$ with $L$) and the slope in the ${\rm FWHM_{\rm mean}}-\sigma_{\rm line, rms}$ (predicting $\sigma_{\rm line, rms}$ with ${\rm FWHM_{\rm mean}}$). In reality, there is substantial scatter in these correlations, and the slope of predicting $Y$ at $X$ could differ significantly from that of predicting $X$ at $Y$. These differences in correlation slopes for scattered samples are apparent in Fig.~\ref{fig:line_width_corr} and Fig.~\ref{fig:rl_relation}, as well as the compiled regression results in Table~\ref{tab:regression}. 

So how do we choose slopes of $\alpha$ and $\beta$ for each line? The SDSS-RM lag sample alone does not constrain the $R-L$ relations particularly well to allow fitting $\alpha$ and $\beta$ as free parameters. Instead, we follow \citet{Vestergaard_Peterson_2006} to fix these slopes, and use the lag sample to determine the normalization $C$ in Eqn.~(\ref{eqn:SE}).

For \hbeta\ and \CIV, we adopt $\alpha=0.5$, the canonical slope in the $R-L$ relation, and $\beta=2$. These choices are similar to the choices in \citet{Vestergaard_Peterson_2006}. Adopting $\beta=2$ is reasonably well justified as the bisector relation between ${\rm FWHM_{mean}}$ and $\sigma_{\rm line,rms}$ does appear to be approximately linear for \hbeta\ and \CIV\ (Fig.~\ref{fig:line_width_corr}) -- recall for an unbiased SE estimator, we require the regression relation of ${\rm FWHM_{\rm mean}}|\sigma_{\rm line,rms}$. Our measured $R-L$ relation for \hbeta\ is slightly shallower than but consistent with $\alpha=0.5$ \citep[e.g.,][]{Bentz_etal_2013}. We are unable to robustly constrain the slope of the $R-L$ relation for \CIV, but $\alpha=0.5$ is roughly consistent with the findings in \citet{Kaspi_etal_2021}. The comparison between the uncalibrated $M_{\rm SE}$ and the measured $M_{\rm RM}$ for \hbeta\ and \CIV\  are shown in the top row of Fig.~\ref{fig:mrm_mse_comp_new}. For \hbeta, the correlation is nearly diagonal, indicating our choices of $\alpha$ and $\beta$ are reasonable. For \CIV, however, the correlation is poor, mainly resulting from the large scatter in the \CIV\ $R-L$ relation. 

For \MgII, we adopt somewhat different slopes of $\alpha$ and $\beta$. The measurement of the \MgII\ $R-L$ relation is still in an early stage, and a slope different from 0.5 is possible. In addition, the ${\rm FWHM_{mean}}-\sigma_{\rm line,rms}$ is more non-linear than those for \hbeta\ and \CIV\ (Fig.~\ref{fig:line_width_corr}). This non-linearity implies $\beta$ should not be 2, since the fiducial RM masses are computed as $\propto \sigma_{\rm line,rms}^2$. We found that in order to produce the most consistent results, we require $\beta>2$ and $\alpha>0.5$ for \MgII. After experimenting, we adopt $\alpha=0.6$ and $\beta=3.0$ for \MgII. This luminosity slope is not too different from the canonical value, and this ${\rm FWHM_{\rm mean}}$ slope is consistent with the measured slope in predicting ${\rm FWHM_{\rm mean}}$ with $\sigma_{\rm line,rms}$ (see Table~\ref{tab:regression}). Compared with the canonical case of $\alpha=0.5$ and $\beta=2$, our fiducial \MgII\ recipe produces: (1) a less biased estimator at fixed $M_{\rm RM}$; (2) more consistent results with \hbeta\ and \CIV\ SE masses for the common objects, over the luminosity range probed by SDSS quasars. 

Given the chosen slopes of $\alpha$ and $\beta$, we show the uncalibrated $\log M_{\rm SE}$ against $\log M_{\rm RM}$ in the top row of Fig.~\ref{fig:mrm_mse_comp_new}. There are good correlations between the SE mass and the RM mass for \hbeta\ and \MgII, and a weak correlation for \CIV. We then constrain the normalization $C$ using a MCMC model that incorporates a term for intrinsic scatter, in addition to the measurement uncertainties in $M_{\rm RM}$. The constrained ranges of $C$ are listed in Table~\ref{tab:regression}. We adopt fiducial $C$ values that are slightly different (but consistent within 2$\sigma$) from the MCMC results. These tweaked normalizations produce the best agreement between the SE masses from two lines for the same quasars.  

The middle row in Fig.~\ref{fig:mrm_mse_comp_new} displays the residuals in $\log$ mass as a function of continuum luminosity, which show no significant luminosity trend. The bottom row displays the residuals as a function of $\log M_{\rm RM}$. By design, the \MgII\ SE estimator is approximately unbiased, with no significant correlation between the mass residual and $M_{\rm RM}$. For \hbeta\ and \CIV, however, the adopted slopes are appropriate for producing unbiased $\log M_{\rm RM}$ at given $\log M_{\rm SE}$. As a consequence, there are systematic biases in $\log M_{\rm SE}$ as a function of $\log M_{\rm RM}$ for \hbeta\ and \CIV, as also seen in \citet{Vestergaard_Peterson_2006}. Nevertheless, this SE mass bias is much less severe for \hbeta\ than for \CIV. It is possible to design less biased SE estimators for \CIV\ and for \hbeta, which we plan to investigate with more RM data from upcoming MOS-RM programs \citep[e.g.,][]{Kollmeier_etal_2017}.

Our results thus signify the severe limitations and caveats of SE BH mass recipes, especially for \CIV. It is our opinion that individual quasar studies that require robust BH mass estimates should not rely too heavily on \CIV\ SE masses. On the other hand, \hbeta\ and \MgII\ SE masses seem to reproduce the RM masses well, with an additional $\sim 0.45$~dex scatter.

\subsection{Caveats}

The dynamical time of the BLR is typically a few years to a few decades for luminous SDSS quasars. While rms quasar variability is on average $\sim 10\%$ over multi-year timescales, a small fraction of quasars display significantly larger variability amplitude, and sometimes even monotonically increasing or decreasing light curves over time. It is possible that the structure of the BLR in some hypervariable quasars undergoes significant changes over multi-year timescales. 

In the SDSS-RM sample, several quasars have notably increased their luminosity over the past few years by more than a factor of a few \citep[e.g., RM017, RM160, etc.][]{Dexter_etal_2019,Fries_etal_2023}. While we have attempted to measure an average lag over the full SDSS-RM baseline, we found that the measured lags are substantially longer than those measured from their earlier light curves when they were at a lower luminosity state \cite[][]{Grier_etal_2017}. It would be interesting to study how the BLR sizes have evolved when luminosity changes significantly over a timescale comparable to the dynamical time of the BLR, and see if there is evidence for structural changes in the BLR, in addition to the anticipated breathing effect of the BLR. 

{A related issue is whether or not one should remove a long-term (typically assumed linear) trend in the light curves before measuring the lag. Linear trends in the light curve do not contribute to the short-term cross-correlated variations of interest but may tilt the resulting CCF, and therefore there is an argument for detrending before measuring the lag. On the other hand, the removal of a linear trend from the data is somewhat arbitrary: the long-term trend may not be linear but using higher-order detrending cannot be easily justified; the RMS spectra are constructed without detrending; and the removed trend may be part of a real long-term signal in the BLR response. For these reasons, we opt to measure lags without detrending, as the multi-year baseline is typically much longer than the damping timescale of the light curve for our targets. Nevertheless, we perform a test of measuring ICCF lags with (linearly) detrended light curves for our fiducial lag sample. We did not find significant differences for the bulk of lags w/ and w/o detrending. }

Another important caveat is that the overall detection rate for SDSS-RM quasars is low (see Table~\ref{tab:lag_sum}), as necessarily limited by the light curve quality of the project. Selection biases from incompleteness in lag detection may have consequences on the inferred $R-L$ relations. In Appendix~\ref{sec:inform_future} we investigate the detection incompleteness for \hbeta, \MgII\ and \CIV. We found that the success of a lag detection primarily depends on how well we can measure the line variability, e.g., via the SNR2 metric discussed in \S\ref{sec:lc_var}. Most of the lags are undetected simply because their line variability is not well measured (e.g., ${\rm SNR2<20}$). 

For \hbeta, we found that the lag detection rate is very high ($\gtrsim 70\%$) once SNR2 reaches $\sim 35$, and there is no obvious difference in quasar properties (e.g., expected lag, quasar luminosity, line equivalent width, line width, etc.) between detected and undetected cases matched in SNR2. These lag non-detections are caused by random fluctuations of the light curves that led to weak cross-correlations in the light curves. Indeed, our visual inspection of non-detection cases with ${\rm SNR2}>40$ suggests that most of the cases still have some hint of a lag, but the lag is usually small and below the $2\sigma$ threshold for our lag detection. There are a handful of cases where the \hbeta\ line is not responding to continuum variations in the expected way, which will be investigated in future work. Thus we conclude there is limited impact on the \hbeta\ $R-L$ relation from detection incompleteness. 

For \MgII\ and \CIV, we found similar trends as for \hbeta, and no obvious differences in quasar properties between the detections and non-detections. However, their lag detection fractions are notably lower than that for \hbeta\ at fixed SNR2. In Appendix~\ref{sec:inform_future} we elaborate on the potential overestimation of SNR2 for \MgII\ and \CIV\ due to the insignificant improvement on the spectrophotometry with PrepSpec for these high-redshift quasars. In addition, it is possible that \MgII\ and \CIV\ in high-redshift and high-luminosity quasars are more often to show abnormal responses to continuum variations, making it more difficult to detect a lag. 

Finally, there is sill room to improve the lag measurement methodology, spectrophotometry refinement with PrepSpec, and scrutiny on individual light curves. While these additional efforts are beyond the scope of the current work, we plan to investigate these issues as we continue monitoring the SDSS-RM field in SDSS-V with the Black Hole Mapper program \citep{Kollmeier_etal_2017}. In particular, we plan to improve the spectrophotometry at the blue end of SDSS spectra, which would improve the lag detection for \MgII\ and \CIV, as well as other UV broad lines, in high-redshift quasars.

\subsection{Applications of SDSS-RM data}\label{sec:app}

The rich data from SDSS-RM enable a broad array of applications. The large number of RM lags and reliable lag-based RM masses have proven valuable to study the evolution of the BH mass-host galaxy correlations toward high redshift \citep[e.g.,][]{Shen_etal_2015b,Matsuoka_etal_2015,Li_etal_2021a,Li_etal_2023}. By extending RM measurements to high redshift and measuring large samples of \MgII\ and \CIV\ lags, we are starting to explore the diversity in the BLR structure for these UV broad lines, and to critically evaluate their potential for designing efficient single-epoch BH mass recipes.  

In the meantime, the multi-year photometric and spectroscopic light curves for the SDSS-RM sample allow investigations of the general variability of quasars \citep{Sun_etal_2015,Dyer_etal_2019}, and of the variability of particular subsets of quasars, such as broad absorption line quasars \citep[][]{Grier_etal_2015,Hemler_etal_2019} and extreme variability quasars \citep{Dexter_etal_2019,Fries_etal_2023}. In addition to the unprecedented spectroscopic monitoring data that allow detailed spectral variability analyses, the SDSS-RM sample provides high-quality multi-wavelength information \citep{Shen_etal_2019b,Liu_etal_2020} for a well defined quasar sample. These data have enabled additional RM science, such as constraining the accretion disk structure using continuum lags \citep[][]{Homayouni_etal_2019} and modeling the propagation of variability in the accretion disk \citep[e.g.,][]{Stone_Shen_2023}.

In this paper we focused on the main science goal of SDSS-RM, i.e., the measurements of lags for major broad emission lines in quasar spectra. We will explore other topics with the final SDSS-RM data in future work. We also invite the community to exploit the rich SDSS-RM data for RM science or general AGN science.  

\section{Summary and Outlook}\label{sec:con}

\subsection{Main Findings}

This work marks the official conclusion of the SDSS-RM program, although we will continue to analyze this data set. We present the final SDSS-RM data, including 7-yr spectroscopy (2014-2020) and 11-yr photometry (2010-2020). Using this data set, we measure velocity-integrated average RM lags for \halpha, \hbeta, \MgII\ and \CIV, over broad ranges of redshifts ($0.1<z<4.5$) and quasar luminosities ($L_{\rm bol}=10^{44-47.5}\,{\rm erg\,s^{-1}}$). We report 23, 81, 125 and 110 lags for each of the four lines, respectively. The main findings are summarized as follows. 

\begin{enumerate}

\item[$\bullet$] The measured lags from different lines in the same quasars reveal BLR stratification, consistent with earlier findings based on low-$z$ RM work. Specifically, \halpha\ and \MgII\ have on average longer lags than \hbeta, while \CIV\ lags are somewhat shorter than \MgII\ lags overall (but the sample size is small). This result suggests increasing distances of the (observable) emitting clouds from the BH in the order of \CIV$<$\hbeta$<$\halpha$\lesssim$\MgII. [See \S\ref{sec:disc_rm}.]

\item[$\bullet$] The measured RM BH masses are consistent between two different lines, albeit with different lags and line widths for both lines (\S\ref{sec:disc_rm}). This validates the basic principle of measuring BH mass with the RM technique. In calculating the RM masses, we have derived a new estimate of the average (geometric mean) virial factor of $\left<\log f\right>=0.62\pm0.07$ (when using $\sigma_{\rm line,rms}$ as line width), based on measurements of dynamical-modeling BH masses in 30 low-$z$ RM AGNs. The intrinsic scatter in individual virial factors is $0.31\pm0.07$~dex, indicating a factor of two systematic uncertainties in RM masses calculated using the average virial factor. [See \S\ref{sec:f_factor}.]

\item[$\bullet$] There is a significant global $R-L$ relation for \hbeta\ and \MgII\ in SDSS-RM quasars, using $R=c\tau_{\rm rest}$. The best-fit slopes and normalizations are consistent with latest measurements using heterogeneous samples \citep{Bentz_etal_2013,Yu_etal_2023}. The slopes of the \hbeta\ and \MgII\ $R-L$ relations are also consistent with the canonical value of 0.5 expected from photoionization. The intrinsic scatter in the $R-L$ relations (in the $R$ direction) is $\sim 0.3$~dex. Because the luminosity dynamic range for the SDSS-RM sample is still somewhat limited, it is possible that the constrained slopes are biased from the global slope across many orders of magnitude in AGN luminosity. [See \S\ref{sec:disc_rm}.]

\item[$\bullet$] Given the limited dynamic range in luminosity for \CIV, we are unable to well constrain the slope of an assumed $R-L$ relation. However, our slope measurements are consistent with the latest results \citep{Kaspi_etal_2021} based on a heterogeneous sample that spans a much larger luminosity range than the SDSS-RM sample. More importantly, the intrinsic scatter in the said $R-L$ relation for \CIV\ is substantially larger than that for \hbeta\ or \MgII, likely caused by the larger dispersion in the ratio between \CIV\ ionizing flux and 1350\,\AA\ continuum flux in quasars. [See \S\ref{sec:disc_rm}.]

\item[$\bullet$] We present re-calibrated single-epoch BH mass recipes for \hbeta, \MgII\ and \CIV, using the SDSS-RM lag sample. The new \hbeta\ and \MgII\ SE recipes are approximately unbiased estimators of RM mass (i.e., expectation value $\left<\log M_{\rm SE} \right>=\log M_{\rm RM}$), with an intrinsic scatter of $\sim 0.45$~dex around RM masses. The \CIV\ SE recipe, on the other hand, shows substantially larger scatter ($\sim 0.58$~dex) and the weakest correlation with RM masses. Nevertheless, these new SE recipes show overall consistency across two lines over the luminosity range probed by SDSS-RM quasars. Considering the systematic uncertainty of $\sim 0.3$~dex in RM masses (\S\ref{sec:f_factor}), one can argue that the absolute uncertainties of SE masses are $\sim 0.35$~dex for \hbeta\ and \MgII, and $\sim 0.5$~dex for \CIV. Our findings continue to support the usage of \hbeta\ and \MgII\ SE masses, but caution on the usage of \CIV\ SE masses for high-redshift quasars. The three SE mass recipes are summarized below. [See \S\ref{sec:se_mass}.]

\end{enumerate}

\begin{eqnarray}
\log\bigg(\displaystyle\frac{M_{\rm SE,H\beta}}{M_\odot}\bigg)=&& \log\bigg[\bigg( \displaystyle\frac{L_{\rm 5100,AGN}}{10^{44}\,{\rm erg\,s^{-1}}}\bigg)^{0.5}\bigg(\displaystyle\frac{\rm FWHM}{\rm km\,s^{-1}}\bigg)^{2}\bigg] \nonumber\\
&&+ 0.85 \\
\nonumber \\
\log\bigg(\displaystyle\frac{M_{\rm SE,MgII}}{M_\odot}\bigg)=&& \log\bigg[\bigg( \displaystyle\frac{L_{\rm 3000}}{10^{45}\,{\rm erg\,s^{-1}}}\bigg)^{0.6}\bigg(\displaystyle\frac{\rm FWHM}{\rm km\,s^{-1}}\bigg)^{3}\bigg] \nonumber\\
&&- 2.05  \\ 
\nonumber \\
\log\bigg(\displaystyle\frac{M_{\rm SE,CIV}}{M_\odot}\bigg)=&& \log\bigg[\bigg( \displaystyle\frac{L_{\rm 1350}}{10^{45}\,{\rm erg\,s^{-1}}}\bigg)^{0.5}\bigg(\displaystyle\frac{\rm FWHM}{\rm km\,s^{-1}}\bigg)^{2}\bigg] \nonumber\\
&&+ 1.40  \
\end{eqnarray}

\subsection{Outlook}

The results from the SDSS-RM project unambiguously confirm the feasibility and potential of performing MOS-RM in the high-redshift regime. \hbeta\ continues to be the most reliable line for RM purposes. We have shown that it is possible to measure \MgII\ lags with optical spectroscopic RM, but the increased \MgII\ lag and diluted RM response make it more difficult to robustly measure the lag. Indeed, the \MgII\ lag detection rate is the lowest among the four lines studied in this work (see Table~\ref{tab:lag_sum}). We also demonstrated that measuring \CIV\ lags in high-redshift, high-luminosity quasars is possible, provided that the monitoring baseline is sufficient to constrain the long time delays. 

The final lag yields from SDSS-RM are broadly consistent with our predictions at the beginning of the project \citep{Shen_etal_2015a}. The forecast was made using simulated light curves of quasars with physical properties and observing cadence/baseline/depth similar to those in SDSS-RM, but the lag detection methodology and criteria in the forecast were somewhat different. Nevertheless, this overall consistency demonstrates the important value of planning a MOS-RM program with tailored simulations. In Appendix~\ref{sec:inform_future}, we provide additional lessons learned from SDSS-RM. 

Given the success of SDSS-RM and other MOS-RM programs, there are several ways to improve RM studies in the high-$z$ regime. First, the typical quality of lag measurements from current MOS-RM programs is still low-to-moderate. This is not a limitation of the principles of the MOS-RM approach, but a limitation from insufficient observing resources. Future MOS-RM programs with larger aperture telescopes \citep[e.g.,][]{Swann_etal_2019,MSE} will provide higher S/N spectroscopy to better measure the reverberation responses in the broad line flux. These future MOS-RM programs will also be able to target fainter quasars (thus extending the sample to lower luminosities), and measure velocity-resolved lags to better constrain dynamical models of the BLR. In anticipation of these future MOS-RM programs, there are also continued efforts to expand the sample size of high-redshift lag measurements with existing MOS facilities, such as the SDSS-V Black Hole Mapper Reverberation Mapping program during 2020--2026 \citep{Kollmeier_etal_2017}. In particular, the SDSS-RM field continues to be monitored in SDSS-V, providing extended light curve baselines for a significant fraction of SDSS-RM quasars. 

Besides the optical MOS-RM approach, there are also important complementary approaches to advance RM in the high-$z$ regime. To extend the $R-L$ relations to the most luminous quasars (e.g., $L_{\rm bol}\gtrsim 10^{47}\,{\rm erg\,s^{-1}}$), it is necessary to perform decade-long monitoring campaigns for individual objects \citep[e.g.,][]{Lira_etal_2018,Czerny_etal_2019,Zajacek_etal_2020,Kaspi_etal_2021}, since these most luminous quasars are rare and sparsely distributed on the sky. The high success rate of measuring \hbeta\ lags at high-$z$ and the overall best reliability of \hbeta\ for RM mass measurements suggests the possibility of IR reverberation mapping for \hbeta\ at $z>1$. Such IR RM programs would require large-aperture telescopes to measure the \hbeta\ response with IR spectroscopy. 

Finally, the advent of high-spatial-resolution spectroastrometric measurements with interferometry, e.g., with the GRAVITY instrument on the VLTs, has enabled a new avenue toward constraining the inner structures of broad-line AGNs and quasars \citep[e.g.,][]{gravity_2018,gravity_2020,gravity_2021}. With the upgraded GRAVITY+ instrument \citep{gravity+}, it will become possible to target relatively faint (e.g., $K\sim 13-15$) quasars at $z>1$. Combining the spectroastrometric measurements \citep[e.g.,][]{Bailey_1998,Bosco_etal_2021} and RM measurements for the same quasar will provide powerful joint constraints on the underlying BLR structure and kinematics \citep[e.g.,][]{LiY_etal_2022}, an approach that can also be used to measure geometric distances toward distant AGNs \citep[e.g.,][]{Elvis02,Honig_2014,Wang_etal_2020NatAs,gravity_2021,Songsheng_etal_2021}. Far down the road, it may eventually become feasible to measure astrometric jitter signals due to reverberation \citep[e.g.,][]{Shen_2012,Li_Wang_2023}. These various techniques motivate continued monitoring programs to constrain AGN inner structures, measure BH masses, and improve our understanding of the growth of SMBHs and the expansion history of the universe. 

\software{astropy \citep{astropy}, celerite \citep{celerite}, emcee \citep{emcee}, {\tt PyCCF} \citep{pyccf}, {\tt Javelin} (v0.33) \citep{Zu_etal_2011}, {\tt PyROA} (v3.1.0) \citep{pyroa}
}

\acknowledgments

We thank the referee for comments that improved the manuscript, and Pu Du, Yan-Rong Li, and Aaron Barth for useful discussions. YS acknowledges support from NSF grants AST-1715579 and  AST-2009947. CJG acknowledges support from NSF grants AST-2009949 and AST-2108667. JIL is supported by the Eric and Wendy Schmidt AI in Science Postdoctoral Fellowship, a Schmidt Futures program. YH was supported as an Eberly Research Fellow by the Eberly College of Science at the Pennsylvania State University. JRT acknowledges support from NSF grants CAREER-1945546, AST-2009539, and AST-2108668. WNB acknowledges support from NSF grant AST-2106990 and the Eberly Endowment at Penn State. LCH was supported by the National Science Foundation of China (11721303, 11991052, 12011540375, 12233001) and the China Manned Space Project (CMS-CSST-2021-A04, CMS-CSST-2021-A06). CT acknowledges Tsinghua University for the support to her work. 

This work is based on observations obtained with MegaPrime/MegaCam, a joint project of CFHT and CEA/DAPNIA, at the Canada-France-Hawaii Telescope (CFHT) which is operated by the National Research Council (NRC) of Canada, the Institut National des Sciences de l'Univers of the Centre National de la Recherche Scientifique of France, and the University of Hawaii.
The authors recognize the cultural importance of the summit of Maunakea to a broad cross section of the Native Hawaiian community. The astronomical community is most fortunate to have the opportunity to conduct observations from this mountain.

SDSS-III is managed by the Astrophysical Research Consortium for the
Participating Institutions of the SDSS-III Collaboration including the
University of Arizona, the Brazilian Participation Group, Brookhaven National
Laboratory, University of Cambridge, Carnegie Mellon University, University
of Florida, the French Participation Group, the German Participation Group,
Harvard University, the Instituto de Astrofisica de Canarias, the Michigan
State/Notre Dame/JINA Participation Group, Johns Hopkins University, Lawrence
Berkeley National Laboratory, Max Planck Institute for Astrophysics, Max
Planck Institute for Extraterrestrial Physics, New Mexico State University,
New York University, Ohio State University, Pennsylvania State University,
University of Portsmouth, Princeton University, the Spanish Participation
Group, University of Tokyo, University of Utah, Vanderbilt University,
University of Virginia, University of Washington, and Yale University.

Funding for the Sloan Digital Sky Survey IV has been provided by the Alfred P. Sloan Foundation, the U.S. Department of Energy Office of Science, and the Participating Institutions. SDSS-IV acknowledges support and resources from the Center for High-Performance Computing at the University of Utah. The SDSS web site is www.sdss.org. SDSS-IV is managed by the Astrophysical Research Consortium for the Participating Institutions of the SDSS Collaboration including the Brazilian Participation Group, the Carnegie Institution for Science, Carnegie Mellon University, the Chilean Participation Group, the French Participation Group, Harvard-Smithsonian Center for Astrophysics, Instituto de Astrof\'isica de Canarias, The Johns Hopkins University, Kavli Institute for the Physics and Mathematics of the Universe (IPMU) / University of Tokyo, the Korean Participation Group, Lawrence Berkeley National Laboratory, Leibniz Institut f\"ur Astrophysik Potsdam (AIP),  
Max-Planck-Institut f\"ur Astronomie (MPIA Heidelberg), 
Max-Planck-Institut f\"ur Astrophysik (MPA Garching), 
Max-Planck-Institut f\"ur Extraterrestrische Physik (MPE), 
National Astronomical Observatories of China, New Mexico State University, 
New York University, University of Notre Dame, 
Observat\'ario Nacional / MCTI, The Ohio State University, 
Pennsylvania State University, Shanghai Astronomical Observatory, 
United Kingdom Participation Group,
Universidad Nacional Aut\'onoma de M\'exico, University of Arizona, 
University of Colorado Boulder, University of Oxford, University of Portsmouth, 
University of Utah, University of Virginia, University of Washington, University of Wisconsin, 
Vanderbilt University, and Yale University.

The PS1 has been made possible through contributions by the Institute for Astronomy, the University of Hawaii, the Pan-STARRS Project Office, the Max-Planck Society and its participating institutes, the Max Planck Institute for Astronomy, Heidelberg and the Max Planck Institute for Extraterrestrial Physics, Garching, The Johns Hopkins University, Durham University, the University of Edinburgh, Queen's University Belfast, the Harvard-Smithsonian Center for Astrophysics, the Las Cumbres Observatory Global Telescope Network Incorporated, the National Central University of Taiwan, the Space Telescope Science Institute, the National Aeronautics and Space Administration under Grant No. NNX08AR22G issued through the Planetary Science Division of the NASA Science Mission Directorate, the National Science Foundation under Grant No. AST-1238877, the University of Maryland, and Eotvos Lorand University (ELTE). 

ZTF: Based on observations obtained with the Samuel Oschin 48-inch Telescope at the Palomar Observatory as part of the Zwicky Transient Facility project. ZTF is supported by the National Science Foundation under grant No. AST-1440341 and a collaboration including Caltech, IPAC, the Weizmann Institute for Science, the Oskar Klein Center at Stockholm University, the University of Maryland, the University of Washington, Deutsches Elektronen-Synchrotron and Humboldt University, Los Alamos National Laboratories, the TANGO Consortium of Taiwan, the University of Wisconsin at Milwaukee, and Lawrence Berkeley National Laboratories. Operations are conducted by COO, IPAC, and UW.

\appendix

\section{Format of Data Products}\label{sec:data_format}

The final SDSS-RM data set and related data products are provided via anonymous ftp and https services\footnote{{\bf ftp://quasar.astro.illinois.edu/public/sdssrm/final\_result/, https://ariel.astro.illinois.edu/sdssrm/final\_result/}}
 (or contact Y.~Shen for latest archival info). We organize the data products by the object ID (RMID). In each object directory (e.g., {\em ./rm000/}), we provide the original spectra, continuum and emission-line light curves, PrepSpec outputs, and lag measurements. Additional information about the SDSS-RM sample is provided in the sample characterization paper \citet{Shen_etal_2019b}. Table~\ref{tab:content} provides a brief description of the data products available. Future updates of these products will also be distributed via this site. Below we describe the main data products released with this paper. 


{\bf SDSS spectra}\quad Customarily reprocessed \citep[e.g.,][]{Shen_etal_2015a} optical spectroscopy from SDSS is provided in the {\em ./spec/} directory in ascii format. A total of 90 epochs are available for each object. These are the original spectra before the PrepSpec run. However, these spectra were customarily reprocessed as described in \citet{Shen_etal_2015a}, thus are slightly different from the spectra included in the SDSS-III/IV public data releases. 

{\bf Continuum and emission-line light curves}\quad The merged continuum light curve and emission-line light curves (after processed by PrepSpec) are provided in plain ascii files with the {\em \_lc.txt} affix in each object directory (e.g., {\em ./rm000/}). These light curves are used in the lag measurements. The emission-line light curve files only include \halpha, \hbeta, \MgII, and \CIV, with prefixes {\em ha, hb, mg2} and {\em c4}, respectively. Additional emission-line light curves, if covered by spectroscopy, will be included in the {\em ./prepspec/} directory in the future. We are aware that in rare occasions, the merging of the continuum light curves failed, with discrepant fluxes from different facilities. These corrupted light curves usually do not produce successful lags, and we did not attempt to resolve these flux discrepancies in this work. 

{\bf PrepSpec outputs}\quad Model light curves, flux calibration scaling factor ({\em \_p0\_t.dat}), mean and rms spectra, and line width measurements from {PrepSpec} run are provided in the {\em ./prepspec/} directory. These outputs include all available lines covered in SDSS spectra. The flux calibration scaling factor {\em p0\_t} is used to scale the SDSS spectra to improve spectrophotometry. A subset of the PrepSpec outputs are also compiled in the summary Table~\ref{tab:sum}.

{\bf Lag measurements}\quad Outputs from our lag analyses are provided in the {\em ./pypetal/} directory in each object directory. These outputs include the DRW fits to the continuum light curves and outlier masks, ICCFs, {\tt Javelin}/\pyroa results, and weights applied to the lag calculations. The essential subset of these measurements are compiled in the summary Table~\ref{tab:sum}. For convenience, we provide a lag summary plot in each object directory (e.g., {\em ./rm000/lag\_summary.pdf}) to display the lag measurements for all available lines. 

{\bf Software}\quad Scripts of running lag detection and post-processing are provided in the {\em ./scripts/} directory.

\begin{table*}
\caption{Content of Data Products}\label{tab:content}
\centering
\begin{tabular}{ll}
\hline\hline
File Path & Description  \\
\hline
{\em ./} & Top directory \\
{\em ./spec/} & Directory for SDSS spectra (before PrepSpec) \\
{\em ./prepspec/} & Directory for PrepSpec outputs \\ 
{\em ./rm*/} & Object directory for each SDSS-RM quasar \\
{\em ./scripts/} & Batch scripts and additional software tools \\
\textbf{{\em ./summary.fits}} & \textbf{Summary table (Table~\ref{tab:sum})} \\
\hline\\
\end{tabular}

Content of {\em ./prepspec/} \\
\begin{tabular}{ll}
\hline\hline
File Path & Description  \\
\hline
{\em ./rm*/} & Object directory \\
\quad{\em -- ./rm*\_p0\_t.dat} &  Scaling factor of flux calibration (tabulated in $\ln p_0(t)$) \\
\quad{\em -- ./rm*\_avg\_w.dat} & Average spectrum \\
\quad{\em -- ./rm*\_rms\_w.dat} & RMS spectrum; column 2 is the `RMSx' estimate \citep{Shen_etal_2016a}  \\
\quad{\em -- ./rm*\_\$cont\$.dat} & Continuum light curve (e.g., `c3000', `c5100') \\
\quad{\em -- ./rm*\_\$cont\$\_t\_stats.dat} & Statistics of the continuum light curve \\
\quad{\em -- ./rm*\_vblr.dat} & Line width measurements from the average spectrum \\
\quad{\em -- ./rm*\_vvblr.dat} & Line width measurements from the rms spectrum \\
\quad{\em -- ./rm*\_\$line\$\_t.dat} & Line light curve \\
\quad{\em -- ./rm*\_\$line\$\_w.dat} & RMS line profile \\
\quad{\em -- ./rm*\_\$line\$\_t\_stats.dat} & Statistics of the line light curve \\
\hline\\
\end{tabular}

Content of {\em ./rm*/} \\
\begin{tabular}{ll}
\hline\hline
File Path & Description  \\
\hline
{\em ./lag\_summary.pdf} & Summary plot of lag measurements \\
{\em ./cont\_lc.txt} & Merged continuum light curve \\
{\em ./\$line\$\_lc.txt} & Original line light curve (first three columns in PrepSpec output {\em rm*\_\$line\$\_t.dat}) \\
{\em ./pypetal/} & Lag measurement directory \\
\quad{\em -- ./light\_curves/} & Copied original light curves with flags for outliers added \\
\quad{\em -- ./processed\_lcs/} & Outlier-rejected light curves and light curves formatted for \javelin and \pyroa \\
\quad{\em -- ./cont/drw\_rej/} & Continuum DRW fitting result and outlier mask \\
\quad{\em -- ./\$line\$/} & Lag measurements for the line \\
\quad\quad{\em -- ./pyccf/} & ICCF results \\
\quad\quad{\em -- ./javelin/} & \javelin results \\
\quad\quad{\em -- ./pyroa/} & \pyroa results \\
\quad\quad{\em -- ./weights/} & Weights for lag posteriors \\
\hline
\hline\\
\end{tabular}

\tablecomments{There are multiple files inside each directory. The content and format of these files are either self-explanatory or explicitly specified in the header. Occasionally, the `lag\_summary.pdf' plot needs to be regenerated for better display ranges using the scripts provided in the {\em ./scripts/} directory. }
\end{table*}

\section{Informing future MOS-RM Programs}\label{sec:inform_future}

\begin{figure*}
  \centering
   \includegraphics[width=0.48\textwidth]{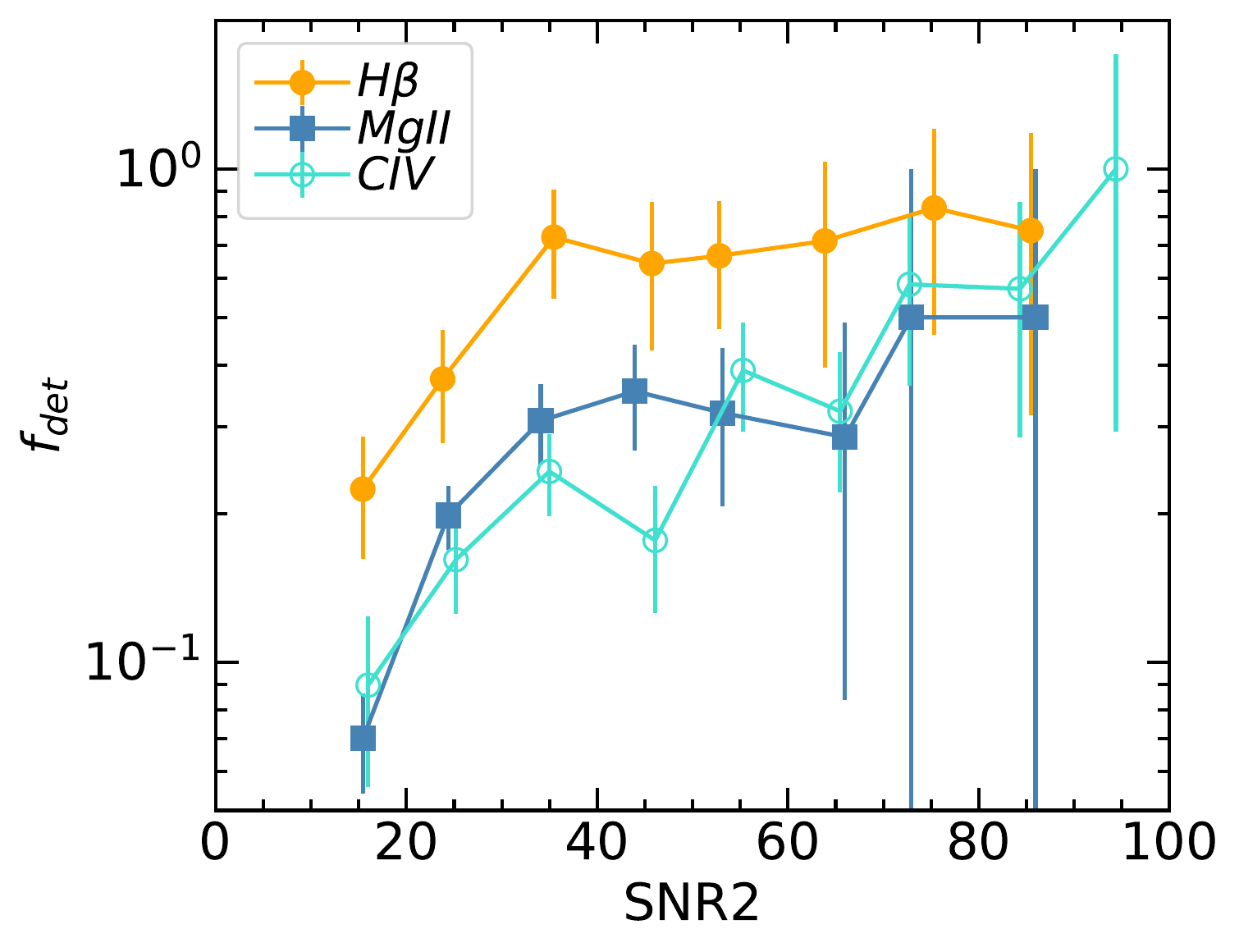}
    \caption{Lag detection fraction as a function of line variability parameter SNR2, for \hbeta, \MgII\ and \CIV, respectively. Error bars are Poisson counting uncertainties. }
    \label{fig:fdet}
\end{figure*}

The lag yields from the final SDSS-RM data are roughly in line with the simulation forecast in \citet{Shen_etal_2015a}, despite the slightly different total baseline and lag detection criteria assumed in the forecast, and the fact that the predictions were based on simulated light curves. In Fig.~\ref{fig:fdet}, we show the final lag detection fraction as a function of the line variability SNR2 parameter described in \S\ref{sec:lc_var} \citep[also see][]{Shen_etal_2019b}. For a future MOS-RM program, it is necessary to have a sufficient number of epochs to sample variability patterns over a sufficiently long (e.g., multi-years) spectroscopic baseline. However, in order to detect the line variability properly, each spectroscopic epoch must also achieve a significant S/N for the line flux measurement. A value of ${\rm SNR2}=20$ with the benchmark SDSS-RM cadence/baseline (with $N_{\rm epoch}\approx 100$) approximately corresponds to ${\rm line\ RMS}/\left<\sigma_i\right>\approx 2$, where line RMS is the rms line flux variability (absolute not fractional), and $\left<\sigma_i\right>$ is the average per-epoch flux uncertainty. Thus for a $20\%$ fractional line variability, i.e., ${\rm RMS/AVG}=0.2$, we require a fractional line flux measurement uncertainty of $\left<\sigma_i\right>/{\rm AVG}=10\%$ in order to reach ${\rm SNR2=20}$. Deeper epoch spectra, e.g., with 5\% flux measurements, will increase SNR2 to $\sim 40$ and subsequently boost the probability of lag detection (Fig.~\ref{fig:fdet}). 

For MOS spectroscopy, there is also a systematic floor of flux uncertainties limited by the spectrophotometry accuracy of the MOS program. For example, in SDSS-RM, we achieved a typical systematic flux uncertainty floor of $\sim 5\%$. Using PrepSpec and for quasars with strong narrow lines (such as \OIII) at low redshift, we were able to further improve the spectrophotometry to $\sim 2-3\%$. This additional improvement is likely partially responsible for the higher lag detection rate for \hbeta\ at fixed SNR2 (computed using statistical flux uncertainties only), compared with \MgII\ and \CIV\ for which the PrepSpec improvement on spectrophotometry is marginal. Moreover, the systematic uncertainty of spectrophotometry increases toward the blue end of SDSS spectra \citep[e.g.,][]{Shen_etal_2015a}, resulting in a larger impact on lag detection for \MgII\ and \CIV\ in high-redshift quasars than for \hbeta\ in the low-redshift subset. 

The outcome from the SDSS-RM final data set thus provides useful guidance for future MOS-RM programs with similar baseline and cadence. Given the typical $\sim 10\%$ RMS variability of broad lines, a low systematic floor of spectrophotometry will greatly improve the lag detection rate. At the same time, we recommend to reach a statistical (per epoch) line flux uncertainty of better than $\sim 10\%$ at the flux limit of the sample. For SDSS-RM, the latter goal is achieved with 2~hr integration per epoch at a limiting magnitude of $i_{\rm psf}<21.7$, for the general quasar population.

\section{Additional Quality Cuts on Lags}\label{sec:quality_cuts}

{As discussed in Sec~\ref{sec:alias}, we impose a set of quantitative cuts for lag detection. Our choice of the $r_{\rm max}>0.4$ cut is more lenient than those used in other high-quality RM studies. However, while imposing this less stringent cut inevitably introduces more false positives in the lag sample, it also ensures that we are not removing lags near the detection boundary that could potentially bias the $R-L$ relation measurements. }

\begin{figure*}
  \centering
   \includegraphics[width=0.8\textwidth]{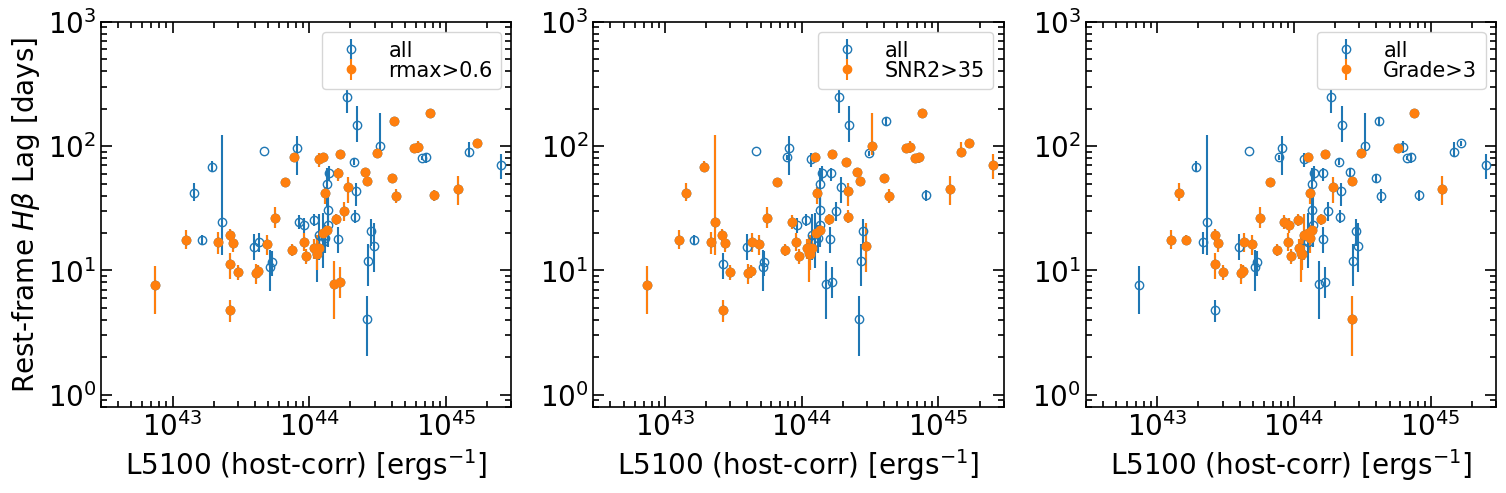}
   \includegraphics[width=0.8\textwidth]{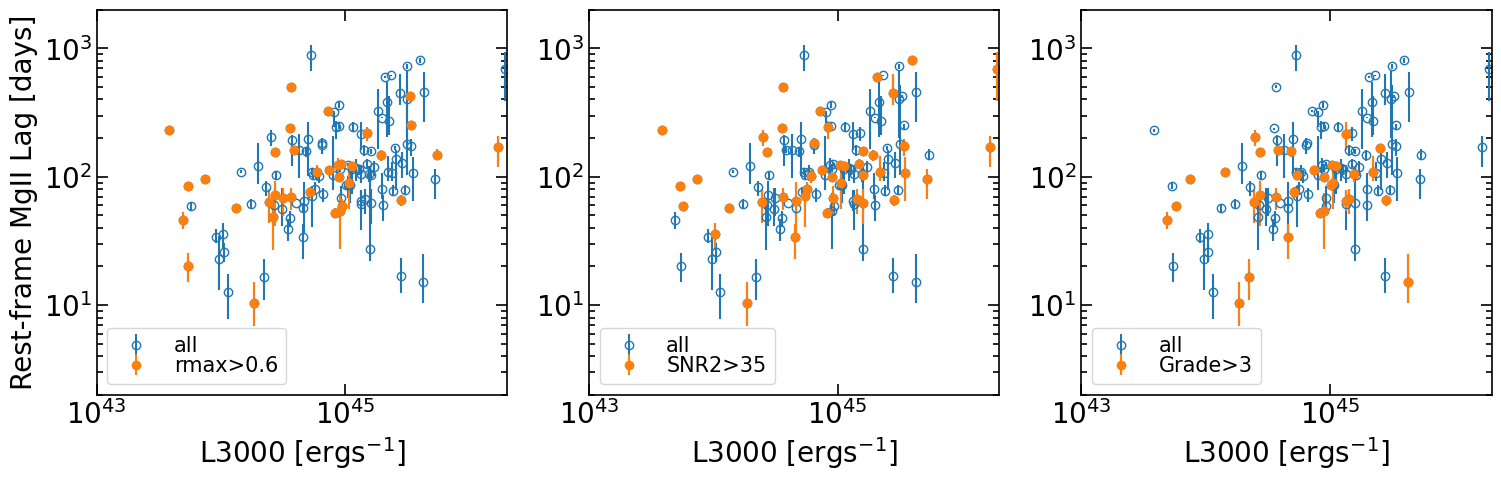}
   \includegraphics[width=0.8\textwidth]{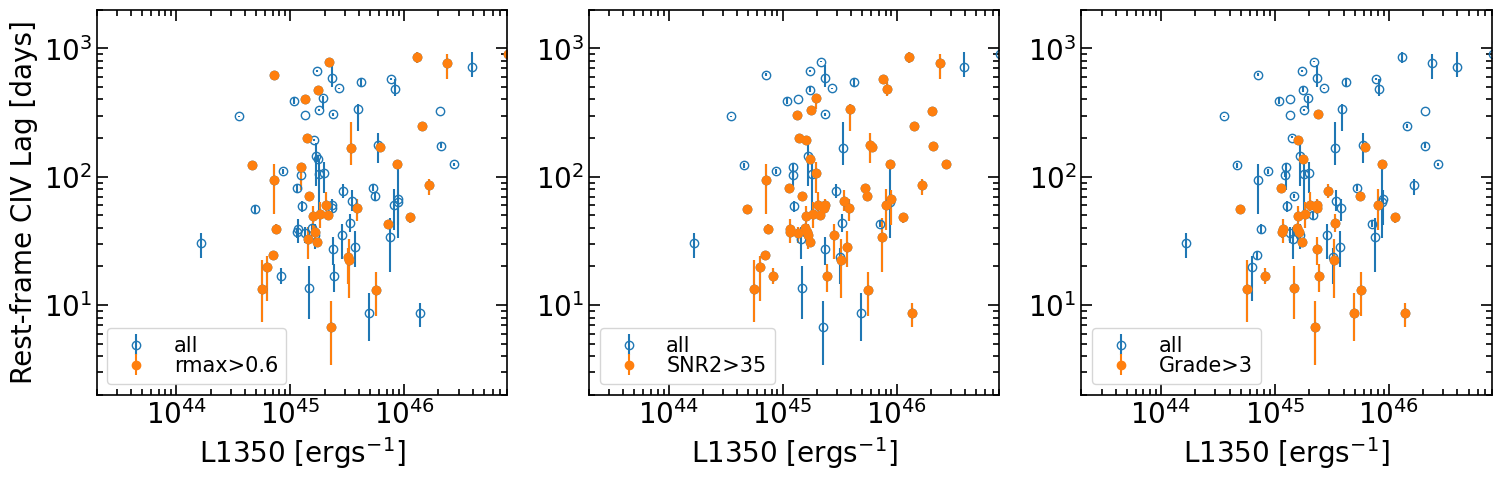}
    \caption{Tests of lag detection with more stringent quality cuts. In each panel, the open symbols represent the fiducial lag sample, and the filled symbols are for the subset. The left column shows results with a $r_{\rm max}>0.6$ cut; the middle column shows results with a ${\rm SNR2}>35$ cut; the right column shows the results with the manual inspection cut. Details regarding these quality cuts and their impact on the measured $R-L$ relations are presented in Appendix~\ref{sec:quality_cuts}. }
    \label{fig:qa_test}
\end{figure*}

{We test the effects of imposing more stringent quality cuts on lag detection. Fig.~\ref{fig:qa_test} shows the comparison in the $R-L$ plane with our fiducial lag results for: (1) a higher $r_{\rm max}>0.6$ cut; (2) a ${\rm SNR2>35}$ cut; (3) high-quality lags from manual inspection as detailed in \S\ref{sec:inspection}. These additional quality cuts ($r_{\rm max}$, SNR2, grade) roughly reduce the fiducial lag sample by half for each line.}

{As shown in Fig.~\ref{fig:qa_test}, for the quantitative cuts with $r_{\rm max}$ and SNR2, we generally found consistent results on the best-fit $R-L$ relation in terms of the slope, normalization and intrinsic scatter. For the visual inspection grade cut, however, the \MgII\ and \CIV\ $R-L$ relations are notably more flattened and shifted to lower normalizations. This is because long lags with limited light curve overlap are generally less likely to be perceived as a good-quality detection. As a result, manual inspection preferentially removes long lags and bias the measurement of the $R-L$ relation. This highlights the danger of cherrypicking the best-quality lags in the $R-L$ relation measurements from a systematic RM program such as ours. }

{For full disclosure, we provide the best-fit $R-L$ relation for these higher-quality lag subsets in Table~\ref{tab:regression_cuts}, but we strongly caution that these relations are not intended for any practical uses for the reasons discussed above. }

\begin{table*} 
\caption{Regression Results for Higher-quality Lags}\label{tab:regression_cuts}
\centering
$Y=a + bX + {\rm scatter}$\\
\begin{tabular}{lccccccc}
\hline\hline
Cut & Y$|$X & X & X range & $a$ & $b$ & $\sigma_{\rm int}$ & $N_{\rm fit}$ \\
\hline
$r_{\rm max}>0.6$& $\log\tau_{\rm rest,H\beta}$ & $\log L_{\rm 5100,host-corr}$ & [42.87, 45.22]  & $1.432_{-0.047}^{+0.044}$ & $0.54_{-0.09}^{+0.08}$  & $0.28_{-0.03}^{+0.04}$  & 44 \\
${\rm SNR2}>35$& $\log\tau_{\rm rest,H\beta}$ & $\log L_{\rm 5100,host-corr}$ & [42.87, 45.40]  & $1.450_{-0.041}^{+0.041}$ & $0.43_{-0.07}^{+0.07}$  & $0.27_{-0.03}^{+0.03}$  & 48 \\
Grade$=4/5$ & $\log\tau_{\rm rest,H\beta}$ & $\log L_{\rm 5100,host-corr}$ & [43.10, 45.08]  & $1.418_{-0.048}^{+0.047}$ & $0.42_{-0.11}^{+0.10}$  & $0.28_{-0.03}^{+0.04}$  & 37 \\ 
\hline
$r_{\rm max}>0.6$ & $\log\tau_{\rm rest,MgII}$ & $\log L_{\rm 3000}$ & [43.58, 46.23]  & $2.079_{-0.061}^{+0.060}$ & $0.21_{-0.09}^{+0.10}$  & $0.32_{-0.04}^{+0.05}$ & 35 \\
${\rm SNR2}>35$ & $\log\tau_{\rm rest,MgII}$ & $\log L_{\rm 3000}$ & [43.58, 46.28]  & $2.111_{-0.053}^{+0.053}$ & $0.22_{-0.09}^{+0.09}$  & $0.33_{-0.04}^{+0.04}$ & 44 \\
Grade$=4/5$ & $\log\tau_{\rm rest,MgII}$ & $\log L_{\rm 3000}$ & [43.69, 45.63]  & $1.935_{-0.058}^{+0.056}$ & $0.09_{-0.11}^{+0.11}$  & $0.27_{-0.04}^{+0.05}$ & 32 \\
\hline
$r_{\rm max}>0.6$ & $\log\tau_{\rm rest,CIV}$ & $\log L_{\rm 1350}$ & [44.66, 46.95] & $1.752_{-0.119}^{+0.121}$ & $0.47_{-0.16}^{+0.16}$ & $0.54_{-0.06}^{+0.08}$ & 36 \\
${\rm SNR2}>35$  & $\log\tau_{\rm rest,CIV}$ & $\log L_{\rm 1350}$ & [44.69, 46.95] & $1.665_{-0.086}^{+0.085}$ & $0.47_{-0.12}^{+0.12}$ & $0.43_{-0.04}^{+0.05}$ & 57 \\
Grade$=4/5$ & $\log\tau_{\rm rest,CIV}$ & $\log L_{\rm 1350}$ & [44.69, 46.14] & $1.634_{-0.112}^{+0.113}$ & $0.005_{-0.219}^{+0.219}$ & $0.41_{-0.05}^{+0.07}$ & 32 \\
\hline\\
\end{tabular}
\tablecomments{Same as Table~\ref{tab:regression} but for subsets of higher-quality lags as described in Appendix~\ref{sec:quality_cuts}. Because these more stringent cuts impose additional selection functions on the resulting lag sample, we caution that these best-fit $R-L$ relations are most likely biased, and therefore are not for practical uses. }
\end{table*}

\section{Lag Examples}\label{sec:lag_examples}

{We provide several additional examples for \hbeta, \MgII\ and \CIV\ lags in our fiducial lag sample that represent the range of detection qualities based on our visual inspection. We caution that our visual inspection is subjective on individual cases, but on average higher-grade lags are more robustly detected. Lower-grade lags not necessarily mean the lag is spurious, and may reflect the complexity of the BLR and the limitations of measuring an average lag. }

\begin{figure*}
  \centering
   \includegraphics[width=0.8\textwidth]{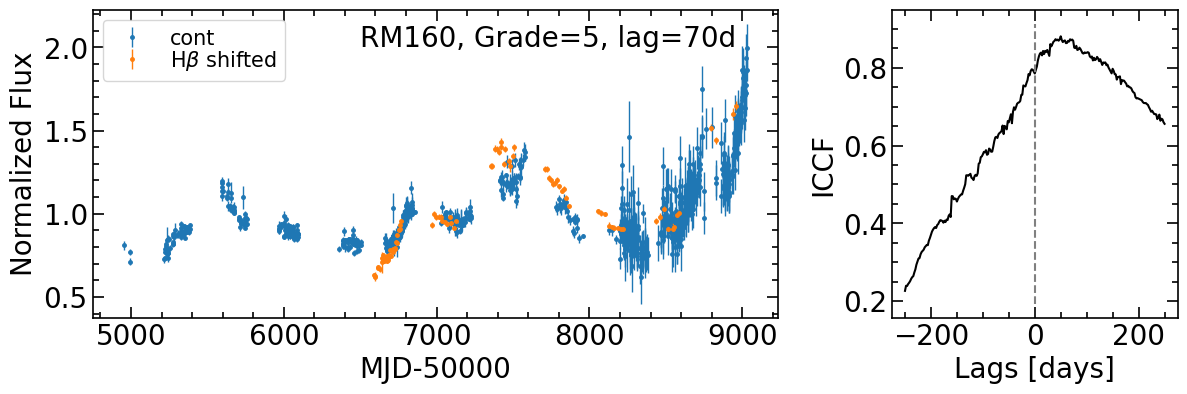}
   \includegraphics[width=0.8\textwidth]{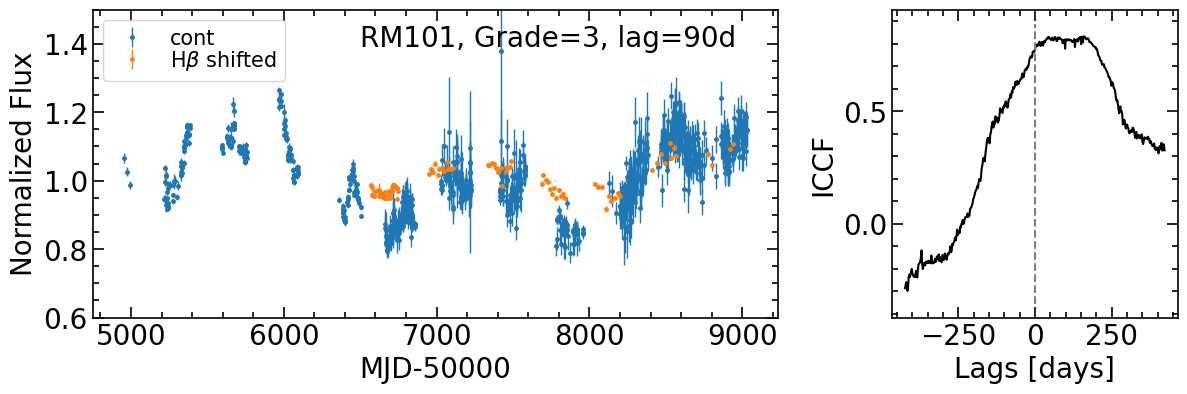}
   \includegraphics[width=0.8\textwidth]{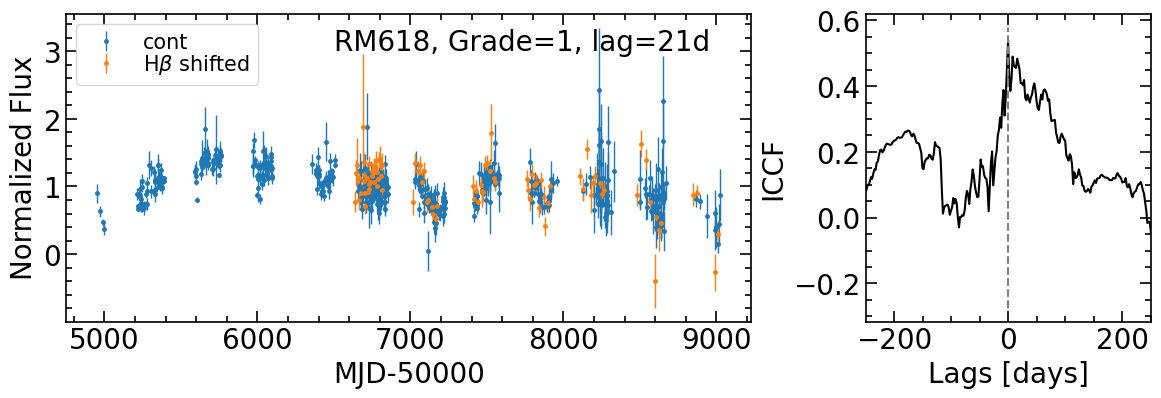}
    \caption{Additional examples of \hbeta\ lags (light curves in the left panel and ICCF in the right panel). Higher (visual inspection) grades correspond to better measured lags. The continuum and broad-line light curves are normalized to have unity mean within their respective baseline, and therefore they do not necessarily line up after shifting the broad-line light curve by the observed-frame \pyroa lag. }
    \label{fig:more_lags_hb}
\end{figure*}

\begin{figure*}
  \centering
   \includegraphics[width=0.8\textwidth]{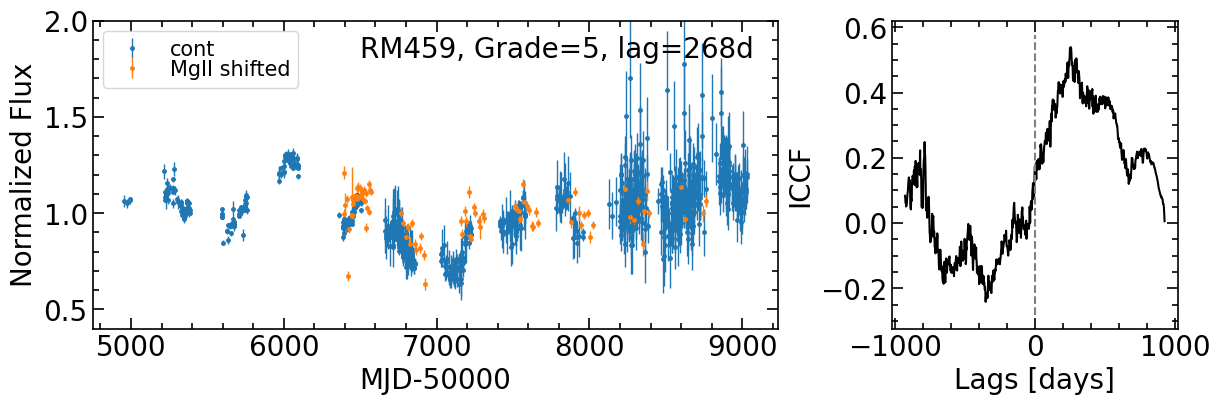}
   \includegraphics[width=0.8\textwidth]{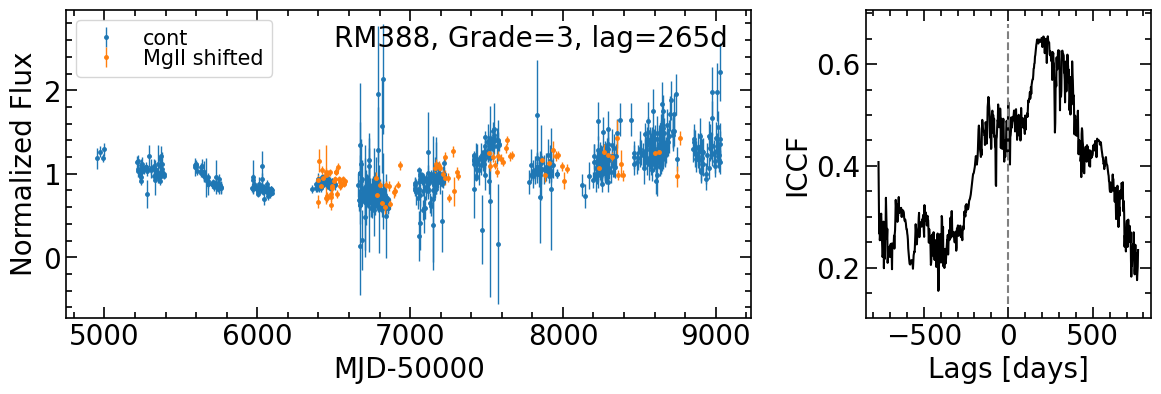}
   \includegraphics[width=0.8\textwidth]{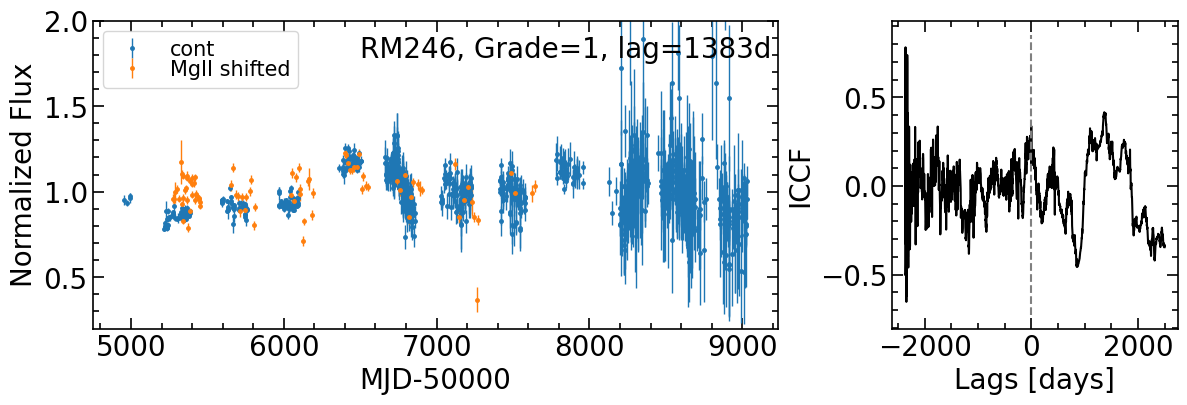}
    \caption{Additional examples of \MgII\ lags. Higher (visual inspection) grades correspond to better measured lags. }
    \label{fig:more_lags_mg2}
\end{figure*}

\begin{figure*}
  \centering
   \includegraphics[width=0.8\textwidth]{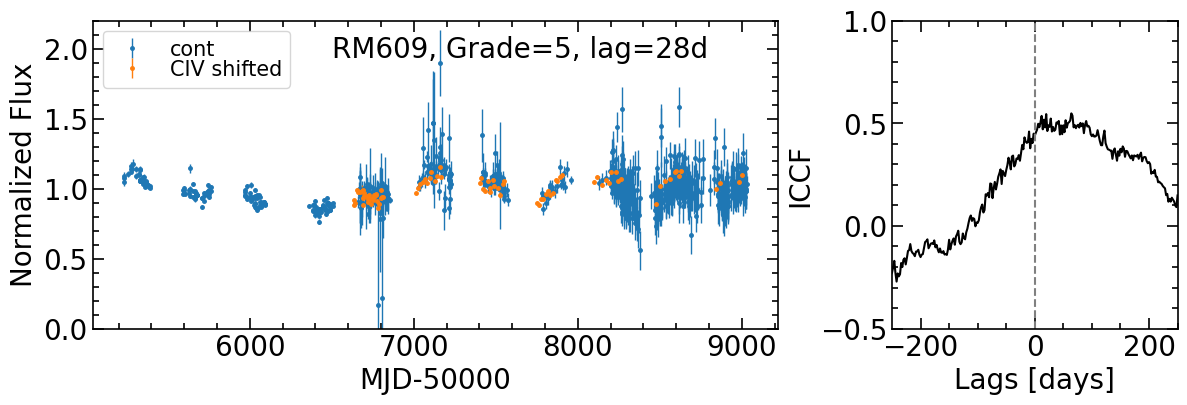}
   \includegraphics[width=0.8\textwidth]{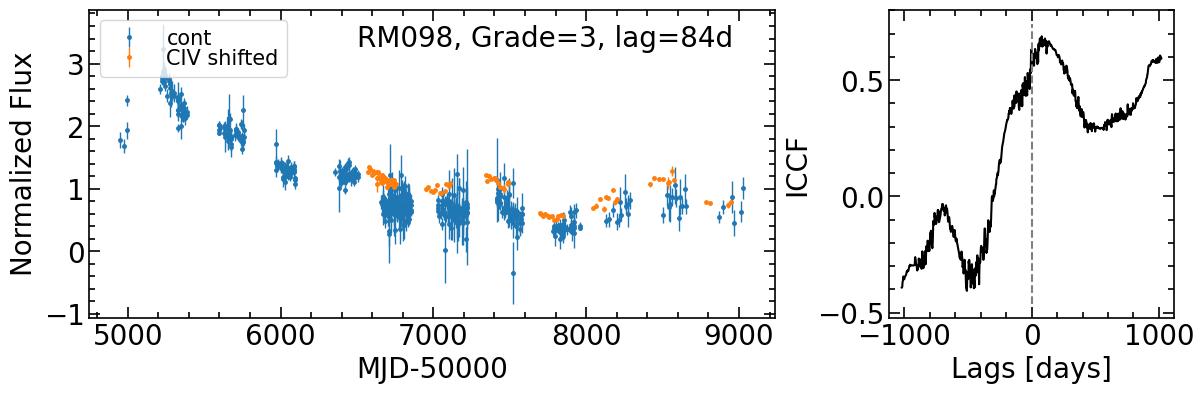}
   \includegraphics[width=0.8\textwidth]{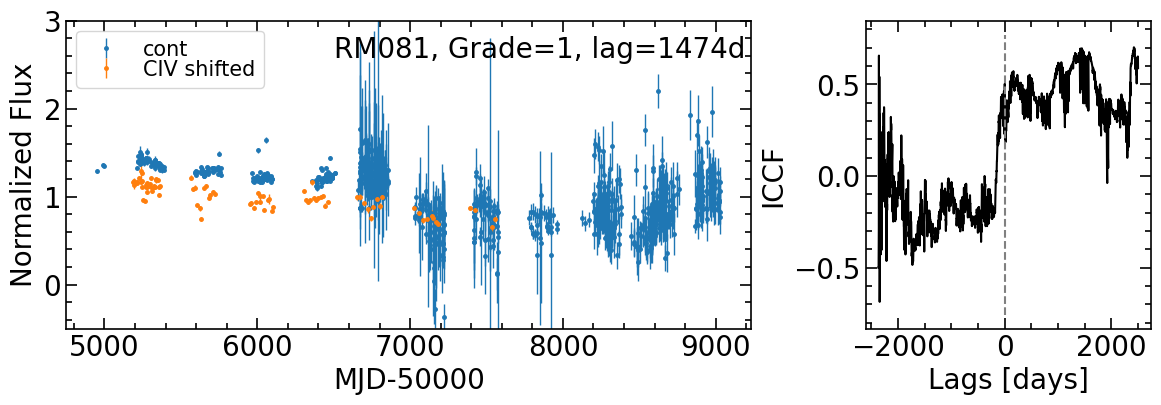}
    \caption{Additional examples of \CIV\ lags. Higher (visual inspection) grades correspond to better measured lags. }
    \label{fig:more_lags_c4}
\end{figure*}

\bibliography{./final_results_submit}

\end{document}